\documentclass{aa}  

\usepackage{graphicx}
\graphicspath{{figs/}}
\usepackage{hyperref}
\usepackage[all]{hypcap}
\usepackage[capitalise]{cleveref}
\usepackage[separate-uncertainty]{siunitx}
\usepackage{txfonts}
\usepackage{xcolor}
\usepackage{ulem}

\begin{document} 

   \title{Examining the local Universe isotropy with galaxy cluster velocity dispersion scaling relations}

   \author{A. Pandya \inst{1}
          \and K. Migkas \inst{2,3}
          \and T. H. Reiprich \inst{1}
          \and A. Stanford \inst{4}
          \and F. Pacaud \inst{1}
          \and G. Schellenberger \inst{5}
          \and L. Lovisari \inst{{6,5}}
          \and M. E. Ramos-Ceja \inst{7}
          \and N. T. Nguyen-Dang \inst{8}
          \and S. Park \inst{{9,10}}
          }

   \institute{Argelander-Institut f\"ur Astronomie (AIfA), Universit\"at Bonn, Auf dem H\"ugel 71, 53121 Bonn, Germany \\ e-mail: \texttt{apandya@astro.uni-bonn.de}
   \and
   Leiden Observatory, Leiden University, PO Box 9513, 2300 RA Leiden, the Netherlands
   \and
   SRON Netherlands Institute for Space Research, Niels Bohrweg 4, NL-2333 CA Leiden, the Netherlands
   \and
   University of California, Davis, CA 95616, USA
   \and 
   Center for Astrophysics $|$ Harvard \& Smithsonian, 60 Garden St., Cambridge, MA 02138, USA
   \and
   INAF, Istituto di Astrofisica Spaziale e Fisica Cosmica di Milano, via A. Corti 12, 20133 Milano, Italy
   \and 
   Max Planck Institute for Extraterrestrial Physics, Giessenbachstrasse 1, 85748 Garching, Germany
   \and 
   Institute for Astronomy and Astrophysics of T\"ubingen (IAAT), Germany.
   \and
   Korea Astronomy and Space Science Institute, Daejeon 34055, Republic of Korea
   \and 
   University of Science and Technology, Daejeon 34113, Republic of Korea
   }

   \date{Received ...; accepted ...}

% \abstract{}{}{}{}{} 
% 5 {} token are mandatory
 
  \abstract
   {In standard cosmology, the late Universe is assumed to be statistically homogeneous and isotropic. This assumption suggests that the expansion rate of the Universe, as measured by the Hubble parameter, should be the same in all directions. However, a recent study based on galaxy clusters by \citet{K.Migkas_21} found an apparent spatial variation of approximately $\sim$$9\%$ in the Hubble constant, $H_0$, across the sky. In the study, the authors utilised galaxy cluster scaling relations between various cosmology-dependent cluster properties and a cosmology-independent property, i.e., the temperature of the intracluster gas ($T$). A position-dependent systematic bias of $T$ measurements can, in principle, result in an overestimation of apparent $H_0$ variations. Therefore, it is crucial to confirm or exclude this possibility.}
   {In this work, we search for directional $T$ measurement biases by examining the relationship between the member galaxy velocity dispersion and gas temperature ($\sigma_\mathrm{v}-T$) of galaxy clusters. Both measurements are independent of any cosmological assumptions and do not suffer from the same potential systematic biases. Additionally, we search for apparent $H_0$ angular variations independently of $T$ by analysing the relations between the X-ray luminosity and Sunyaev-Zeldovich signal with the velocity dispersion, $L_\mathrm{X}-\sigma_\mathrm{v}$ and $Y_\mathrm{SZ}-\sigma_\mathrm{v}$.}
   {To study the angular variation of scaling relation parameters, we determine the latter for different sky patches across the extra-galactic sky. We constrain the possible directional $T$ bias using the $\sigma_\mathrm{v}-T$ relation, as well as the apparent $H_0$ variations using the $L_\mathrm{X}-\sigma_\mathrm{v}$ and $Y_\mathrm{SZ}-\sigma_\mathrm{v}$ relations. We utilise Monte Carlo simulations of isotropic cluster samples to quantify the statistical significance of any observed anisotropies. We find and rigorously take into account a correlation of $L_\mathrm{X}$ and $Y_\mathrm{SZ}$ residuals.}
   {No significant directional $T$ measurement biases are found from the $\sigma_\mathrm{v}-T$ anisotropy study. The probability that the previously observed $H_0$ anisotropy is caused by a directional $T$ bias is only $0.002\%$. On the other hand, from the joint analysis of the $L_\mathrm{X}-\sigma_\mathrm{v}$ and $Y_\mathrm{SZ}-\sigma_\mathrm{v}$ relations, the maximum variation of $H_0$ is found in the direction of $(295^\circ\pm71^\circ, -30^\circ\pm71^\circ)$ with a statistical significance of $3.64\sigma$, fully consistent with \citet{K.Migkas_21}.}
  % conclusions
   {Our findings, based on the analysis of new scaling relations utilising a completely independent cluster property, $\sigma_\mathrm{v}$, strongly corroborate the previously detected spatial anisotropy of galaxy cluster scaling relations. The underlying cause, e.g., a spatial $H_0$ variation or large-scale bulk flows of matter, remains to be identified.}

   \keywords{Galaxies: clusters: general --
                X-rays: galaxies: clusters --
                Cosmology: observations --
                large-scale structure of Universe
               }

   \maketitle
%
%-------------------------------------------------------------------

\section{Introduction}
Galaxy clusters are the largest gravitationally virialised systems in the Universe. They are crucial tools for astrophysical and cosmological studies as they provide valuable information about the large-scale structure and evolution of the Universe \citep{Pratt_19}. Galaxy clusters can be observed in multiple wavelengths, providing insights into different cluster properties. Galaxy cluster scaling relations describe the correlation between various important cluster properties using simple power laws \citep{Giodini_13, Lovisari_22}. These relations were predicted first by \citet{Kaiser_86}, and the model predicts that the objects of different sizes are merely scaled versions of each other. Due to this, it is also referred to as the self-similar model.

The evaluation of certain cluster observable properties involves cosmological assumptions. For instance, the X-ray luminosity ($L_\mathrm{X}$) and the total integrated Compton parameter ($Y_\mathrm{SZ}$) rely on cosmological distances inferred from the estimated cluster redshift. The relation between these two quantities is a function of the assumed cosmological model. However, the measurement of some properties, such as intracluster gas temperature, $T$, and galaxy velocity dispersion, $\sigma_\mathrm{v}$, depend very weakly on such cosmological assumptions. Valuable insights about different aspects of cosmology can be obtained by utilising the scaling relations between these two types of cluster properties (for a recent review, see \citet{Migkas_24}).

The cosmological principle is a fundamental assumption of the standard cosmological model, and it states that the Universe is homogeneous and isotropic on sufficiently large scales. 
The Friedmann equations -- the basic set of equations that underpins most cosmological models -- are derived under this assumption.
The $\Lambda$CDM model assumes that matter should converge to an isotropic behaviour for cosmic volumes greater than $\SI{150}{Mpc}$. This implies that the cosmic expansion rate, cosmological parameters, and the distances to extra-galactic objects depend solely on redshift, $z$, regardless of direction. Therefore, given the fundamental importance of this assumption, it is necessary to assess the validity of it.

The cosmic microwave background (CMB) observations support the cosmological principle within the CMB rest frame, exhibiting remarkable isotropy at small angular scales \citep{Bennett_94, Bennett_13, Planck_20}. However, the CMB provides us with the rest frame of cosmic radiation in the early Universe, whereas examining the (an)isotropic behaviour of the matter rest frame in the local Universe (e.g., galaxies, clusters, and supernovae at $z\lesssim 0.2-0.3$) through CMB data is particularly challenging and not extensively explored, with very few exceptions \citep[e.g.,][]{Yeung_22}. Therefore, assessing if cosmic matter behaves isotropically in the late Universe is crucial. 
% (more information in \citet{Peebles_22}). 
Thus, additional cosmological probes are required to test the cosmological principle in the late Universe. Some of these probes include the use of Type Ia supernovae (SNIa) \citep{Appleby_15}, infrared quasars \citep{Secrest_22}, the distribution of optical \citep{Javanmardi_17, Sarkar_19} and infrared galaxies \citep{Yoon_14, Rameez_18} and the distribution of distant radio sources \citep{Singal_11, Rubart_13, Tiwari_16, Colin_19} and gamma-ray bursts \citep{Ripa_17, Andrade_19}. 

Galaxy clusters have become a useful tool to test the cosmological principle in recent years.
A new method based on cluster scaling relations was introduced by \citet{Migkas_18}.
\citet{K.Migkas_20} (hereafter M20) and \citet{K.Migkas_21} (hereafter M21) then applied various scaling relations for galaxy clusters to examine the isotropy of the local Universe quantitatively. The key principle behind these studies involved pairing a cosmology-dependent quantity (e.g., $L_\mathrm{X}$ and $Y_\mathrm{SY}$) with the cosmology-independent $T$. This pairing allowed them to draw conclusions on the directionality of cosmological parameters. 
They found a $(9.0\pm1.7)\%$ variation in the Hubble constant ($H_0$) across the sky which could be alternatively attributed to a $\sim$$\SI{900}{km\ s^{-1}}$ cluster bulk flow extending up to $\sim$$\SI{500}{Mpc}$ ($z\sim0.12$). These results strongly disagree with the isotropy assumption underlying the standard model of cosmology ($\Lambda$CDM) (for a recent review, see \citet{Migkas_24}).

Galaxy cluster $T$ is a measure of the average kinetic energy of the electrons in the hot intra-cluster medium (ICM) and is measured using X-ray spectroscopy. However, there is a possibility that such $T$ measurements could be affected by systematic biases and, eventually, bias cosmological conclusions. Indeed, it has long been known that X-ray $T$ measurements may be biased in various ways \citep[see, e.g., Chapter~4 in the review by][]{Reiprich_13}. However, a directionality or spatial variation of such a bias is not expected. 
Possible reasons for directional-dependent systematic biases include under or overestimation of absorption corrections in $T$ in certain directions. However, this effect would likely be very small, as we use a range of $0.7-7\,\unit{keV}$ for the $T$ measurement. Another potential factor could be the presence of a hot, diffuse cloud that is not accounted for in the X-ray background. Regardless, if present, such a direction-dependent bias may lead to variations in best-fit parameters of scaling relations that depend on sky position. If the variations observed by M20 and M21 were indeed due to such systematic biases, a $\sim 10-13\%$ overestimation of $T$ towards the most anisotropic direction identified by M21 could alleviate the tension between their findings and the $\Lambda$CDM model\footnote{However, this scenario was tested in M20 and M21 and strongly disfavoured}. Therefore, it is essential to carefully investigate systematic $T$ biases in the data used by M21 throughout the whole sky.

In this work, we look for potential $T$ measurement biases across the sky by pairing the cluster $T$ with the $\sigma_\mathrm{v}$ since its measurement also does not involve any cosmological assumptions. We convert the variations obtained in the  $\sigma_\mathrm{v}-T$ relation's best-fit parameters into $T$ over- or underestimations across the sky. $\sigma_\mathrm{v}$ has completely different systematics as compared to $T$ but is unaffected by Galactic absorption. Therefore, if the latter is the reason for a possible $T$ bias, this will, in principle, show up in this scaling relation's anisotropy. A significant variation in the best-fit normalisation across the sky that results in a noticeable overestimation of $T$ in a particular region would suggest that systematic biases influence the results reported in M21. Conversely, if the best-fit parameters are uniform across the sky, it implies that a systematic $T$ bias is a highly unlikely explanation for the observed anisotropies in M21.

We also explore cosmic isotropy by using $\sigma_\mathrm{v}$ in combination with cluster properties, whose measurements rely on cosmological parameters. Two scaling relations, the X-ray luminosity-velocity dispersion ($L_\mathrm{X}-\sigma_\mathrm{v}$) and the total integrated Comptonization parameter-velocity dispersion ($Y_\mathrm{SZ}-\sigma_\mathrm{v}$), are studied for this purpose. The variations in the best-fit parameters of these relations are converted to variations in $H_0$. This test provides new insights into a potential $H_0$ angular variation and acts as a cross-check of the M21 results. For this work, a flat $\Lambda$CDM model is assumed with $H_0 = \SI{70}{km\ s^{-1}\ Mpc^{-1}}$, $\Omega_\mathrm{m} = 0.3$ and $\Omega_\Lambda = 0.7$.

The paper is structured as follows: Section~\ref{sec:sample} describes the various data samples and cluster properties used in this work. Section~\ref{sec:scaling_relations} explains the fitting procedure for scaling relations and the techniques used to study their variations across different parts of the sky. In Sect.~\ref{sec:general_behaviour}, we present the general behaviour of the scaling relations used in this work. Section~\ref{sec:temp_bias} provides detailed information on the variations in the best-fit parameters of the $\sigma_\mathrm{v}-T$ relation, $T$ bias across the sky, and the isotropic Monte Carlo simulation results. Section~\ref{sec:cosmic_isotropy} explores the variations of $L_\mathrm{X}-\sigma_\mathrm{v}$ and $Y_\mathrm{SZ}-\sigma_\mathrm{v}$ scaling relations, $H_0$ variations across the sky from their joint analysis, and their comparison with isotropic Monte Carlo simulations. In Sect.~\ref{sec:discussion}, we discuss possible systematics and compare our results with M21. Finally, in Sect.~\ref{sec:summary}, the conclusions of this work are given.

%--------------------------------------------------------------------
\section{Data samples}
\label{sec:sample}
This work utilises four key cluster parameters: $T$, $\sigma_\mathrm{v}$, $L_\mathrm{X}$, and $Y_\mathrm{SZ}$. These parameters are obtained for clusters present in three different data catalogues: the Meta Catalogue of X-ray detected Clusters of galaxies (MCXC, \citealt{Piffaretti_11}), the extremely expanded HIghest X-ray FLUx Galaxy Cluster Sample (eeHIFLUGCS, Pacaud et al., in preparation), and the Euclid velocity dispersion metacatalog (Euclid Collaboration: Melin et al. 2024, in preparation). 

MCXC is a comprehensive catalogue of X-ray-detected galaxy clusters, which is based on the ROSAT All-Sky Survey (RASS). The RASS mapped the entire sky in the X-ray band with the ROSAT PSPC detector in the energy range of $0.1 - 2.4\ \unit{keV}$ \citep{Voges_99}. The catalogue contains 1743 clusters with virtually no duplicate entries. For each cluster, the catalogue provides the position, redshift, standardised $0.1 - 2.4\ \unit{keV}$ band luminosity $L_\mathrm{X}=L_{500}$, total mass $M_{500}$\footnote{$M_{500}$ obtained from $L_\mathrm{X}-M_{500}$ scaling relation.}, and radius $R_{500}$\footnote{The suffix 500 refers to the radius within which the average density is 500 times the critical density of the Universe.}. 

eeHIFLUGCS is a homogeneously selected sample of galaxy clusters that has been created by imposing several selection criteria on the MCXC. The selection criteria include a flux limit on the unabsorbed X-ray flux of $f_\mathrm{0.1-2.4\ \unit{keV}} \geq 5 \times 10^{-12}\ \unit{erg\ s^{-1}\ cm^{-2}}$. Clusters in the Galactic plane ($|b| \leq 20^\circ$), the Magellanic clouds, and the Virgo cluster area are excluded. Another selection criterion used here is the availability of a high quality \textit{Chandra} \citep{Chandra_00} or \textit{XMM-Newton} \citep{XMM_01} observation.

The Euclid velocity dispersion metacatalog is a collection of homogeneous velocity dispersion measurements from various sources compiled by the Euclid Collaboration. The catalogue includes 614 clusters from previous catalogues such as the Planck cluster sample \citep{Aguado_22}, Abell \citep{Girardi_98}, ACO \citep{Mazure_96}, XDCP \citep{Nastasi_14}, SDSS/Abell \citep{Popesso_07}, and other samples.

\subsection{X-ray luminosity $L_\mathrm{X}$ and Temperature $T$}
We utilise the $L_\mathrm{X}$ from the eeHIFLUGCS catalogue, which contains the luminosity measurements from the RASS \citep{Voges_99} covering the entire $R_{500}$ of a cluster. This is different from most of the \textit{XMM-Newton} and \textit{Chandra} clusters available in this catalogue. The luminosities of the MCXC have been corrected for the Galactic absorption based on the \citet{Willingale_13} $N_\mathrm{Htot}$ values as described in M21. 

The cluster $T$ data is sourced from the eeHIFLUGCS catalogue, as provided in \citet{K.Migkas_20}. The $T$ values are measured within the range of $0.2-0.5\ R_{500}$ for each cluster and are derived from observations made by \textit{Chandra} and \textit{XMM-Newton}. In total, $T$ values are available for 313 eeHIFLUGCS clusters.

\subsection{Total integrated Compton parameter $Y_\mathrm{SZ}$}
The Sunyaev-Zeldovich (SZ) effect provides information about the total gas pressure of the ICM and is characterised by the Compton parameter denoted by the symbol $y$. The integrated Compton parameter $Y$ is obtained by integrating $y$ over the cluster's solid angle. This quantity is multiplied by the square of the angular diameter distance to obtain the total integrated Compton parameter $Y_\mathrm{SZ}$, which has units of $\unit{kpc^2}$, and scales with other cluster properties (detailed description available in M21).

We determine $Y_\mathrm{SZ}$ values for all MCXC clusters using the final data release of \citet{Planck_20a} as explained in M21. We apply several signal-to-noise (S/N) cuts to the $Y_\mathrm{SZ}$ measurements to see how our results depend on the $Y_\mathrm{SZ}$ selection. For the default analysis, we use an S/N cut of $\geq2$, which results in 1093 clusters being included. Results in Sect. \ref{sec:comparison} show that no systematic trends are observed when we vary the S/N cut. We are already aware of the existence of clusters in these areas. Therefore, there is no necessity to increase the S/N cut for detection purposes.

\subsection{Velocity dispersion $\sigma_\mathrm{v}$}
The velocity dispersion of a galaxy cluster is a measure of the spread of velocities among the member galaxies of the cluster. It is obtained by measuring the line-of-sight velocities of the member galaxies in optical bands. We utilise the Euclid velocity dispersion metacatalog for the $\sigma_\mathrm{v}$ measurements.
 This catalogue has several important properties: Firstly, all the clusters have at least ten member galaxies with spectroscopic redshift measurements, and interlopers have been adequately rejected while defining cluster members. Secondly, the velocity dispersion calculations are based on the methodology of \citet{Beers_90} with a S/N $\geq 4$ and a minimum aperture of $\SI{0.5}{Mpc}$. In addition to the velocity dispersion errors from the parent catalogues, dispersion uncertainties are also calculated in a standard way (\citet{Sereno_15}).

\subsection{Matching different catalogues}
\label{sec:catalog_matching}
We cross-matched the $\sigma_\mathrm{v}$ catalogue to eeHIFLUGCS and the $Y_\mathrm{SZ}$ data (MCXC clusters) to study various scaling relations. Our primary matching criteria were that the cluster positions should be within $5^\prime$ of each other, and the difference in redshifts $\Delta z$ of the matching clusters should not exceed $0.01$. We discovered 165 matching clusters with both $\sigma_\mathrm{v}$ and $T$ measurements and 200 clusters with both $L_\mathrm{X}$ and $\sigma_\mathrm{v}$ measurements for the eeHIFLUGCS sample. We identified 284 matching clusters between the $\sigma_\mathrm{v}$ catalogue and the $Y_\mathrm{SZ}$ data (S/N $\geq 2$).

\begin{figure*}[htbp]
    \centering
    \includegraphics[width=0.33\textwidth]{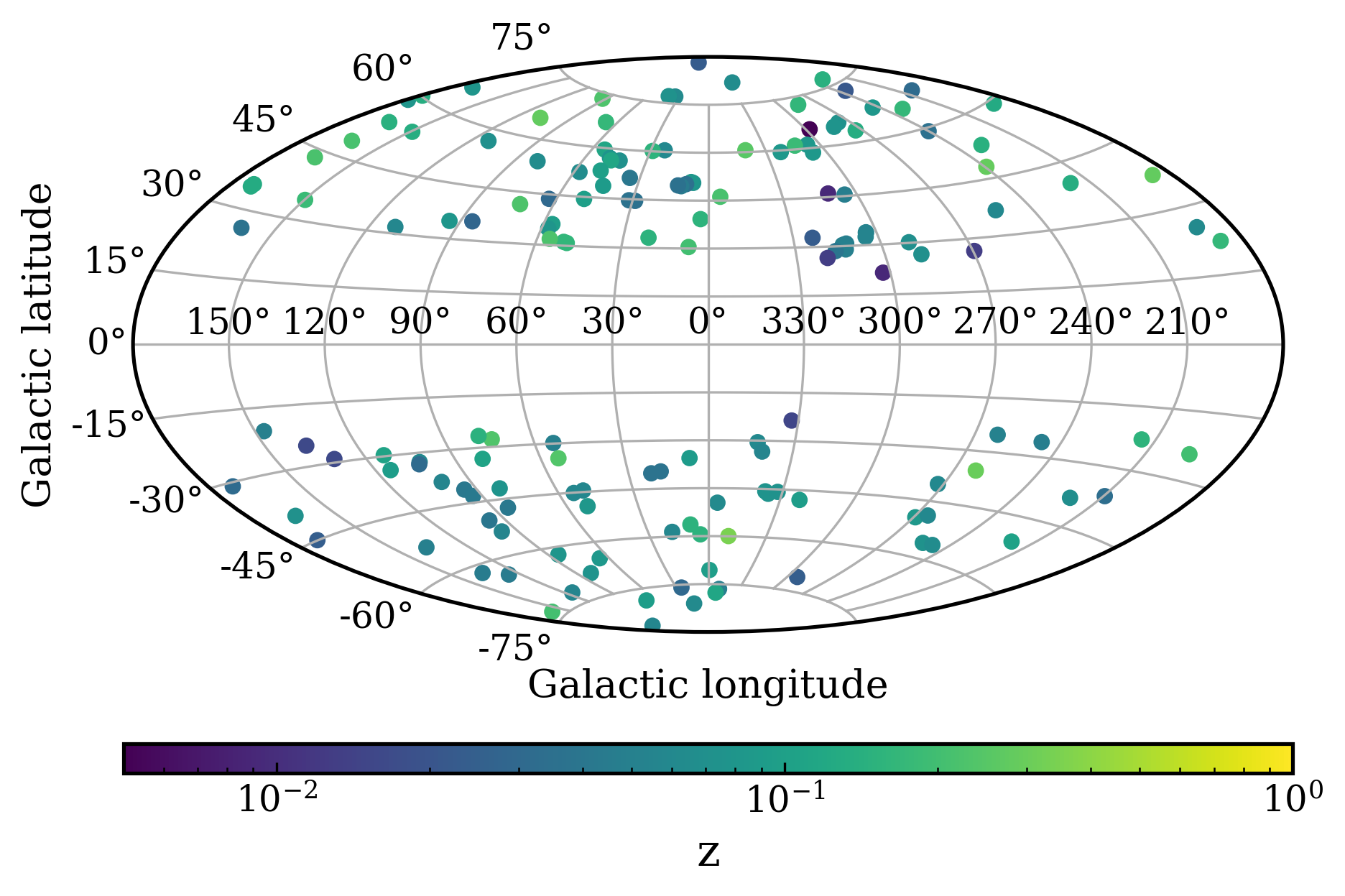}
    \includegraphics[width=0.33\textwidth]{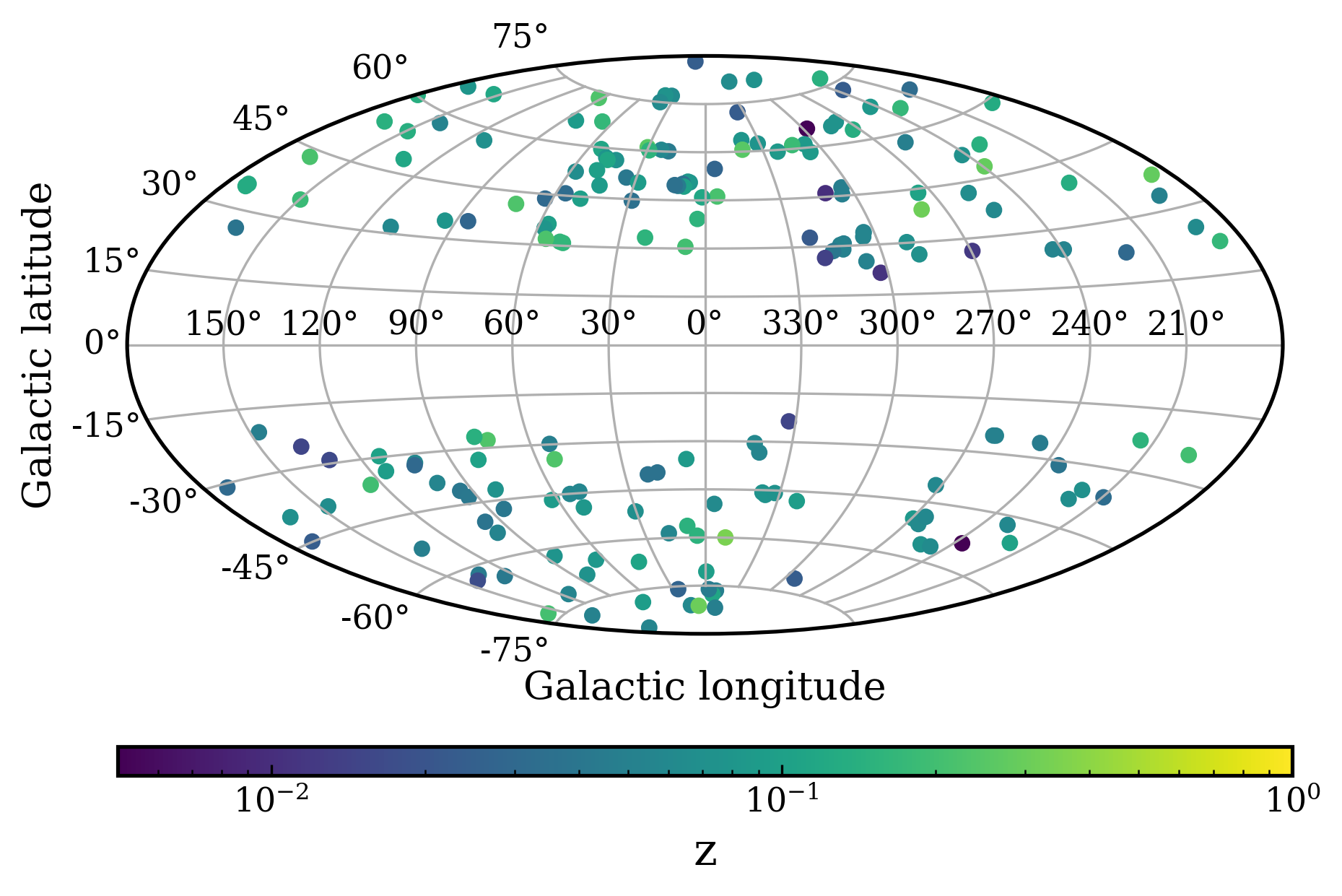}
    \includegraphics[width=0.33\textwidth]{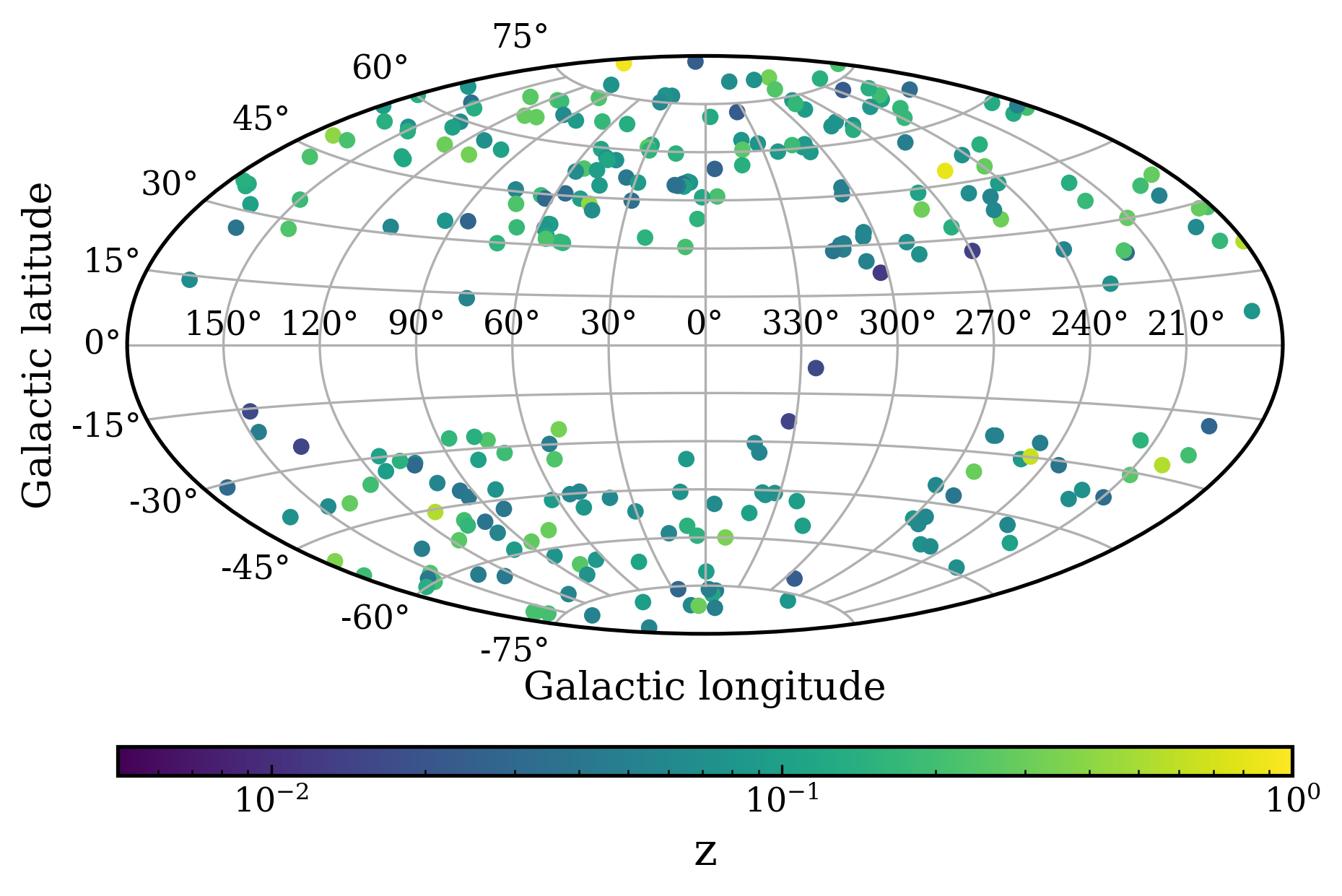}

    \caption{Sky distribution of the galaxy clusters used in the scaling relations $\sigma_\mathrm{v}-T$ (\textit{left}), $L_\mathrm{X}-\sigma_\mathrm{v}$ (\textit{middle}), and $Y_\mathrm{SZ}-\sigma_\mathrm{v}$ (\textit{right}). The clusters' redshifts are colour-coded.}
    \label{fig:sky_distribution}
\end{figure*}

%--------------------------------------------------------------------
\section{Analysing the scaling relations}
\label{sec:scaling_relations}

To study the three distinct scaling relations, a general form of the scaling relation is adopted. The scaling relation between two variables, $Y$ and $X$, is expressed as

\begin{equation}
    \label{eq:scaling_relation_notation}
    \frac{Y}{C_Y} \times E(z)^{\gamma_{YX}} = A_{YX} \times \left( \frac{X}{C_X} \right)^{B_{YX}}. 
\end{equation}

Here $C_Y$ and $C_X$ are the pivot points for the $Y$ and $X$ quantity respectively, $E(z)$ is the redshift evolution factor $\left( E(z) = \sqrt{\Omega_\mathrm{m} (1+z)^3 + \Omega_\Lambda}\right)$ with $\gamma_{YX}$ being its power law index. $A_{YX}$ and $B_{YX}$ are the normalisation and slope of the scaling relation, respectively.
The pivot points $C_Y$ and $C_X$ are chosen close to the median of the $Y$ and $X$, respectively. This is done to minimise the correlation between the best-fit parameters. These terms and the $\gamma_{YX}$ for all three scaling relations are mentioned in the Table \ref{tab:scaling_relations}.

\subsection{Bayesian linear regression}
\label{sec:bayesian_linear_regression}
A linear regression is performed using Bayesian statistics to find the best-fit parameters for scaling relations in the logarithmic space. We constrain the best-fit parameters by maximising the posterior probability distribution of the parameters. The Markov Chain Monte Carlo (MCMC) method is used for sampling these distributions. We take its logarithm to convert the power law relation to a linear relation. The general form of scaling relations in terms of a linear model is given as

\begin{equation}
    \label{eq:scaling_relation_linear}
    y = m\times x + c,
\end{equation}
where $y = \log_{10} \left( \frac{Y}{C_Y} \times E(z)^{\gamma_{YX}} \right)$, $x = \log_{10} \left( \frac{X}{C_X} \right)$, $m = B_{YX}$, and $c = \log_{10} \left( A_{YX} \right)$.

To speed up the calculation process, the logarithm of the posterior probability distribution is used. The log-likelihood function is given as

\begin{equation}
    \label{eq:likelihood}
    \log \left( \mathcal{L} \right) = -\frac{1}{2} \sum_{i=1}^{N} \left( \frac{[y_i - mx_i - c]^2}{\sigma_i^2}  + \log \left( 2\pi \sigma_i^2 \right)\right),
\end{equation}
where $N$ is the number of data points, $y_i$ and $x_i$ are the $i^\mathrm{th}$ data points of $Y$ and $X$ respectively, and $\sigma_i$ is the total uncertainty given by:

\begin{equation}
    \label{eq:total_uncertainty}
    \sigma_i = \sqrt{\sigma_{y,i}^2 + m^2\sigma_{x,i}^2 + \sigma_\mathrm{intr}^2}.
\end{equation}
Here $\sigma_{y, i}$ and $\sigma_{x, i}$ are the uncertainties\footnote{Uncertainties in the linear space are converted to the logarithmic space using $\sigma_{\log_{10} a} = \log_{10}(e)\times \frac{a^+-a^-}{2a}$. Here, $a^+$ and $a^-$ are the upper and lower limits of the parameter $a$, respectively.} for the $y_i$ and $x_i$, respectively, and $\sigma_\mathrm{intr}$ is the intrinsic scatter of the scaling relation measured in $Y$ direction.
Thus, the three parameters to be constrained are $m$, $c$, and $\sigma_\mathrm{intr}$. Flat priors are used for $m$ and $c$ with upper and lower bounds of $+10$ and $-10$, respectively. Since $\sigma_\mathrm{intr}$ is a positive definite quantity, a flat prior with a lower bound of $0$ and an upper bound of $+10$ is selected. Even though the prior choices are uninformative, the resulting posterior distribution always converges to a normal distribution.

The chain is initialised with a random set of parameters. It is run for at least 20,000 iterations in 4 chains to ensure it has converged\footnote{The convergence is tested by discarding burn-in samples and checking the acceptance fraction.}. The best-fit parameters are obtained by taking the median of the parameter space distribution. Lower and upper bounds are then determined by using the $16^\mathrm{th}$ and $84^\mathrm{th}$ percentile values of the distribution, respectively. The implementation of Bayesian linear regression is coded in \texttt{Python}\footnote{The code is available at \url{https://github.com/AdiPandya/Fast_Bayesian_Regression}} and the \texttt{numba} \citep{numba} package is used to improve computational time. The validity of the code is verified by comparing the results with the \texttt{LINMIX} \citep{linmix} and \texttt{PyMC} \citep{PyMC} packages, and we find results that are consistent.

This method for Bayesian linear regression is Y|X since the $y$-axis distance of the data points from the best-fit line is minimised, and $x$ is treated as the independent variable. For certain scaling relations, X|Y best-fit implementation was used (more details in Sect. \ref{sec:general_behaviour}). In this method, the $x$-axis distance of the data points from the best-fit line is minimised, and $y$ is treated as the independent variable. The log-likelihood function for X|Y is given as 

\begin{equation}
    \label{eq:likelihood_x_y}
    \log \left( \mathcal{L} \right)_\mathrm{X|Y} = -\frac{1}{2} \sum_{i=1}^{N} \left( \frac{\left[x_i - \left(\frac{y_i - c}{m}\right)\right]^2}{\sigma_{i, \mathrm{X|Y}}^2}  + \log \left( 2\pi \sigma_{i, \mathrm{X|Y}}^2 \right)\right),
\end{equation}

where $\sigma_{i, \mathrm{X|Y}}$ is $\sqrt{\sigma_{x,i}^2 + \left(\frac{\sigma_{y,i}}{m}\right)^2 + {\sigma_{{\mathrm{intr}}_\mathrm{X|Y}}}^2}$. The $\sigma_{\mathrm{intr}_\mathrm{X|Y}}$ is the intrinsic scatter of the scaling relation measured in $X$ direction. Note that the slope $m$ and the normalisation ($10^c$) refer to the best-fit parameters of the Y|X form of the relation ($L_\mathrm{X}-\sigma_\mathrm{v}$), and not to the X|Y form ($\sigma_\mathrm{v}-L_\mathrm{X}$).

\subsection{Removing outliers}
We use an iterative 3$\sigma$ clipping method to detect and remove outliers in the data. This method removes outliers based on their residual distance from the best-fit line. When using the Y|X fitting method, we consider residuals in the Y direction, and for the X|Y method, we consider residuals in the X direction. The method assumes that residuals follow a normal distribution, and points lying outside the 3$\sigma$ of this distribution are removed as outliers. Using the 3$\sigma$ cut, we expect approximately one out of every 370 clusters to fall outside this range purely due to chance. Since the number of clusters in all relations is less than 370, we can safely use the 3$\sigma$ cut. This approach does not eliminate any clusters that adhere to normal scaling relation behaviour and helps eliminate problematic measurements.

It is important to note that this method may remove additional outliers when repeated because the new sample will have different best-fit parameters, resulting in different residual distributions. The method is repeated several times until no new outliers are found. Typically, this method removes 2--5 outliers from each scaling relation sample. Table \ref{tab:scaling_relations} lists the final sample sizes for the scaling relations. Fig. \ref{fig:sky_distribution} displays the sky distribution of the clusters used in the scaling relations $\sigma_\mathrm{v}-T$, $L_\mathrm{X}-\sigma_\mathrm{v}$, and $Y_\mathrm{SZ}-\sigma_\mathrm{v}$, along with their redshifts.

%---------------------------------------------------------------------------------------------

\subsection{Scanning the sky}
\label{sec:sky_scanning}

\begin{figure}[htbp]
    \centering
    \includegraphics[width=0.9\hsize]{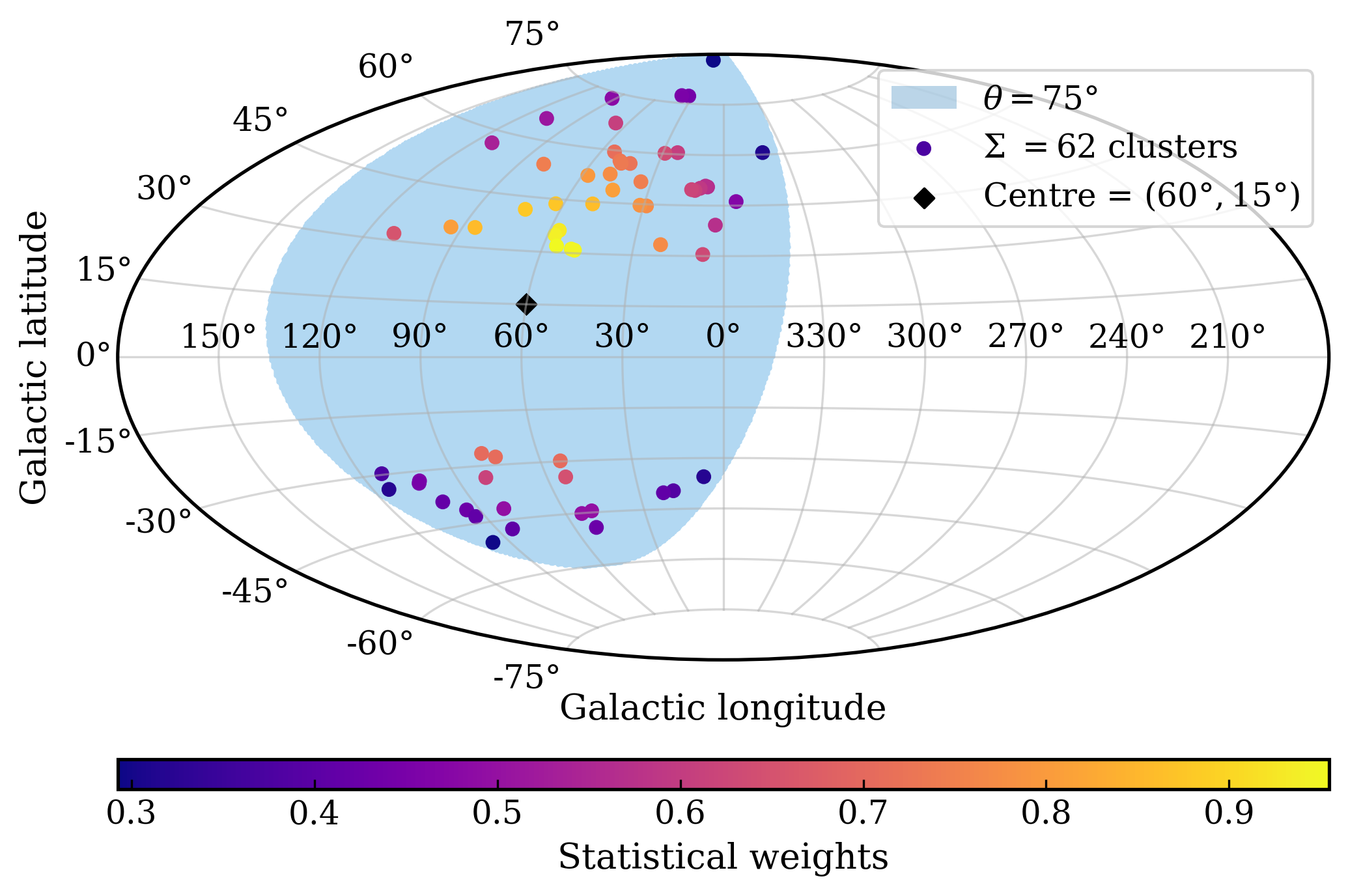}
    \caption{Example of a region used for 2-dimensional scanning centred at $(l,b)=(60^\circ,15^\circ)$ with cone size, $\theta$, of radius $75^\circ$ for $\sigma_\mathrm{v}-T$ relation.
    Different colours represent the statistical weights assigned to the clusters inside these regions. $\sum$ represents the number of clusters inside this region.}
    \label{fig:2d_example}
\end{figure}

We study the consistency of the scaling relations in different sky directions by considering sections of the sky and calculating the best-fit parameters of the clusters inside those regions. We use the two-dimensional scanning method adopted from M21. 
In this method, a cone of radius $\theta$ is constructed for a given galactic longitude $l$ and latitude $b$, and the scaling relation best-fit parameters of the clusters inside this cone are constrained.
To cover the entire sky, the central longitudes and latitudes of the cones are varied with a step size of $5^\circ$ for their whole range. This results in $72\times37=2664$ different cones covering the entire sky\footnote{The final statistical directional uncertainties are much larger than the step size; thus, the step size of $5^\circ$ is sufficiently small.}.
Ideally, the cone sizes should be as small as possible to minimise overlap. However, the cone size should also be large enough to have sufficient clusters\footnote{We ensure that there are at least 30 clusters in each cone.} inside the cone to efficiently constrain the best-fit parameters. We tested various cone sizes and found a cone size of radius $\theta = 75^\circ$ to be optimal for this analysis (additional details in Appendix \ref{sec:cone_interval_sizes}). An example of such a cone, along with the statistical weights of clusters inside this cone for the $\sigma_\mathrm{v}-T$ sample, is shown in the bottom panel of Fig. \ref{fig:2d_example}.

We also apply statistical weights to clusters inside the scanning regions. This is done by dividing the statistical uncertainties of the observables ($\sigma_y$ and $\sigma_x$) by the cosine of the angular separation of the cluster from the cone centre\footnote{Use of this method underestimates the intrinsic scatter and the statistical significance since the statistical uncertainties of the observables are increased. However, the isotropic MC simulations capture this information, and its significance is unbiased.}. This results in a lower statistical weight for clusters that are further from the cone centre.
2-D maps are created to visualise the results of 2-D scanning. These maps are a grid plot with a box size of $5^\circ$ on the Aitoff projection of the sky using the \texttt{Python} package \texttt{desiutil}\footnote{Package available at \url{https://github.com/desihub/desiutil}}. The colour of each box in these maps represents the parameter value of interest for the cone centred at a given $l$ and $b$.

\subsection{Constraining $H_0$ angular variations}
\label{sec:constraining_H0}
In the analysis of $L_\mathrm{X}-\sigma_\mathrm{v}$ and $Y_\mathrm{SZ}-\sigma_\mathrm{v}$ scaling relations, we assume that the intracluster physics remain unchanged for different directions, and variations of cosmological parameters cause the apparent anisotropy of scaling relations. According to this assumption, $H_0$ can vary across the sky while the true normalisation of the scaling relations remains constant. Alternatively, the $A$ variations can also be interpreted in other ways, such as the presence of large-scale bulk flows, as shown in M21. However, due to the large scatter in the scaling relations, we restrict our analysis to constraining $H_0$ angular variations.

We use best-fit $A/A_\mathrm{all}$ of every cone to constrain $H_0$ angular variations using the relation
\begin{equation}
    \label{eq:H0_conversion}
    H_0 = H_{70} \times \left(\frac{A}{A_\mathrm{all}}\right)^\frac{1}{2}.
\end{equation}
Here $H_{70}=\SI{70}{km\ s^{-1}\ Mpc^{-1}}$ and $A_\mathrm{all}$ is the best-fit $A$ of the entire sample. $A$ and $H_0$ are degenerate, and absolute constraints can only be put on the quantity $A \times H_0^2$. One of these parameters is assumed to be fixed to constrain the other (more details in Appendix \ref{sec:A_to_H0}). 

This method cannot be used to put absolute constraints on $H_0$ because a value of $H_0$ was assumed to calibrate the relation initially.
The calculated value of $H_0$ expresses relative differences between the regions; thus, the variations are always around the assumed value of $H_0$.

\subsection{Statistical significance of the variations}
We use the best-fit parameters and their uncertainties in different parts of the sky to quote the statistical significance of the variations between the two regions. For two independent sub-samples $i$ and $j$, the statistical significance of their deviation in terms of number of sigma is given by
\begin{equation}
    \label{eq:sigma_map}
    \mathrm{No.\ of\ }\sigma = \frac{\mathrm{p}_i - \mathrm{p}_j}{\sqrt{\sigma_{\mathrm{p}_i}^2 + \sigma_{\mathrm{p}_j}^2}},
\end{equation}
where $\mathrm{p}_i$ and $\mathrm{p}_j$ are the best-fit parameters of the two sub-samples, and $\sigma_{\mathrm{p}_i}$ and $\sigma_{\mathrm{p}_j}$ are their 
uncertainties.
For a cone centred at ($l$, $b$), all the clusters inside it are considered one sub-sample, and the clusters outside the cone are considered the other sub-sample. We find the number of sigma or significance values for each cone to create a significance (sigma) map. The colour of each box in this map represents the statistical significance of deviation compared to the rest of the sky at that location.

\begin{figure*}
    \centering
    \includegraphics[width=0.33\textwidth]{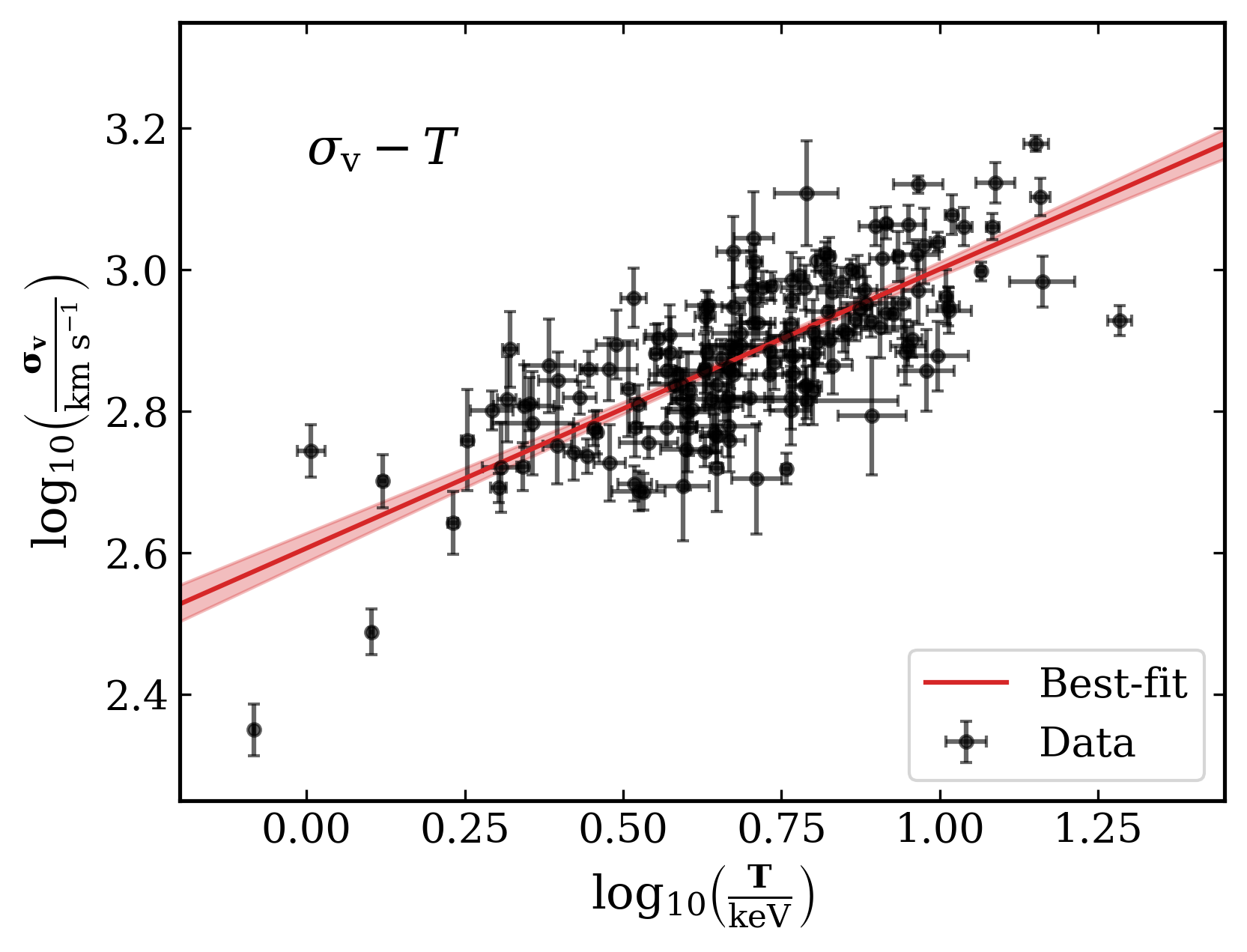}
    \includegraphics[width=0.33\textwidth]{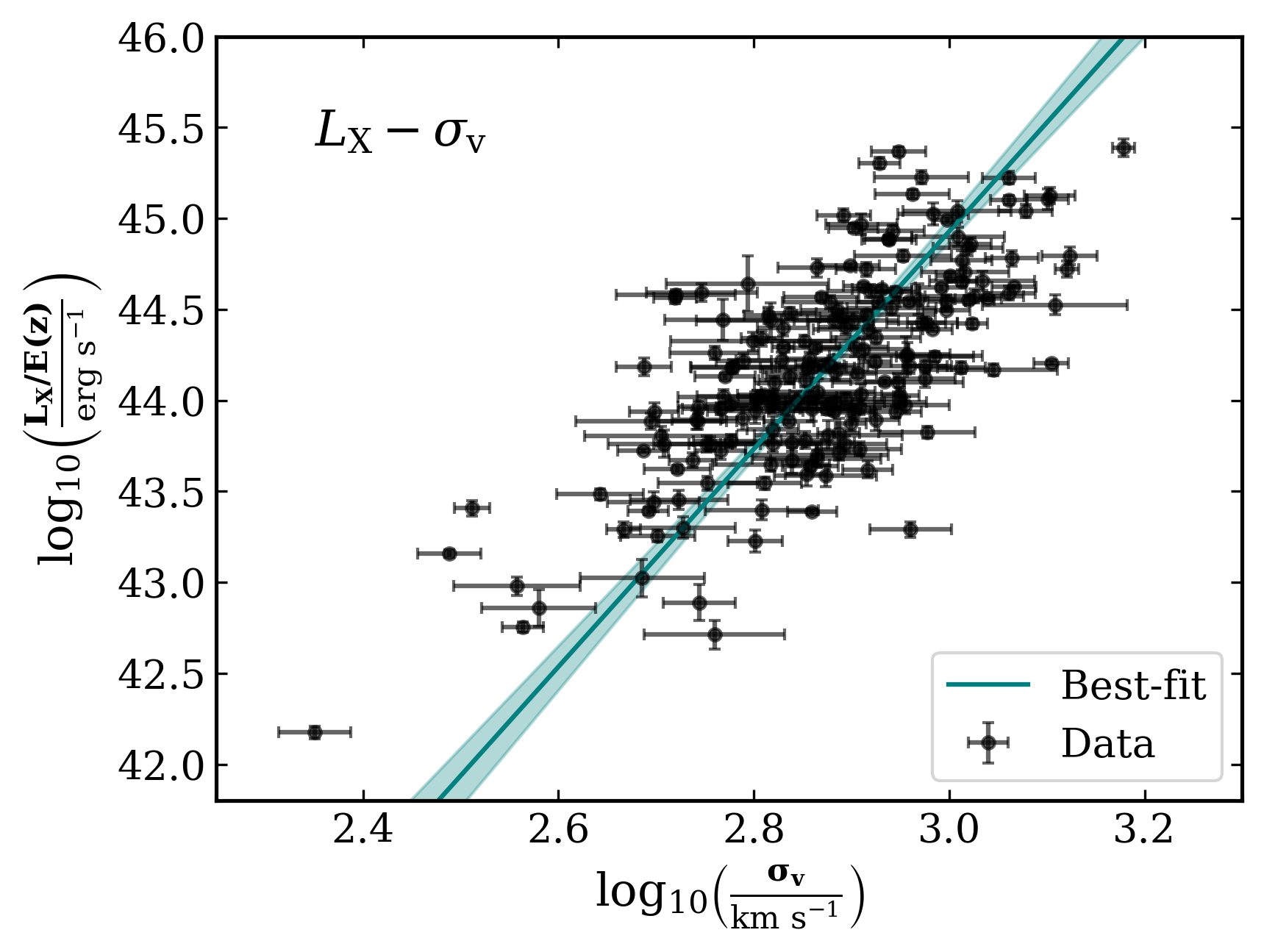}
    \includegraphics[width=0.33\textwidth]{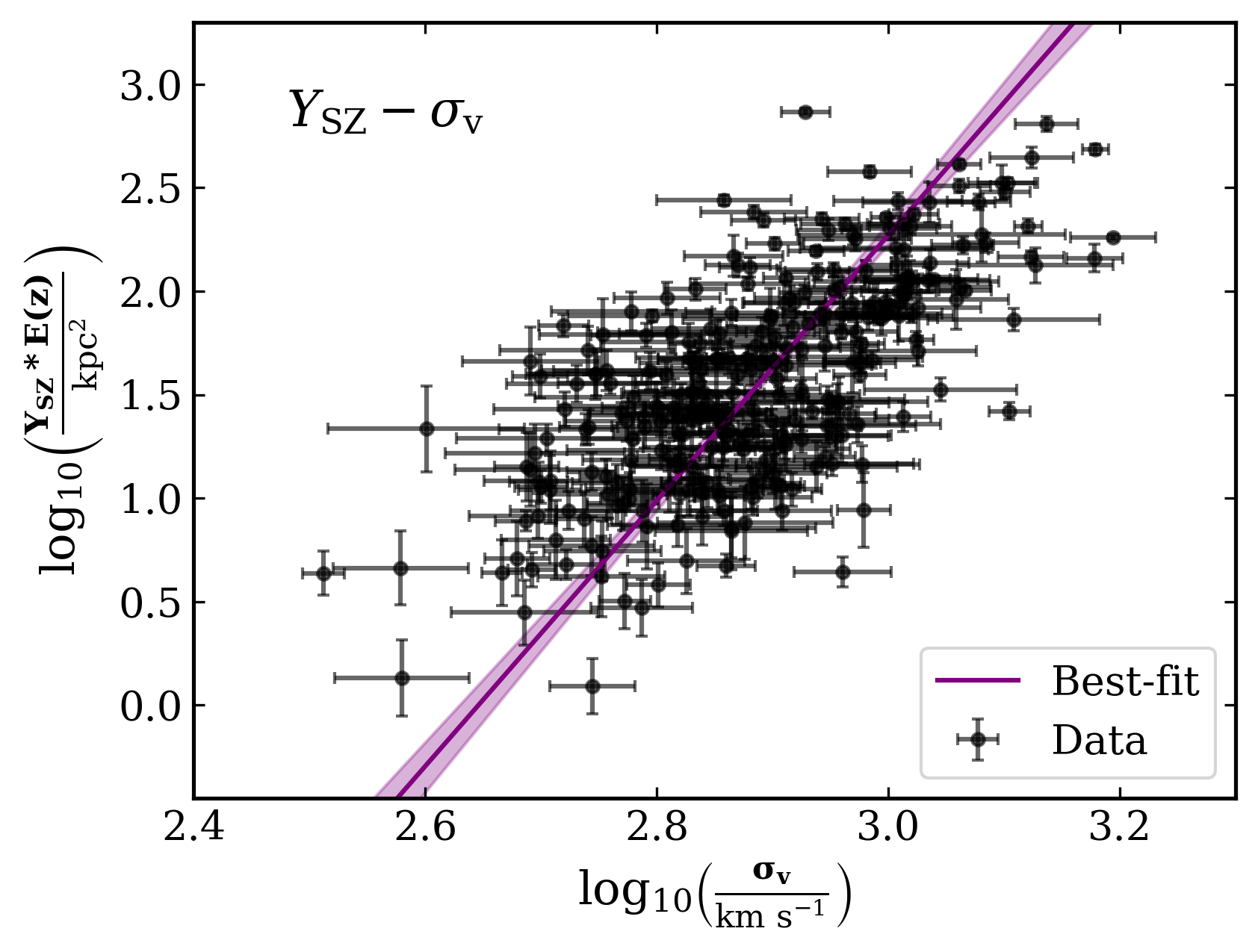}
    \caption{\textit{From left to right:} Best-fit plots for the scaling relation $\sigma_\mathrm{v}-T$ (Y|X), $L_\mathrm{X}-\sigma_\mathrm{v}$ (X|Y), and $Y_\mathrm{SZ}-\sigma_\mathrm{v}$ (X|Y). Shaded regions represent $1\sigma$ uncertainties of the best-fit parameters.}
    \label{fig:scaling_relations}
\end{figure*}

\begin{table*}[htbp]
    \caption{Best-fit parameters for the three scaling relations for the full sample. The table contains number of matching clusters ($N$), pivot points $C_Y$ and $C_X$, redshift evolution power $\gamma$ of the relation, best-fit normalisation ($A$), best-fit slope ($B$) and the intrinsic scatter ($\sigma_\mathrm{intr}$). For the $L_\mathrm{X}-\sigma_\mathrm{v}$ and $Y_\mathrm{SZ}-\sigma_\mathrm{v}$ relations, results are shown using the X|Y method. The intrinsic scatter mentioned here is measured in the direction in which the scatter is minimised.}
    \label{tab:scaling_relations}
    \centering
    \begin{tabular}{lcccc|cccc}
    \hline
    \hline
        {Relation} & {$N$} & {$C_Y$} & {$C_X$} & {$\gamma$} & {Method} & {$A$} & {$B$} & {$\sigma_\mathrm{intr}$ (dex)} \bigstrut \\
    \hline
        &&&&&&&&\\
        $\sigma_\mathrm{v}-T$ & {160} & {$\SI{750}{km\ s^{-1}}$} & {$\SI{5}{keV}$} & {$-$} & Y|X & {$1.016^{+0.014}_{-0.014}$} & {$0.394^{+0.025}_{-0.026}$} & {$0.068^{+0.005}_{-0.005}$}\\
        &&&&&&&&\\
        {$L_\mathrm{X}-\sigma_\mathrm{v}$} & {195} & {$10^{44}\ \unit{erg\ s^{-1}}$} & {$\SI{750}{km\ s^{-1}}$} & {$-1$} & X|Y & {$1.515^{+0.134}_{-0.122}$} & {$5.988^{+0.452}_{-0.394}$} & {$0.080^{+0.005}_{-0.005}$}\\
        &&&&&&&&\\
        {$Y_\mathrm{SZ}-\sigma_\mathrm{v}$} & {284} & {$\SI{30}{kpc^2}$} & {$\SI{800}{km\ s^{-1}}$} & {$+1$} & X|Y & {$1.490^{+0.112}_{-0.106}$} & {$6.419^{+0.423}_{-0.384}$} & {$0.073^{+0.004}_{-0.004}$}\\
        &&&&&&&&\\
    \hline
    \end{tabular}
\end{table*}

\subsection{Isotropic Monte Carlo simulations}
\label{sec:isotropic_MC}
The statistical methods used above provide a way to study the anisotropies in the scaling relations. However, unknown statistical biases could still be present in the analysis methods, which could lead to under- or overestimation of the significance of anisotropies. We perform isotropic Monte Carlo simulations to test the method's validity and provide robust significance. To create an isotropic sample for a scaling relation, we perform the following steps:

\begin{enumerate}
    \item We start by fixing the cluster coordinates, redshifts, and $\sigma_\mathrm{v}$ (including the uncertainties) to their actual values. This is done to encompass potential effects that may give rise to anisotropies, including the spatial arrangement of real clusters.
    \item The cluster $T$, $L_\mathrm{X}$, and $Y_\mathrm{SZ}$ are simulated based on the best-fit parameters and scatter of the real data. First, their predicted values corresponding to the fixed $\sigma_\mathrm{v}$ are calculated using the respective best-fit parameters.
    \item Once all the points lie on the best-fit line, a random offset is added to these predicted values to obtain a simulated sample. This is done by drawing random values from a log-normal distribution centred at this value with a standard deviation equal to the total scatter (intrinsic + statistical) of the respective scaling relation and given cluster.
    \item Step 3 is repeated 1000 times, each time adding a random offset to the predicted values to create 1000 simulated samples.
\end{enumerate}

We compare the maximum variations in scaling relations across the sky for these samples with those obtained from the observations to quantify the statistical significance of the observed differences against random chance.

%--------------------------------------------------------------------------
\section{General behaviour of the three scaling relations}
\label{sec:general_behaviour}
The three scaling relations used are $\sigma_\mathrm{v}-T$, $L_\mathrm{X}-\sigma_\mathrm{v}$, and $Y_\mathrm{SZ}-\sigma_\mathrm{v}$.
This section presents best-fit parameters for these scaling relations for the entire sample. These relations are plotted in Fig. \ref{fig:scaling_relations}, and Table \ref{tab:scaling_relations} shows an overview of the best-fit results. 

\subsection{$\sigma_\mathrm{v}-T$ relation}
To study the $\sigma_\mathrm{v}-T$ relation, 160 galaxy clusters with measured velocity dispersion and temperatures are used.
We obtain a best-fit slope of $B=0.394^{+0.025}_{-0.026}$, which is lower than the predicted self-similar value of $0.5$ \citep{Lovisari_21}. Recent studies \citep{Xue_00, Ortiz-Gil_04, Nastasi_14, Wilson_16} obtain a slightly higher slope than the predicted self-similar scaling relation.
One important thing to note is that most previous studies use the Orthogonal Distance Regression (ODR) method to fit the data. In contrast, Y|X is used for this analysis, which usually returns a flatter slope than ODR. Using the ODR method for our sample, a slope of $B=0.434^{+0.030}_{-0.029}$ is obtained, similar to the results from Y|X but slightly larger.

\subsection{$L_\mathrm{X}-\sigma_\mathrm{v}$ relation}
195 clusters are used to study $L_\mathrm{X}-\sigma_\mathrm{v}$ relation. The X|Y method is used to get best-fit parameters since the Y|X method results in residual trends when plotted against the cluster redshifts (refer to Appendix \ref{sec:z_trends} for more details). However, as shown in Sect. \ref{sec:comparison}, the choice of fitting method does not significantly influence the final results.

Using the X|Y method we find $B=5.988^{+0.452}_{-0.394}$ while the Y|X leads to a shallower slope $B=2.722^{+0.214}_{-0.213}$. The slope from the Y|X method is similar to the self-similar slope $B_\mathrm{self}=2.7-2.8$ when considering $0.1-2.4\,\unit{keV}$ luminosities for clusters \citep{Lovisari_21}. 
Previous studies \citep{Mahdavi_01, Ortiz-Gil_04, Zhang_11, Nastasi_14, Sohn_19} using bolometric luminosities for clusters have obtained slopes consistent with the predicted slope of $B_\mathrm{self}=4$. Most of these studies use the ODR method to fit the data. We obtain similar results using X|Y and ODR methods. 

\subsection{$Y_\mathrm{SZ}-\sigma_\mathrm{v}$ relation}
For the relation $Y_\mathrm{SZ}-\sigma_\mathrm{v}$, 284 clusters are used. These clusters are selected based on S/N $\geq2$ for the $Y_\mathrm{SZ}$ parameter. The theoretical slope using the self-similar model is $B_\mathrm{self}=5$ (derived from \citet{Lovisari_21}). This is different from what is observed as, using X|Y, we find a steeper slope $B=6.419^{+0.423}_{-0.384}$, and Y|X leads to a shallow slope $B=2.915^{+0.194}_{-0.195}$. 
These trends are similar to what was observed in the $L_\mathrm{X}-\sigma_\mathrm{v}$ relation. Previous results from \citet{Rines_16} are quite similar to those obtained for X|Y and Y|X with a slope of $5.68\pm0.64$ and $2.36\pm0.30$, respectively.

%--------------------------------------------------------------------
\begin{figure*}[htbp]
	\centering
	\includegraphics[width=0.49\textwidth]{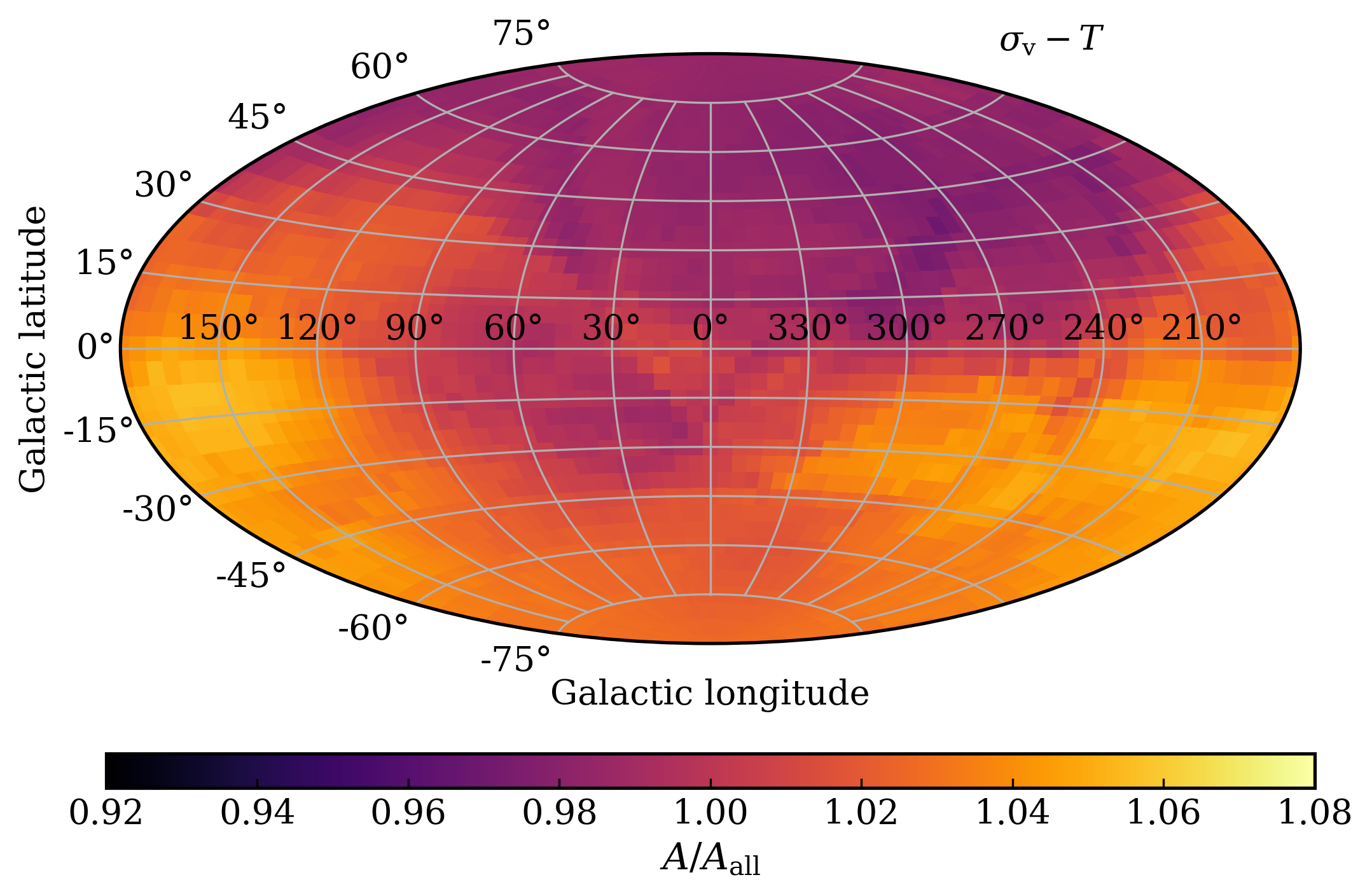}
    \includegraphics[width=0.49\textwidth]{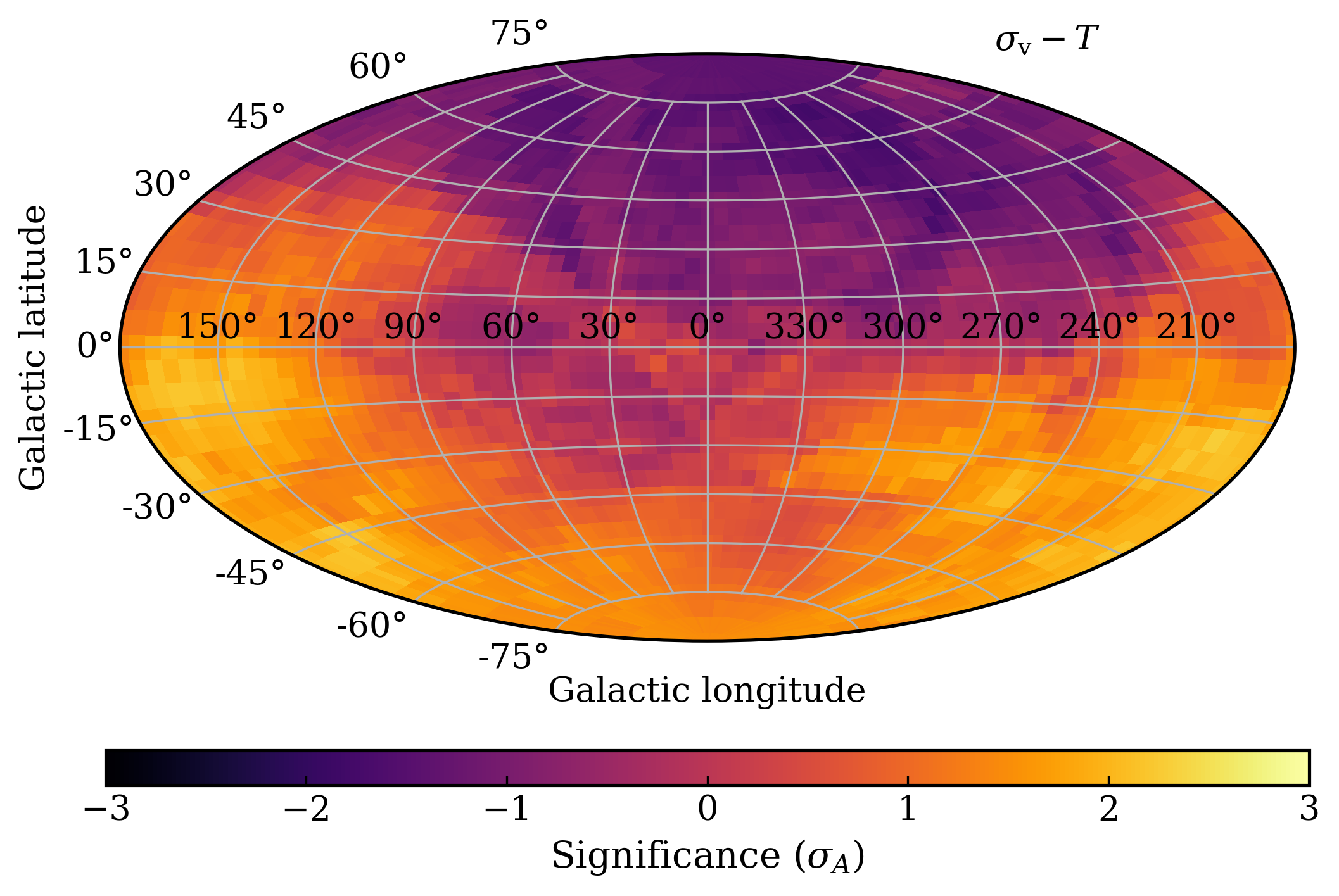}
    \includegraphics[width=0.49\textwidth]{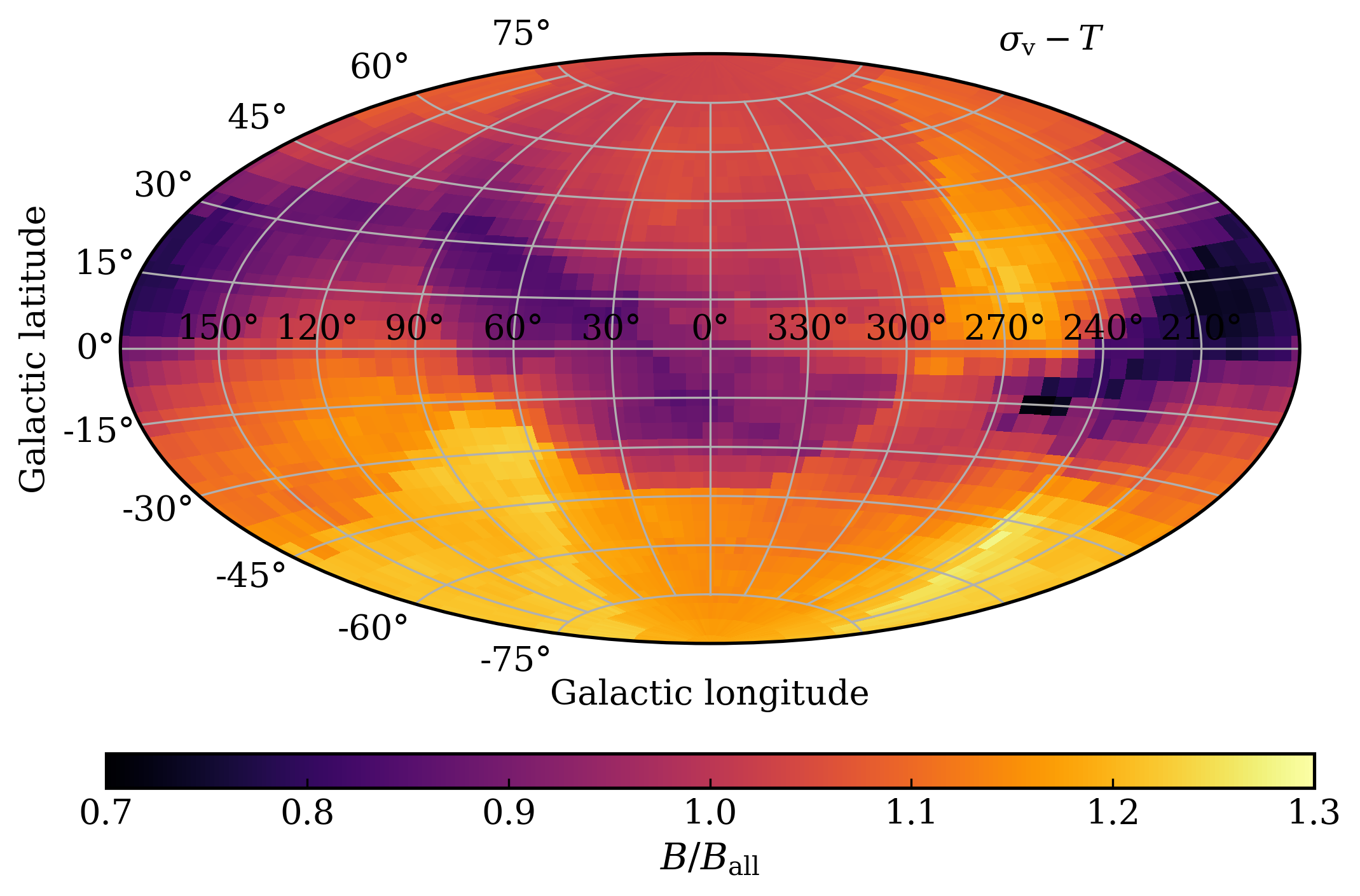}
    \includegraphics[width=0.49\textwidth]{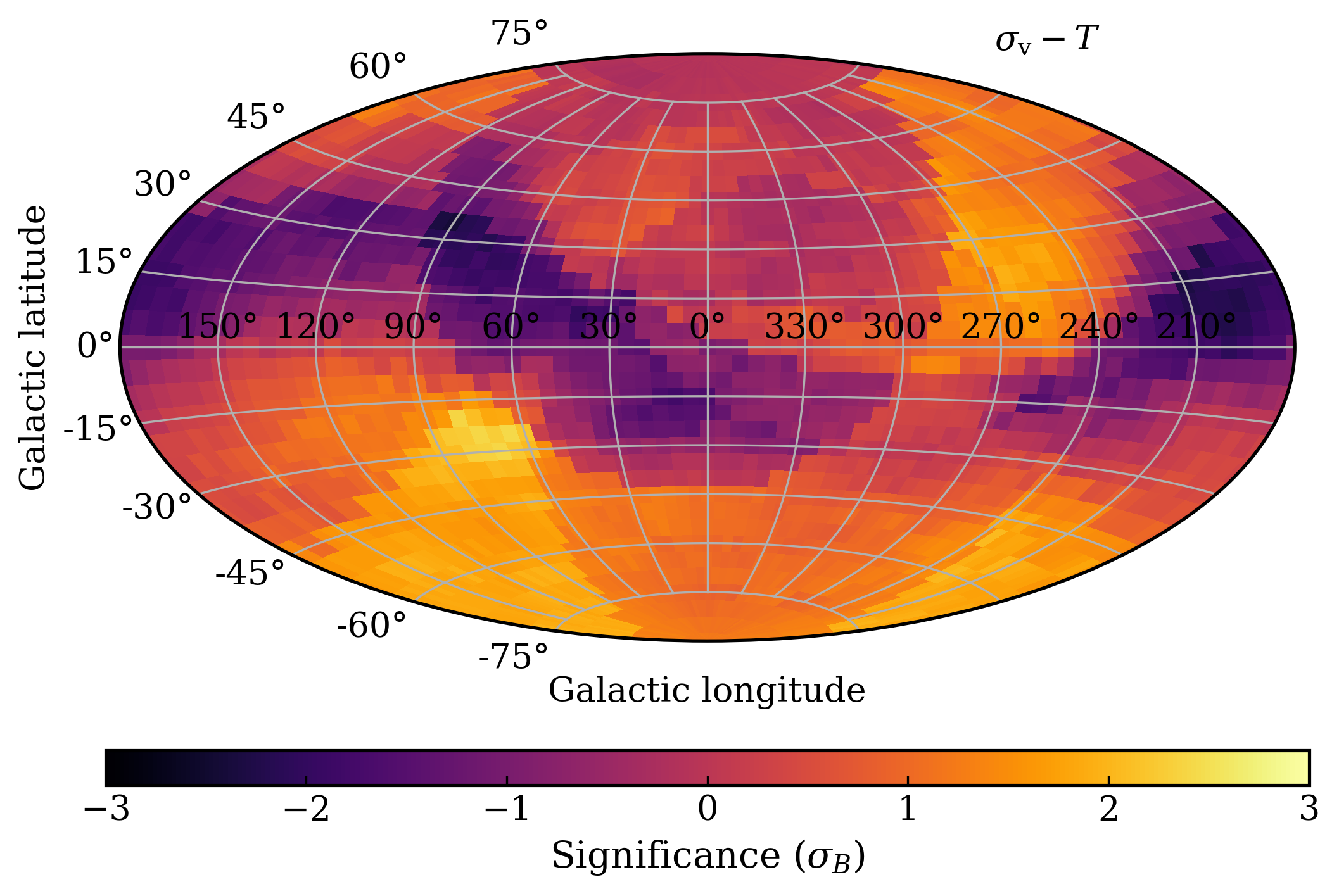}
	\caption{\textit{Left panel:} Maps of best-fit normalisation (\textit{top})
		and the slope (\textit{bottom}) compared to the full sample for the $\sigma_\mathrm{v}-T$ relation. Note that the colour scale is different for both maps. This highlights the small variations in the $A/A_\mathrm{all}$ map. \\
        \textit{Right panel:} Significance maps for the normalisation (\textit{top})
        and slope (\textit{bottom}). Both maps have the same colour scale ($-3\sigma$ to $+3\sigma$). A negative sigma value refers to a value lower than the rest of the sky.}
	\label{fig:2d_anisotropies_Sigma-T}
\end{figure*}

\section{Measuring the systematic temperature bias}
\label{sec:temp_bias}

To perform systematic checks on the $T$ measurements using $\sigma_\mathrm{v}-T$, we study the 2-D variations, and the best-fit results are converted to $T$ over- or underestimation. The statistical significance of the analysis is checked using the isotropic Monte Carlo simulations. 

\subsection{2-dimensional variations}
\label{sec:Sigma-T_2D}
The 2-D variations of best-fit parameters are studied by shifting a cone of radius $75^\circ$ across the sky. The left panel of Fig. \ref{fig:2d_anisotropies_Sigma-T} shows maps of the best-fit $A$ and $B$ in different regions compared to the full sample ($A_\mathrm{all}$ and $B_\mathrm{all}$ respectively). Maps for $A/A_\mathrm{all}$ show values within a few per cent of 1, suggesting no substantial variations. To comment on the statistical significance of the variations, significance maps for $A$ and $B$ are created (right panel of Fig. \ref{fig:2d_anisotropies_Sigma-T}). The significance maps for $A$ show that the most deviating region is $(l,b)=(195^\circ, -20^\circ)$ with a significance of $2.27\sigma$. This region deviates from the rest of the sky by $7.20\pm 3.49\%$.

The sigma maps for $B$ show that the most deviating region is $(l,b)=(65^\circ, -25^\circ)$ with a significance of $2.42\sigma$. The variations obtained in the best-fit slopes have a slightly higher significance, but the most deviating region is still well below $3\sigma$. These simple significance estimates already lead us to conclude that no significant variations are detected in the $\sigma_\mathrm{v}-T$ relation. The statistical \emph{insignificance} of the $\sigma_\mathrm{v}-T$ angular variations is further confirmed in Sect. \ref{sec:isotropic_MC_1}.

\subsection{Temperature bias}
\label{sec:temp_bias_calc}
We perform the following steps to convert the variations in best-fit normalisation into $T$ bias. To quantify the change in $T$ for a given $\sigma_\mathrm{v}=A\times T^B$ across the sky, we use 
\begin{equation}
    \label{eq:temperature_variation}
    \frac{A_2}{A_1}  = \left(\frac{T_1}{T_2}\right)^{B}.
\end{equation}
We take Region One as the region of interest and Region Two as the rest of the sky. The value of $B$ is the best-fit slope for the rest of the sky and is assumed to be the same for both regions. The quantity $\Delta T=T_1-T_2$ is the temperature bias between a region and the rest of the sky. We quantify this in terms of percentage bias using the relation 
\begin{equation}
    \label{eq:temperature_variation_percentage}
    \%\ \mathrm{bias} = \frac{\Delta T}{T_2} = \left(\frac{A_2}{A_1}\right)^{\frac{1}{B}}-1.
\end{equation}

% Consider a cluster with $T$ and $\sigma_\mathrm{v}$ equal to $T_1$ and $\sigma_\mathrm{v}$ located in a cone with $A=A_1$. The best-fit normalisation in this cone varies from the rest of the sky by $\Delta A$. We calculate the change in $T$ for this cluster if it were present outside the cone such that it has the same $\sigma_\mathrm{v}$. Let $A$ for the rest of the sky be $A_2$ such that $\Delta A = A_1 - A_2 $ and the new $T$ be given as $T_2$ such that $\Delta T = T_1 - T_2$. The ratio of scaling relation in both regions is given as

% \begin{equation}
%     \label{eq:temperature_variation}
%     \frac{A_2}{A_1}  = \left(\frac{T_1}{T_2}\right)^{B}.
% \end{equation}
% The value of $B$ is the best-fit slope for the rest of the sky and is assumed to be the same for both regions.

% We find temperature bias ($\Delta T$) in terms of percentage bias by dividing $\Delta T$ with the $T$ from the rest of the sky and is given as
% \begin{equation}
%     \label{eq:temperature_variation_percentage}
%     \%\ \mathrm{bias} = \frac{\Delta T}{T_2} = \left(\frac{A_2}{A_1}\right)^{\frac{1}{B}}-1.
% \end{equation}
% When compared to the rest of the sky, this $\Delta T$ will be the same for all clusters inside a cone.

\begin{figure}[htbp]
    \centering
    \includegraphics[width=\hsize]{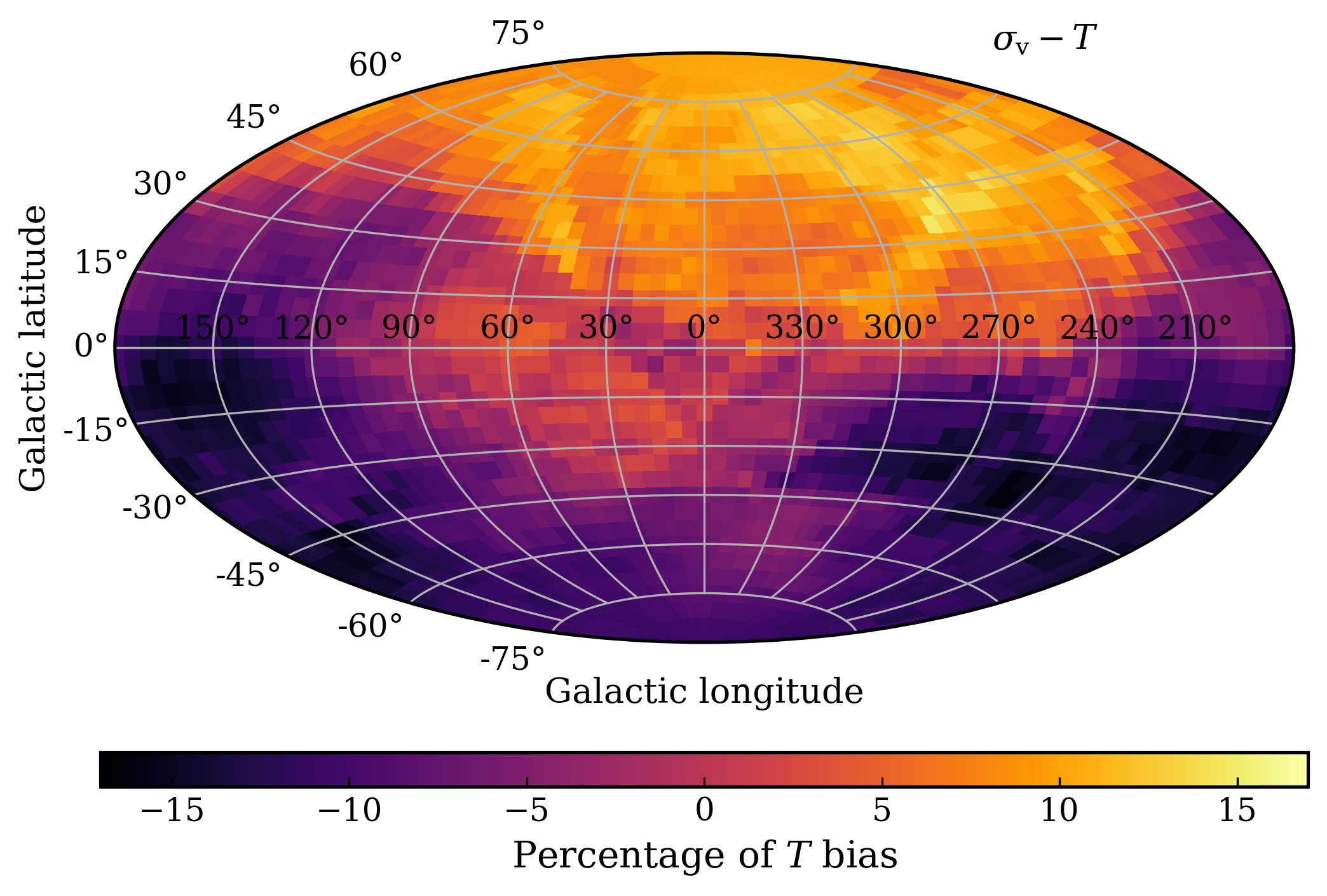}
    \caption{Map of percentage $T$ bias between a region and rest of the sky. Negative $T$ variations refer to an underestimation of $T$ compared to the rest of the sky. To alleviate the observed anisotropies in M21, the behaviour required would have to be opposite to what is observed.}
    \label{fig:2d_temp_var}
\end{figure}

A 2-D map of $\% T$ bias is created to identify regions with over- or underestimation of $T$ (Fig.\ \ref{fig:2d_temp_var}). 
Since $\Delta T$ is calculated from $A$, the region of most bias remains the same. This region shows a $(16.3\pm7.1)\%$ underestimation in $T$ compared to the rest of the sky. There are notable $T$ biases across the sky in the 2-D map, indicating anisotropy. However, these are also accompanied by high uncertainty and already the simple significance analyses above have shown they are not significant. Note that here, the region with the largest deviation from isotropy in M21 has an insignificant but nonetheless \emph{negative} $T$ bias. In contrast, a \emph{positive} bias would be needed to alleviate the anisotropy in M21.

\subsection{Isotropic Monte Carlo simulations}
\label{sec:isotropic_MC_1}
\begin{figure}[htbp]
    \centering
    \includegraphics[width=\hsize]{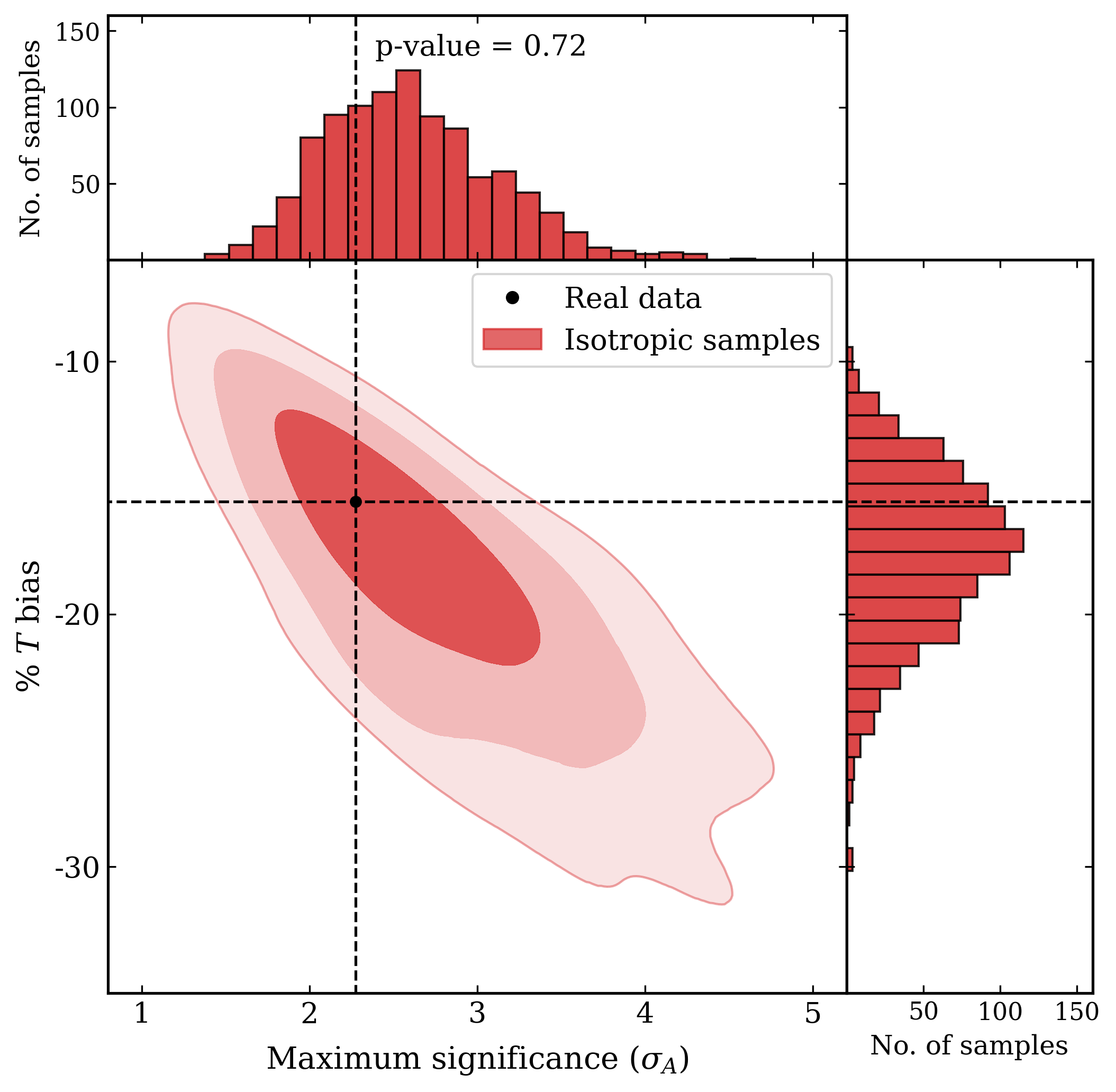}
    \caption{Distribution of maximum sigma values and its $\% T$ bias obtained from the 1000 isotropic Monte Carlo simulated samples of the $\sigma_\mathrm{v}-T$ relation. Shaded contours represent $68\%$, $95\%$, and $99\%$ confidence levels. The value obtained in the real data is shown by a black dashed line on the histograms and by a point on the contour plot. The $p$-value in the histogram of maximum sigma values represents the probability of getting higher significance than the real data.}
    \label{fig:isotropic_random_T}
\end{figure}

Isotropic Monte Carlo simulations are performed for the $\sigma_\mathrm{v}-T$ scaling relation to check the validity of the methodology used to obtain the results from the 2-D analysis and to obtain improved significance estimates. Isotropic samples are created using the method mentioned in Sect. \ref{sec:isotropic_MC}. For these simulated samples, significance maps are created using the same procedure as the real data, and the maximum significance value is noted. We also take note of its corresponding percentage $T$ bias to understand the expected $T$ bias solely due to the statistical scatter of the relation. It is important to note that an isotropic sample with a large scatter around the best-fit line will exhibit a $T$ bias. We can assess the statistical significance of the obtained bias by comparing the expected $T$ bias from the scatter and the real data. The distribution of these quantities obtained from the simulations is shown in Fig. \ref{fig:isotropic_random_T}. Since the maximum variation in the observed data corresponds to a negative $T$ bias, we plot only the negative $T$ bias values from the simulations for their comparison.

The figure shows that the probability of obtaining a sigma value of $2.27\sigma$ or higher is $72\%$ ($p=0.72$), and $\sim18\%$ bias in $T$ is expected purely due to the scatter of the relations. The 2-D distribution of these simulated results demonstrates that real data lies well within the $68\%$ confidence interval, indicating a high probability of such results occurring due to statistical fluctuations.

These results further confirm that the variations in the best-fit normalisation (and, therefore, $T$ biases) are not statistically significant.

%---------------------------------------------------------------------------------------------
\section{Probing cosmic anisotropies}
\label{sec:cosmic_isotropy}
In this section, we present the constraints of the apparent $H_0$ anisotropies using the $L_\mathrm{X}-\sigma_\mathrm{v}$ and $Y_\mathrm{SZ}-\sigma_\mathrm{v}$ relations. We cross-check the results of M21 with an independent cluster property that is different from that used in the original study. 

\subsection{$L_\mathrm{X}-\sigma_\mathrm{v}$ relation}
\begin{figure}[htbp]
    \centering
    \includegraphics[width=0.495\hsize]{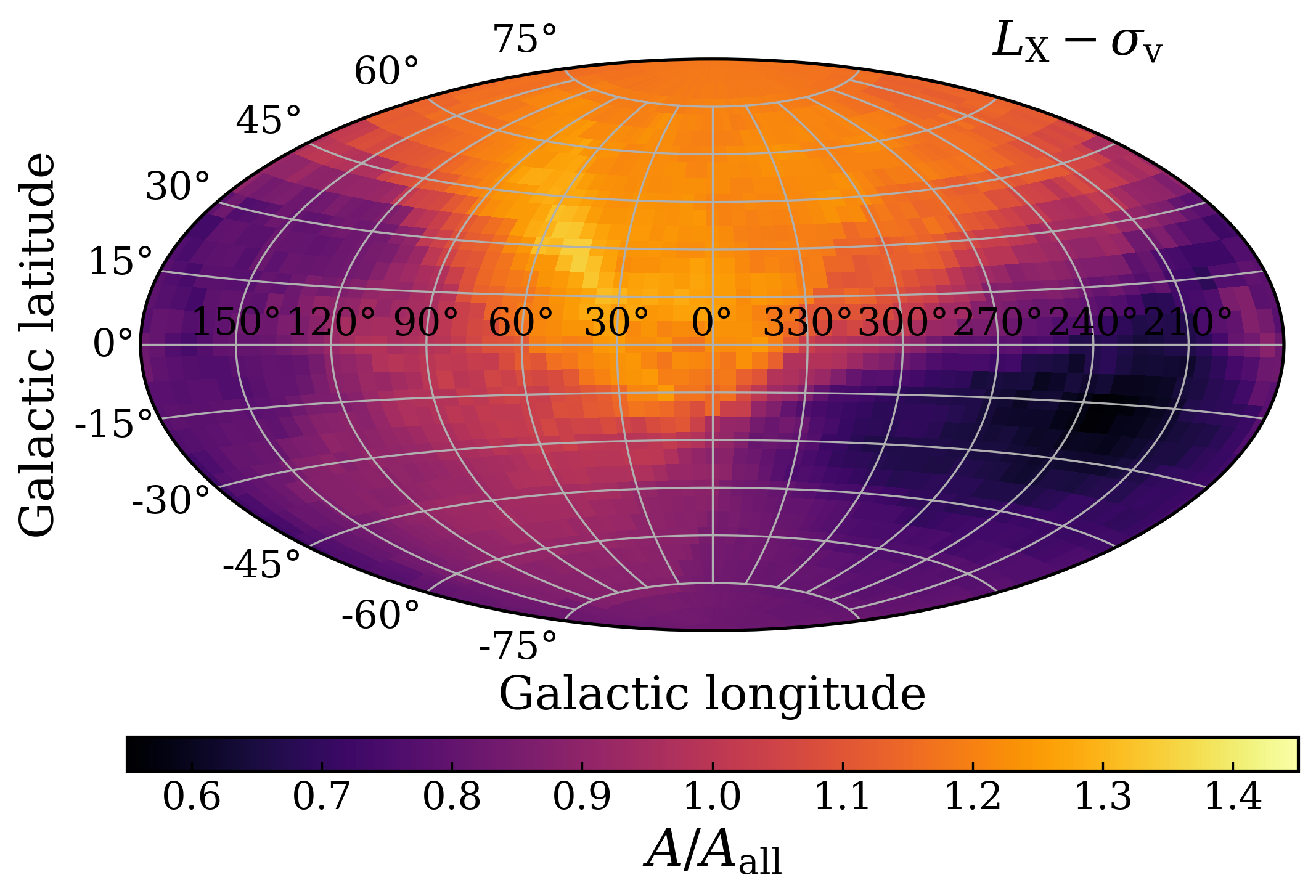}
    \includegraphics[width=0.495\hsize]{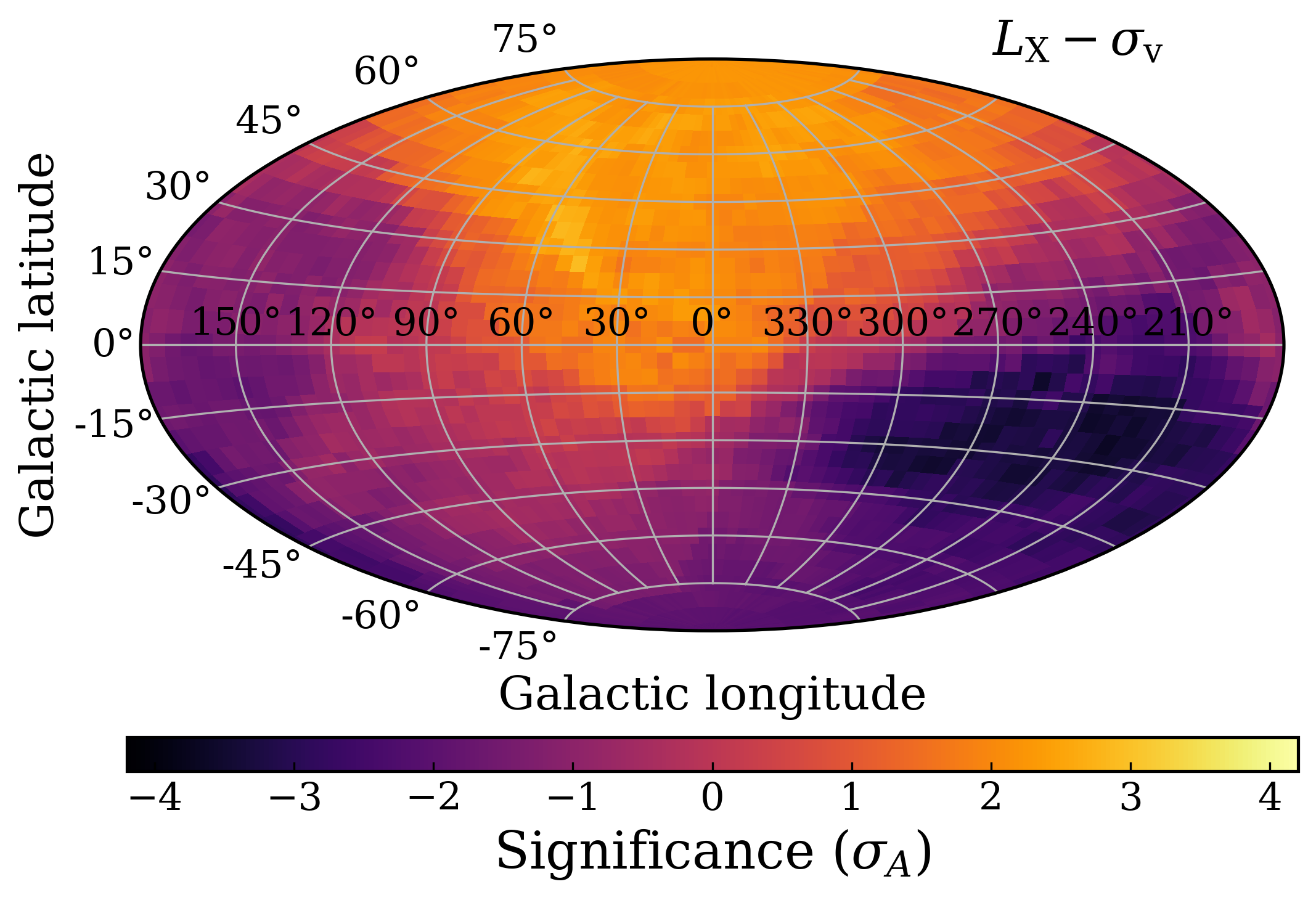}
    \caption{Map of best-fit $A$ compared to the full sample (\textit{left}) and its significance map (\textit{right}) for the $L_\mathrm{X}-\sigma_\mathrm{v}$ relation.}
    \label{fig:2d_norm_Lx-Sigma}
\end{figure}

Using 2-D scanning, we find $A$ variations across the sky and compare them with $A$ of the full sample ($A_\mathrm{all}$). The 2-D maps of $A/A_\mathrm{all}$ along with significance maps are shown in Fig. \ref{fig:2d_norm_Lx-Sigma}. The sigma maps for $A$ show that the most deviating region is $(l,b)=(225^\circ\pm41^\circ, -25^\circ\pm41^\circ)$ with a significance of $-3.66\sigma$. Some variations in the best-fit slopes are also observed in certain regions but at a lower significance. 
% (see Appendix \ref{sec:}). 
However, slope variations are not a focal point of this test as they do not strongly affect the inferred $H_0$ variations (no strong correlation exists between slope and $H_0$ constraints).

\begin{figure}[htbp]
    \label{fig:2d_H0_Lx-Sigma}
    \centering
    \includegraphics[width=\hsize]{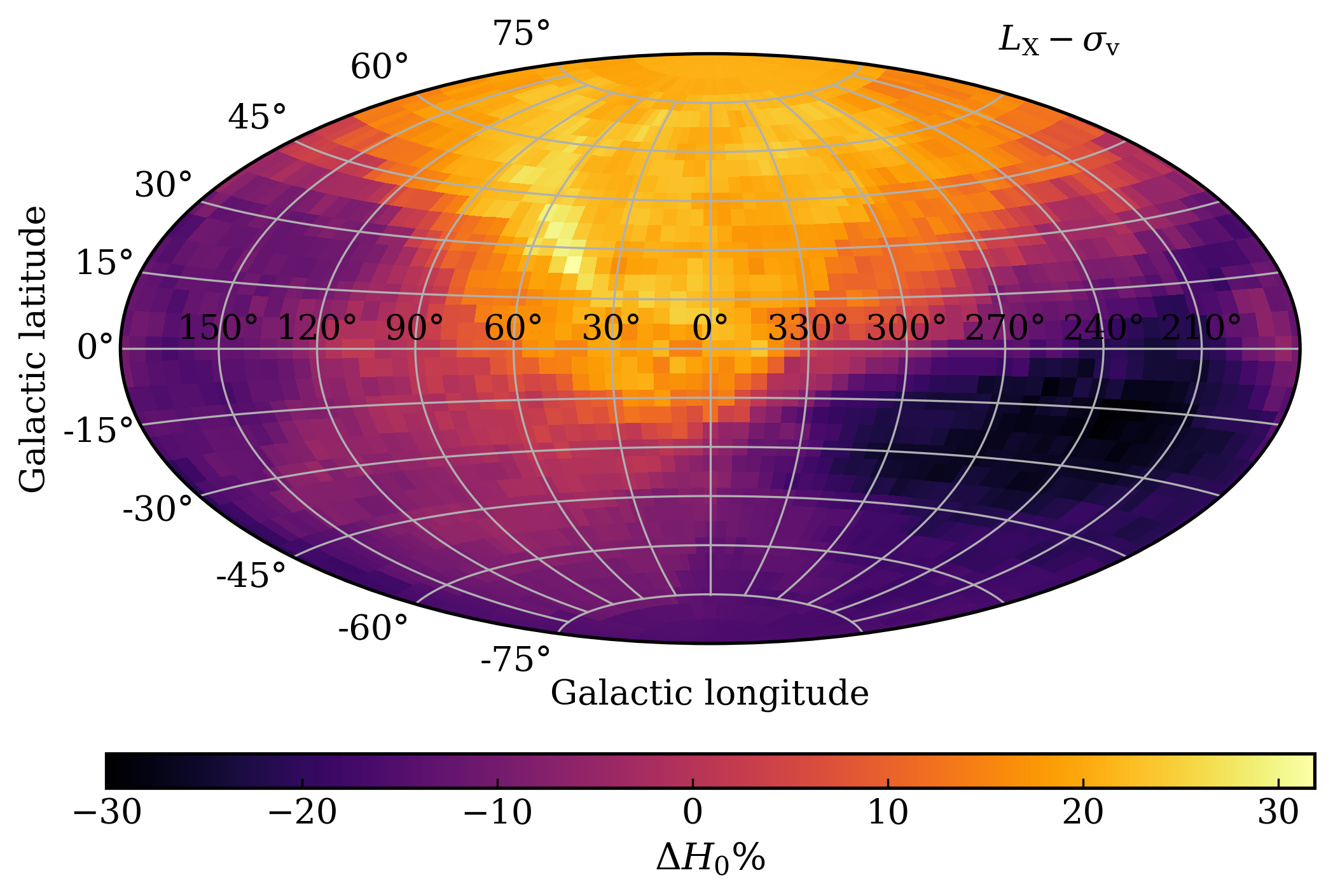}
    \caption{$H_0$ angular variation map created from the $L_\mathrm{X}-\sigma_\mathrm{v}$ relation.}
\end{figure}

The $H_0$ angular variation map obtained from the $L_\mathrm{X}-\sigma_\mathrm{v}$ relation is shown in Fig. \ref{fig:2d_H0_Lx-Sigma}. The region exhibiting the most statistically significant deviation remains consistent with the $A$ maps. In terms of $H_0$, this region differs from the rest of the sky by $\Delta H_0=28.3\pm6.6\%$.

\begin{figure}[htbp]
    \centering
    \includegraphics[width=0.65\hsize]{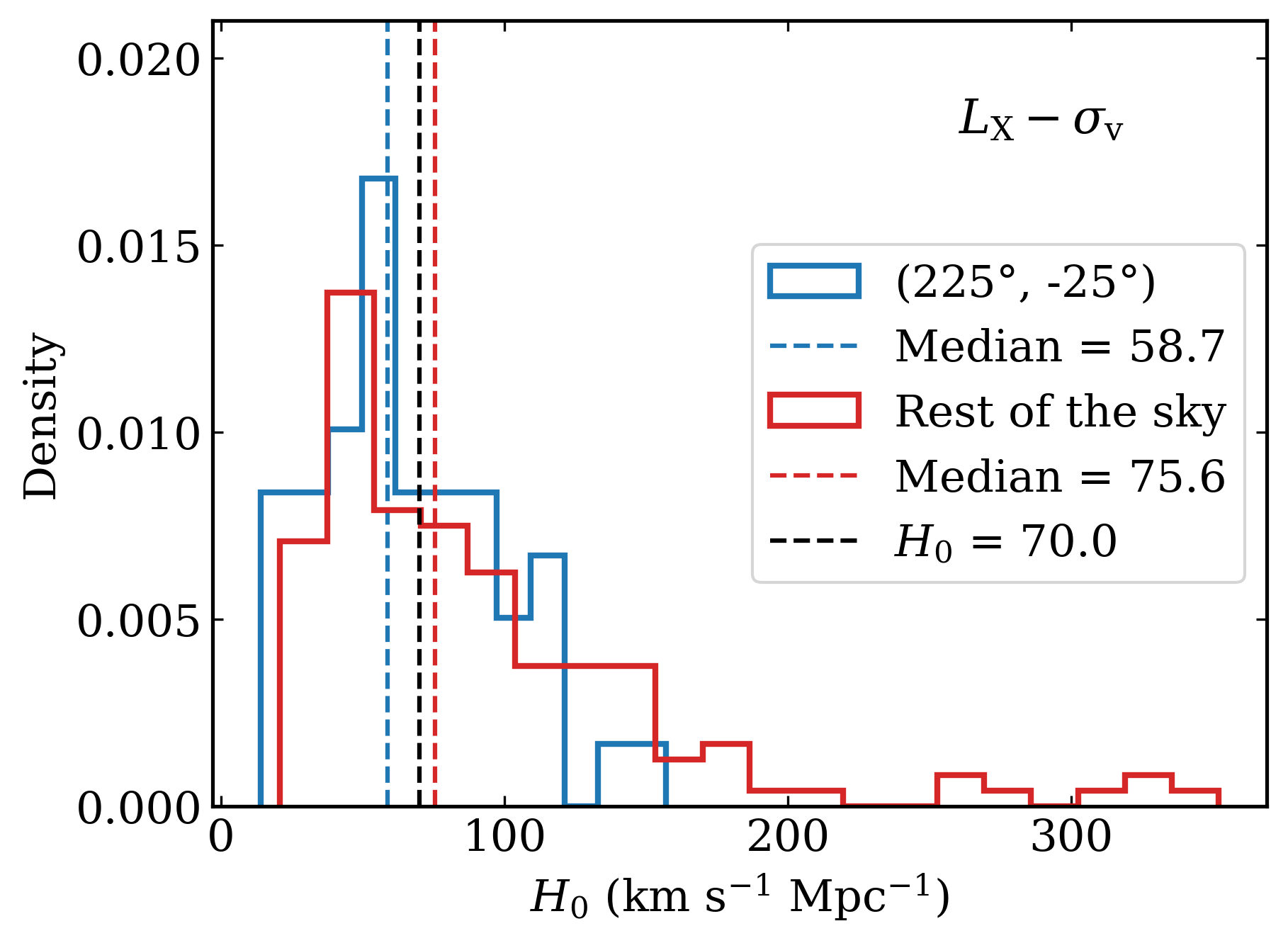}
    \includegraphics[width=0.65\hsize]{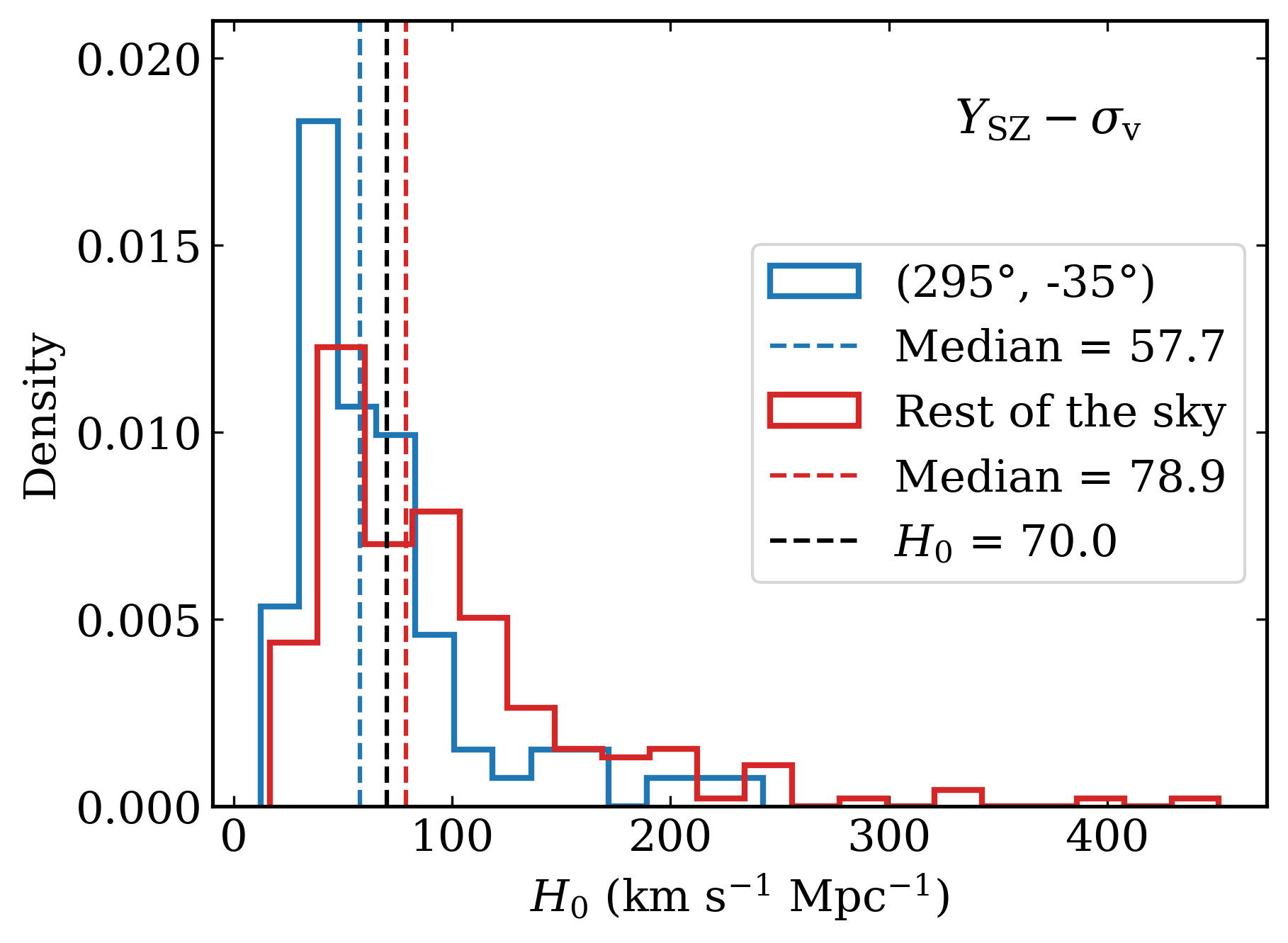}
    \caption{Distribution of $H_0$ values corresponding to each data point in the region of maximum anisotropy and rest of the sky for the relations $L_\mathrm{X}-\sigma_\mathrm{v}$ (\textit{top}) and $Y_\mathrm{SZ}-\sigma_\mathrm{v}$ (\textit{bottom}). The median of both distributions is shown by dashed lines along with $H_0 = \SI{70}{km\ s^{-1}\ Mpc^{-1}}$.\\
    Note that both regions have different numbers of clusters, and thus, to compare both, the density of the distribution is plotted.}
    \label{fig:H0_dist}
\end{figure}

In order to better understand the variation in data points (i.e., individual clusters) between the region of maximum anisotropy $(l=225^\circ, b=-25^\circ)$ and the rest of the sky, we created a distribution of $H_0$ values corresponding to the data points in these regions and compared them with each other (see Fig. \ref{fig:H0_dist}). We assigned a value of $H_0$ to each data point based on its vertical distance from the best-fit line. Points on the line were assigned $H_0 = \SI{70}{km\ s^{-1}\ Mpc^{-1} }$. The $H_0$ value for a point with a $y$ value of $y_i$ and the best-fit $y=y_\mathrm{fit}$ is calculated using the formula $H_{0,\ i} = 70\times \left(\frac{y_i}{y_\mathrm{fit}}\right)^\frac{1}{2}$. On average, data points in the region $(l=225^\circ, b=-25^\circ)$ tend to have lower $H_0$ and thus lie below the best-fit line. The rest of the sky contains a larger number of data points that lie above the best-fit line, as observed from the distribution tail.

\subsection{$Y_\mathrm{SZ}-\sigma_\mathrm{v}$ relation}
\begin{figure}[htbp]
    \centering
    \includegraphics[width=0.495\hsize]{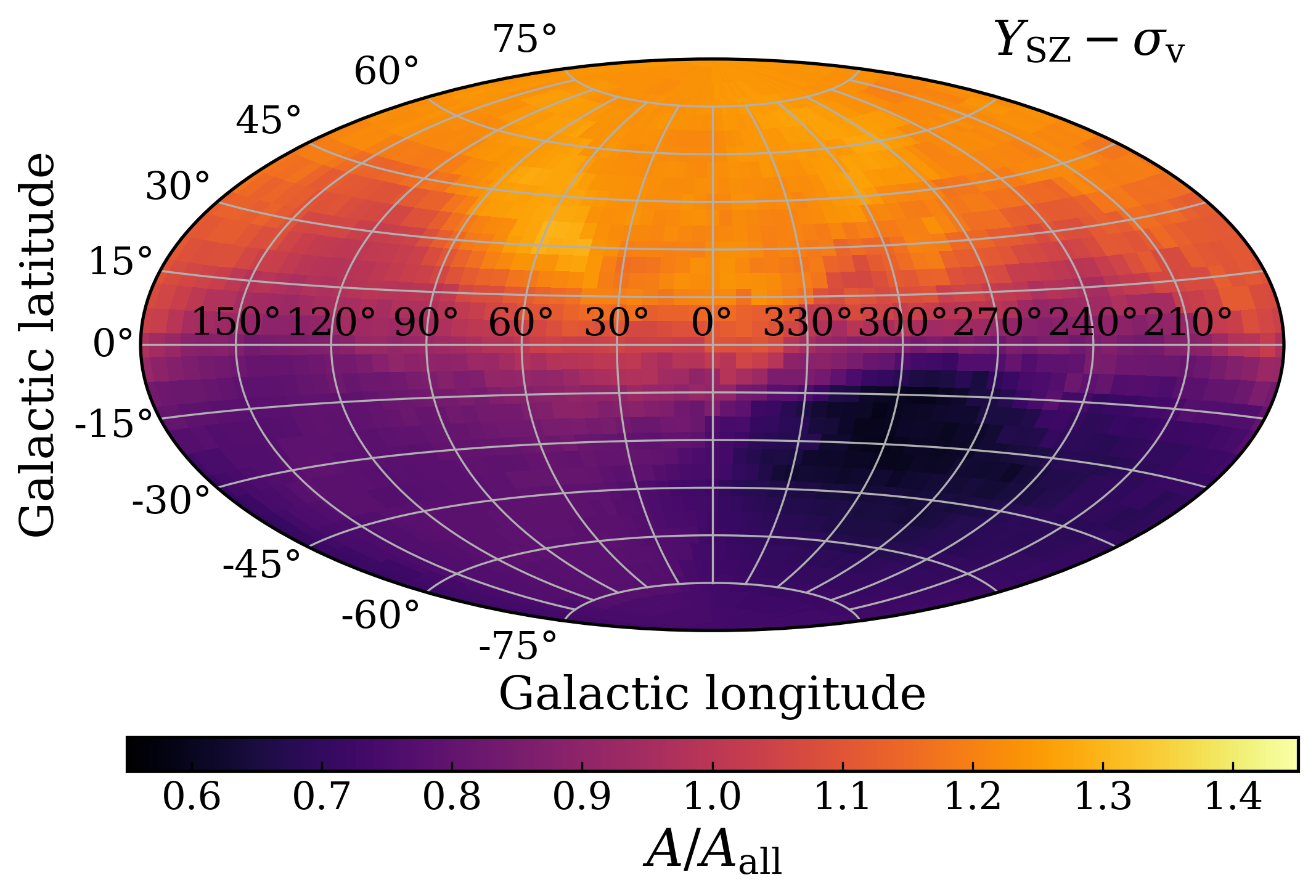}
    \includegraphics[width=0.495\hsize]{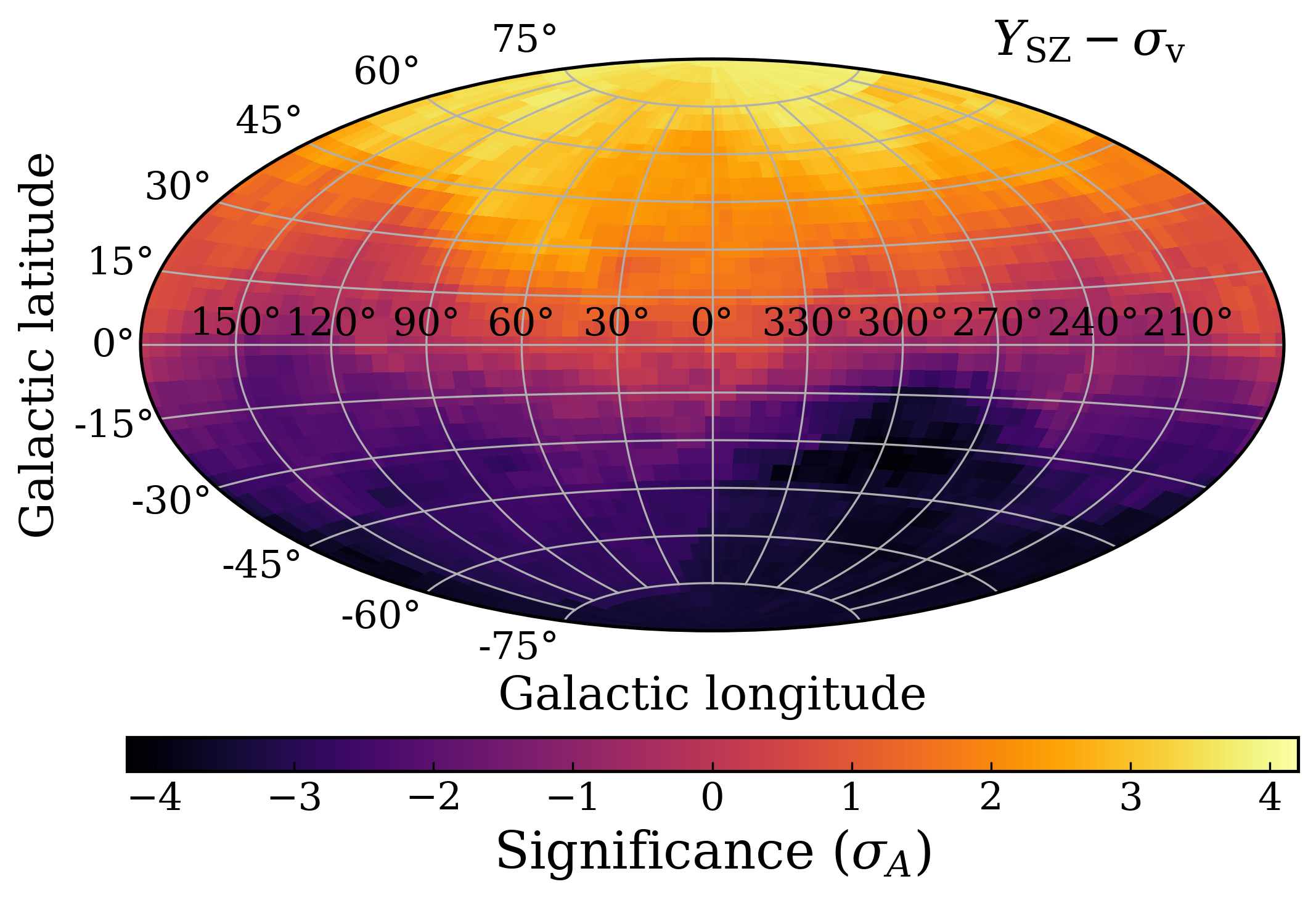}
    \caption{Map of best-fit $A$ compared to the full sample (\textit{left}) and its significance map (\textit{right}) for the $Y_\mathrm{SZ}-\sigma_\mathrm{v}$ relation. Both maps have the same colour scale as Fig. \ref{fig:2d_norm_Lx-Sigma}}
    \label{fig:2d_norm_Ysz-Sigma}
\end{figure}

Fig. \ref{fig:2d_norm_Ysz-Sigma} shows the maps of best-fit $A$ of different regions compared to the full sample and their sigma maps.
The direction showing the largest $A$ deviation is $(l,b)\sim(295^\circ \pm58^\circ, -35^\circ\pm58^\circ)$ with a statistical significance of $-4.13\sigma$. This is in a similar direction to the most deviating region in the $L_\mathrm{X}-\sigma_\mathrm{v}$ analysis.

One interesting feature of this map is that, on average, the Northern Galactic Hemisphere shows higher $A$ values than the Southern one. The best-fit $A$ of all the clusters in the two Galactic Hemispheres differs by $3.35\sigma$. We were unable to identify a systematic bias causing this phenomenon; therefore, it is plausible that this discrepancy may be a random occurrence. We test the possibility of assuming an incorrect redshift evolution factor and its effects on detecting anisotropy and this feature in Sect. \ref{sec:z_cuts_evolution}

\begin{figure}[htbp]
    \centering
    \includegraphics[width=\hsize]{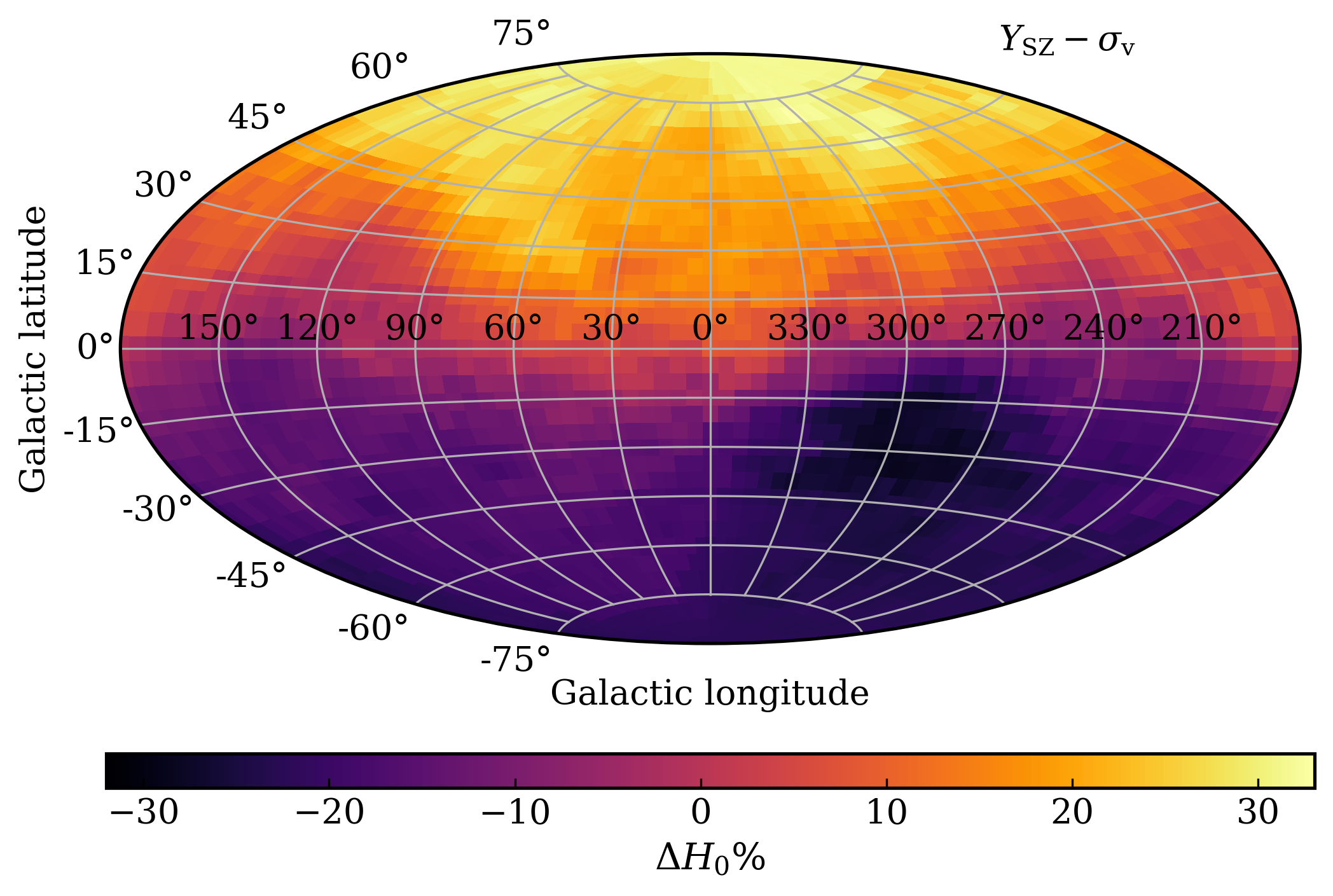}
    \caption{$H_0$ angular variation map created from the $Y_\mathrm{SZ}-\sigma_\mathrm{v}$ relation. The map has the same colour scale as Fig. \ref{fig:2d_H0_Lx-Sigma}}
    % \includegraphics[width=\hsize]{figs/11.b_H0sigma_Ysz-Sigma.png}
    % \caption{Same as Fig. \ref{fig:2d_H0_Lx-Sigma}, for $Y_\mathrm{SZ}-\sigma_\mathrm{v}$ relation. Both maps have the same colour scale as Fig. \ref{fig:2d_H0_Lx-Sigma}}
    \label{fig:2d_H0_Ysz-Sigma}
\end{figure}

The $H_0$ angular variation maps that are derived from the $Y_\mathrm{SZ}-\sigma_\mathrm{v}$ relation are shown in the Fig. \ref{fig:2d_H0_Ysz-Sigma}. The region exhibiting the most statistically significant deviation differs from the rest of the sky by $\Delta H_0=28.5\pm6.4\%$.
The $H_0$ distributions in the region of maximum anisotropy $(l=295^\circ, b=-35^\circ)$ and the rest of the sky (Fig. \ref{fig:H0_dist}) reveal a similar distribution to the ones obtained from $L_\mathrm{X}-\sigma_\mathrm{v}$ analysis. 

\subsection{Joint analysis of $L_\mathrm{X}-\sigma_\mathrm{v}$ and $Y_\mathrm{SZ}-\sigma_\mathrm{v}$ relations}

The apparent anisotropy information of the $L_\mathrm{X}-\sigma_\mathrm{v}$ and $Y_\mathrm{SZ}-\sigma_\mathrm{v}$ scaling relations can be combined into one map that shows $H_0$ angular variations. By multiplying the $H_0$ posterior probability distributions of the two relations, the combined $H_0$ posterior probability distribution can be obtained.
By using the combined posterior probability distribution to constrain the best-fit $H_0$ for each cone, we generate a combined map showing the percentage change in $H_0$ and its significance.

\subsubsection{Limitations of joint analysis}
\label{sec:isotropic_liminations}
There are a couple of things to note here that could overestimate the significance of the detected anisotropies. First, we assume there is no bias in $\sigma_\mathrm{v}$ measurements. Since it is used in both the scaling relations, a bias could overestimate the significance. Second, this method of combining the likelihoods assumes there is no correlation between the scatter of $L_\mathrm{X}$ and $Y_\mathrm{SZ}$ at fixed $\sigma_\mathrm{v}$.

\begin{figure}[htbp]
    \centering
    \includegraphics[width=0.75\hsize]{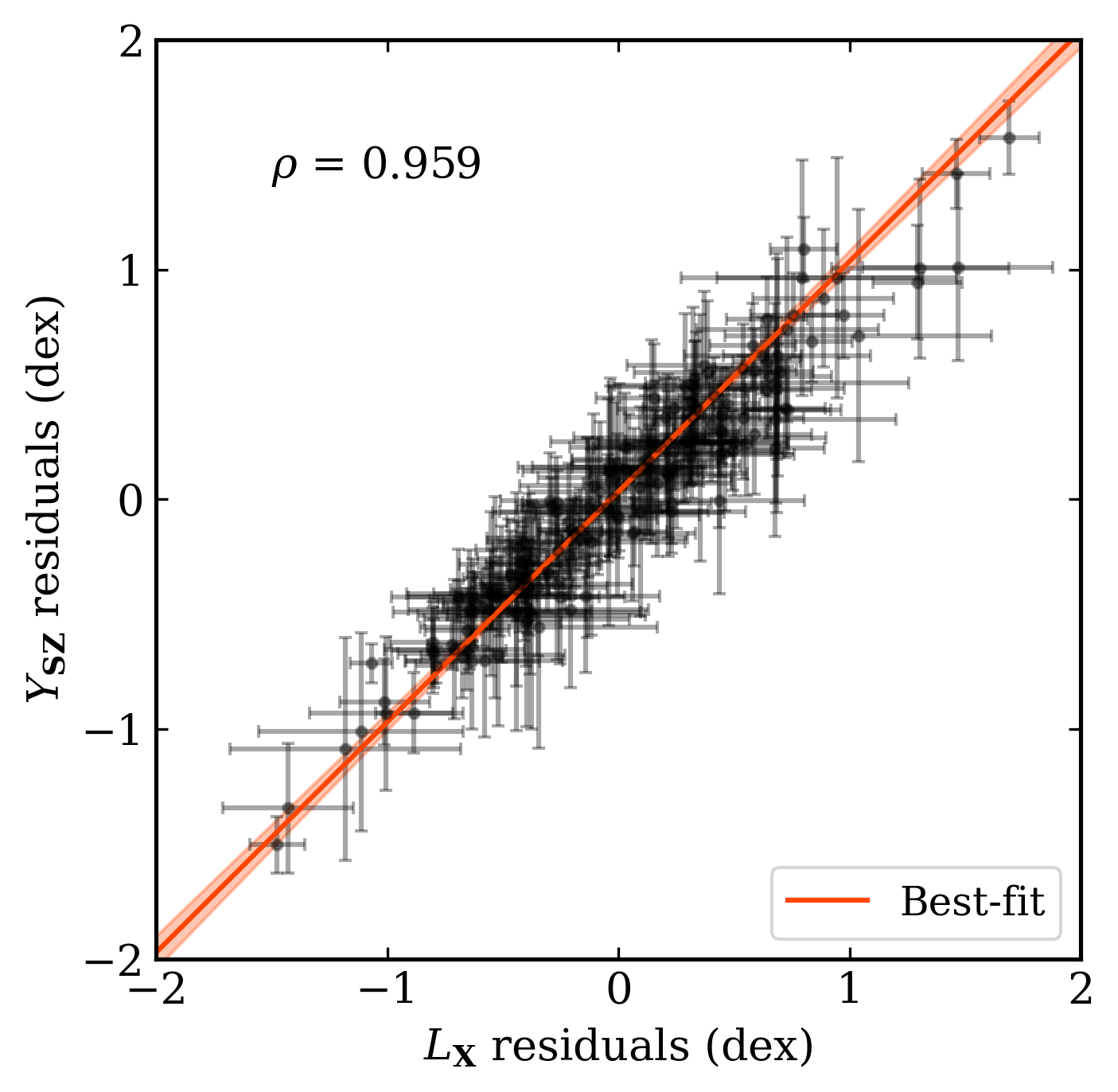}
    \caption{Correlation between the residuals of the
    relations $L_\mathrm{X}-\sigma_\mathrm{v}$ and $Y_\mathrm{SZ}-\sigma_\mathrm{v}$. The best-fit line is shown in orange, along with the correlation coefficient.}
    \label{fig:Residual_corr}
\end{figure}

Previous studies like \citet{Nagarajan_19} have shown a positive correlation between the scatter. We have 180 clusters with both $L_\mathrm{X}$ and $Y_\mathrm{SZ}$ measurements, and we use them to find the correlation between their scatter from their respective scaling relation. Fig. \ref{fig:Residual_corr} shows a plot of residuals of the two relations along with their best-fit line. The correlation coefficient between the residuals is $0.957$, and the best-fit slope of $1.01\pm0.04$. The strong correlation could be the result of a large scatter/uncertainty in $\sigma_\mathrm{v}$, causing the cluster to shift in the same direction for both the $L_\mathrm{X}-\sigma_\mathrm{v}$ and $Y_\mathrm{SZ}-\sigma_\mathrm{v}$ relations.

Despite a correlation, $\sim 37 \%$ clusters present in $Y_\mathrm{SZ}-\sigma_\mathrm{v}$ are unique to this relation and are completely independent of the $L_\mathrm{X}-\sigma_\mathrm{v}$ relation.

\subsubsection{Apparent $H_0$ anisotropy from joint analysis}
\begin{figure}[htbp]
	\centering
	\includegraphics[width=1\hsize]{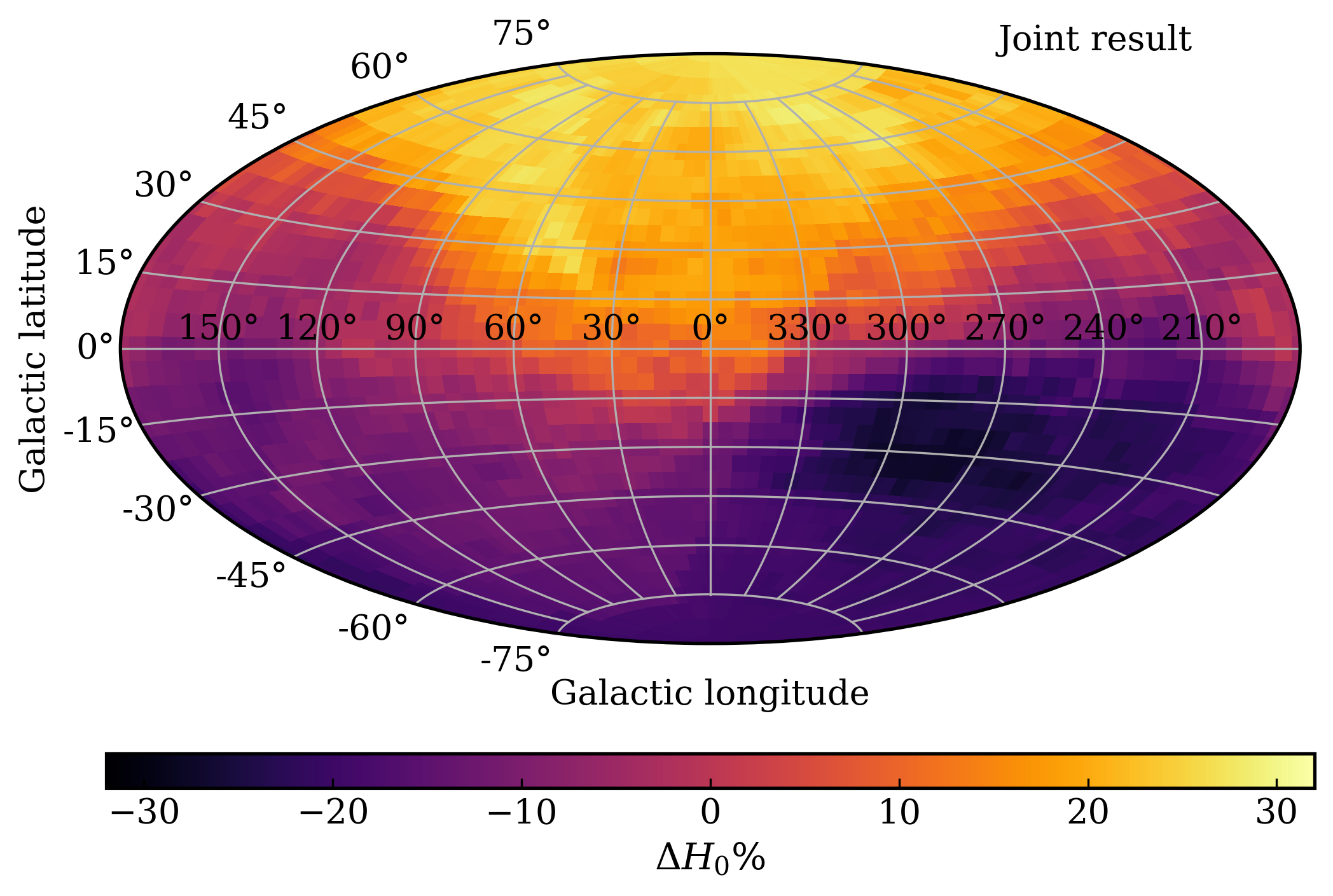}
	\includegraphics[width=1\hsize]{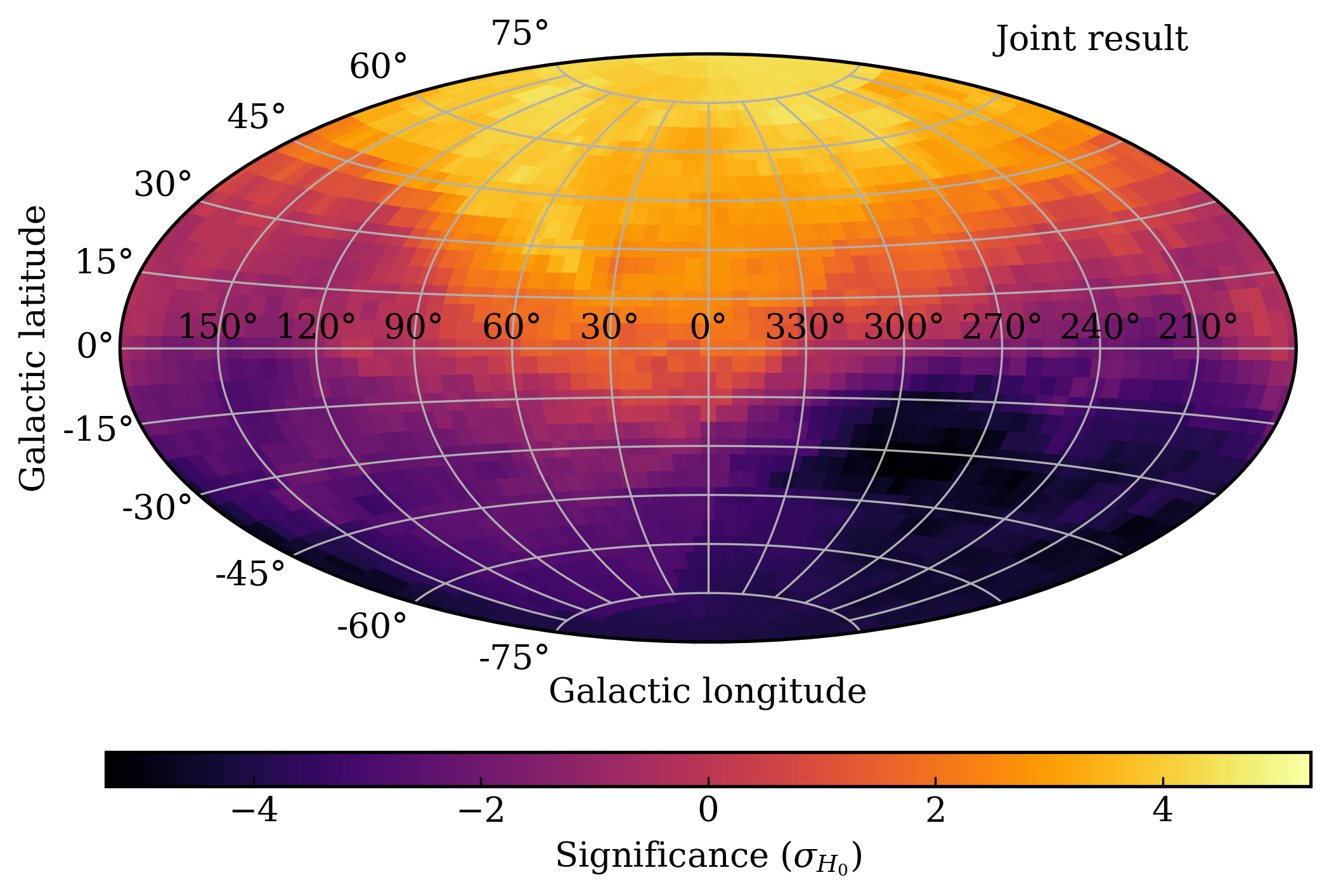}
	\caption{$H_0$ angular variation maps created from the joint analysis of relations $L_\mathrm{X}-\sigma_\mathrm{v}$ and $Y_\mathrm{SZ}-\sigma_\mathrm{v}$ (\textit{top}) and its significance map (\textit{bottom}). Colour scales for the $H_0$ angular variation map are the same as before, but the colour scales for the sigma map have been increased to account for increased significance.}
	\label{fig:Joint_H0_maps}
\end{figure}

The joint $H_0$ map and its significance maps are shown in Fig. \ref{fig:Joint_H0_maps}.
The region with the most deviation is found to be in the direction of $(l,b)\sim(295^\circ \pm71^\circ, -30^\circ\pm71^\circ)$ with $\Delta H_0=27.6\pm4.4\%$ and a statistical significance of $-5.27\sigma$ compared to the rest of the sky. Notice that the significance of the variation has increased compared to the individual analysis, which is a result of smaller $H_0$ uncertainties. The strong correlation between the residuals suggests that the two relations are clearly not independent and that simply combining the likelihoods will undoubtedly result in an overestimation of the significance. 

\subsection{Isotropic Monte Carlo simulations}
\label{sec:isotropic_MC_results}

We perform isotropic Monte Carlo simulations by considering the residual correlation to calculate realistic significance. The method for creating an isotropic MC simulated sample is described in the Sect. \ref{sec:isotropic_MC}.  

First, we generate simulated $L_\mathrm{X}$ and $Y_\mathrm{SZ}$ values based on the best-fit line, taking into account the total scatter of their respective relations. However, simulated $Y_\mathrm{SZ}$ values are calculated \textit{only} for clusters that are unique in the $Y_\mathrm{SZ}-\sigma_\mathrm{v}$ dataset. For the 180 clusters that have both $L_\mathrm{X}$ and $Y_\mathrm{SZ}$ values, we obtain simulated $Y_\mathrm{SZ}$ by taking into account the correlated scatter of $L_\mathrm{X}$ and $Y_\mathrm{SZ}$ with $\sigma_\mathrm{v}$. To do so, we start by determining the residuals of $Y_\mathrm{SZ}$ that correspond to residuals of the simulated $L_\mathrm{X}$ using the correlation between the two quantities (Sect. \ref{sec:isotropic_liminations}).
These residuals are then used to predict $Y_\mathrm{SZ}$ by applying the best-fit parameters of the $Y_\mathrm{SZ}-\sigma_\mathrm{v}$ relation. Finally, a random offset is added to these values, which is drawn from a log-normal distribution centred at $Y_\mathrm{SZ, pred}$ and the standard deviation of
$\sqrt{\sigma^2_{Y_\mathrm{SZ}} + B^2_{Y_\mathrm{SZ}\sigma_\mathrm{v}} \times \sigma^2_{\sigma_\mathrm{v}}}$.

\begin{figure*}[htbp]
    \centering
    \includegraphics[width=0.49\textwidth]{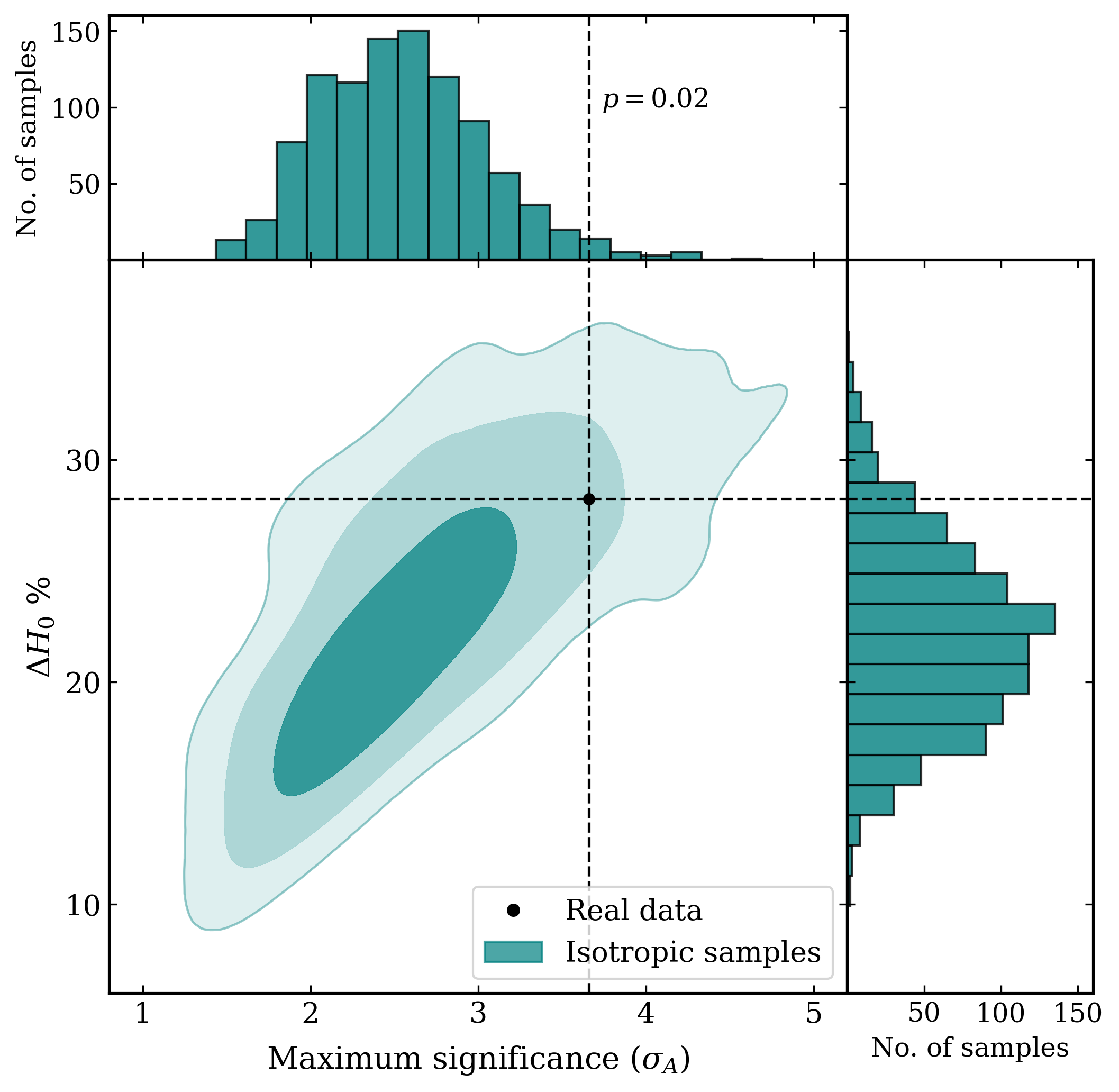}
    \includegraphics[width=0.49\textwidth]{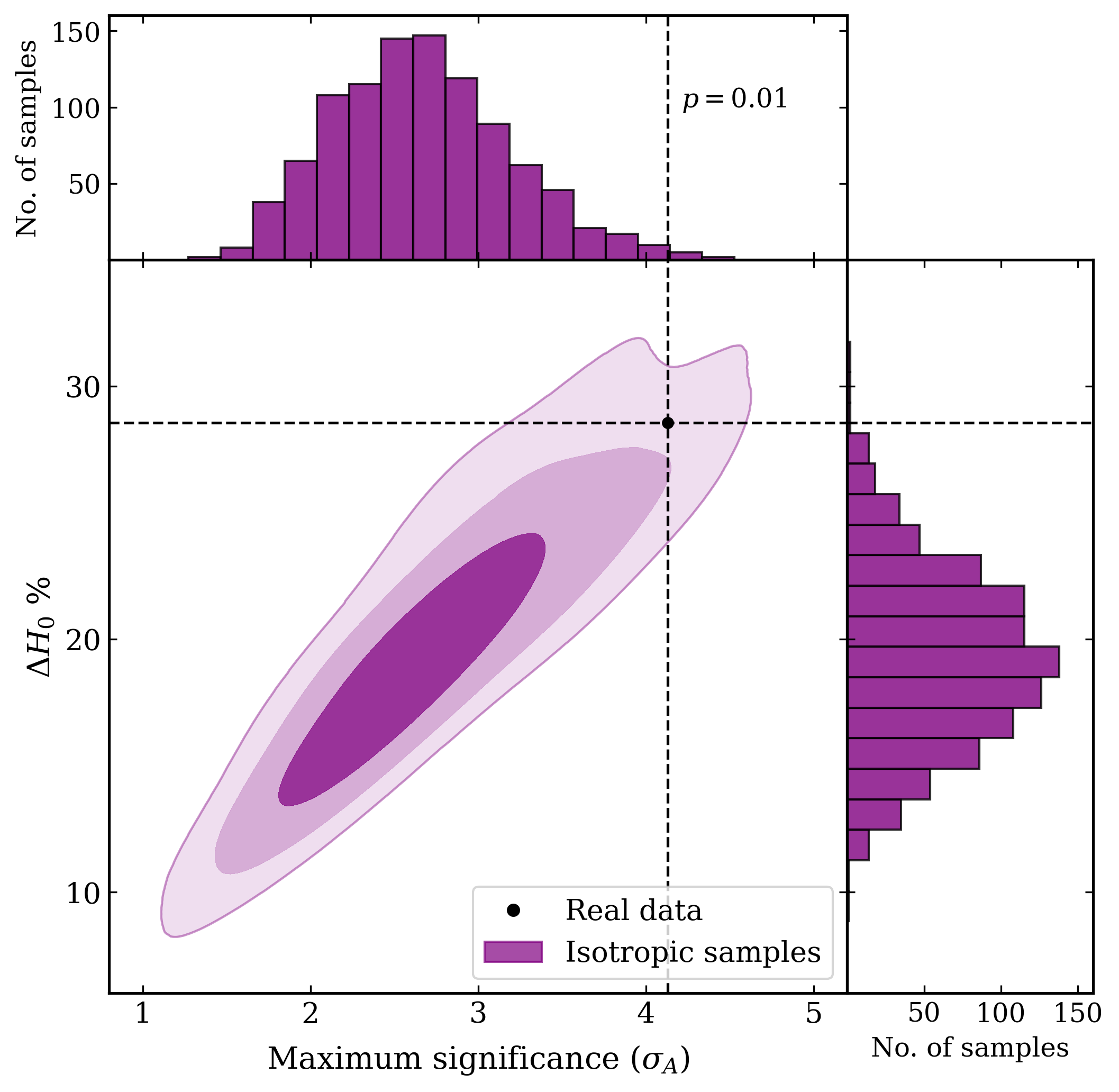}
    \caption{Distribution of maximum sigma values and its $\Delta H_0$ variations obtained from the 1000 isotropic Monte Carlo simulated samples of the $L_\mathrm{X}-\sigma_\mathrm{v}$ (\textit{left}) and $Y_\mathrm{SZ}-\sigma_\mathrm{v}$ (\textit{right}) relations. Shaded contours represent $68\%$, $95\%$, and $99\%$ confidence levels. The value obtained in the real data is shown by a black dashed line on the histograms and by a point on the contour plot. The $p$-value in the histogram of maximum sigma values represents the probability of getting higher significance than the real data.}
    \label{fig:Isotropic_random_Lx_Ysz}
\end{figure*}

This is done 1000 times to create 1000 simulated $L_\mathrm{X}$ and $Y_\mathrm{SZ}$ samples. The distribution of maximum sigma values and $\Delta H_0$ variations obtained from the simulations are compared with the real data in Fig. \ref{fig:Isotropic_random_Lx_Ysz}\footnote{Absolute values for both these quantities are considered.}. We examine the isotropic sample's $\Delta H_0 \%$ corresponding to maximum sigma values to know the expected $\Delta H_0$ caused by the spread in the scaling relation. A large scatter in the scaling relation leads to significant $\Delta H_0$ when considering its sub-samples. Therefore, by comparing $\Delta H_0$ obtained from isotropic samples and the real data, we can assess the significance of these variations given the scatter in the relation.

For the $L_\mathrm{X}-\sigma_\mathrm{v}$, 20 out of the 1000 simulated samples have higher maximum deviation than the real data. Thus, using this method, there is $2\%$ probability ($p=0.02$) of observing a $\geq 3.66 \sigma$ anisotropy in an isotropic Universe. The observed variations in $H_0$ are higher than the expected $\Delta H_0$ of $22.0^{+4.3}_{-4.2}\%$ due to scatter in the relation. The 2-D distribution of these quantities shows that the real data lies within the $95\%$ confidence level. This suggests that significance estimates from Eq. \ref{eq:sigma_map} lead to an overestimation of significance due to the large scatter in the scaling relation.

For the $Y_\mathrm{SZ}-\sigma_\mathrm{v}$ relation, only 11 out of the 1000 simulated samples had higher maximum deviation than the real data. This gives a probability of $1.1\%$ ($p=0.011$) of observing a $\geq 4.13\sigma$ anisotropy in an isotropic Universe. The expected $\Delta H_0$ due to scatter in the relation is $19.2^{+3.7}_{-3.5} \%$, and the real data lies between the $95\%$ and $99\%$ confidence levels in the 2-D distribution of these quantities. 
Again,  significance estimates from Eq. \ref{eq:sigma_map} lead to overestimating the significance. However, due to lower scatter, the $\Delta H_0$ for the $Y_\mathrm{SZ}-\sigma_\mathrm{v}$ relation is more significant than that obtained from the $L_\mathrm{X}-\sigma_\mathrm{v}$ relation.

If the two relations were independent, the joint probability would be the product of the two probabilities. However, as the $\sigma_\mathrm{v}$ data in both the relations are the same and their residuals are correlated, the two relations are not entirely independent.
The joint probability of observing $L_\mathrm{X}-\sigma_\mathrm{v}$ and $Y_\mathrm{SZ}-\sigma_\mathrm{v}$ anisotropies simultaneously is calculated by noting down number of simulated samples that simultaneously satisfy the criteria of having $\geq3.66\sigma$ and $\geq 4.13\sigma$ anisotropies in their respective relations. Using this method, only 1 out of 1000 samples simultaneously have higher anisotropy in both relations. Thus, the probability is given as $p = 0.001$.

In real data, the anisotropies in the two relations are separated by $60.2^\circ$. From the simulated samples, roughly $27.6\%$ samples show lower separation in detected anisotropies than the real data.
The probability of observing higher anisotropies with lower separation is obtained by multiplying the probabilities of the two events. Thus, the probability is $p=2.76\times 10^{-4}$ and it corresponds to a Gaussian $\sigma$ of $3.64\sigma$.

\begin{table*}[htbp]
    \caption{Maximum anisotropy direction and its absolute amplitude for the $L_\mathrm{X}-\sigma_\mathrm{v}$ and $Y_\mathrm{SZ}-\sigma_\mathrm{v}$ relations along with absolute $H_0$ variation compared to the rest of the sky. Results from isotropic MC simulations are shown in terms of the $p$-value, their corresponding Gaussian $\sigma$ values, and the expected $H_0$ variation due to scatter in the relation.}
    \label{tab:H0_results}
    \centering
    \begin{tabular}{lccc|cc}
        \hline
        \hline
        {Scaling} & {Max. anisotropy} & {$H_0$} & {Maximum} & {Isotropic MC} & {Expected} \bigstrut[t] \\
        {relation} & {direction ($l,b$)} & {variation (\%)} & {significance ($\sigma$)} & {$p$-value $(\sigma)$} & {$H_0$ variation (\%)} \bigstrut[b]\\
        \hline
        $L_\mathrm{X}-\sigma_\mathrm{v}$ & {$(225^\circ\pm41^\circ, -25^\circ\pm41^\circ)$} & {$28.3\pm 6.6\%$} & {$3.66\sigma$} & {$p=0.020\ (2.33\sigma)$} & {$22.0^{+4.3}_{-4.2}\%$} \bigstrut\\
        $Y_\mathrm{SZ}-\sigma_\mathrm{v}$ & {$(295^\circ\pm58^\circ, -35^\circ\pm58^\circ)$} & {$28.5\pm 6.4\%$} & {$4.13\sigma$} & {$p=0.011\ (2.54\sigma)$} & {$19.2^{+3.7}_{-3.5} \%$}\bigstrut\\
        \hline
        Joint results & {$(295^\circ\pm71^\circ, -30^\circ\pm71^\circ)$} & {$27.6\pm 4.4\%$} & {$5.27\sigma$} & {$p=2.76\times10^{-4}\ (3.64\sigma)$} & {$19.7^{+2.9}_{-2.5} \%$}\bigstrut\\
        \hline
    \end{tabular}
\end{table*}

The results of the simulations suggest that even in isotropic samples, some variation in $\Delta H_0$ is expected due to the scatter of the relations being used, which biases the measured $\Delta H_0$. It is essential to correct this bias before determining the actual $\Delta H_0$. As a result, the observed $ \Delta H_0 $ in M21 is also somewhat overestimated. However, this overestimation would not affect the statistical significance of the $\Delta H_0$ they detected (at a level of $5.4\sigma$) because this effect is considered in the Monte Carlo simulations they performed.

A summary of results obtained from the relations $L_\mathrm{X}-\sigma_\mathrm{v}$ and $Y_\mathrm{SZ}-\sigma_\mathrm{v}$ along with their joint analysis is presented in Table \ref{tab:H0_results}.

\begin{table*}[htbp]
    \caption{Maximum anisotropy direction for the $L_\mathrm{X}-\sigma_\mathrm{v}$ and $Y_\mathrm{SZ}-\sigma_\mathrm{v}$ relations along with
    $H_0$ angular variation compared to the rest of the sky and the absolute significance of the variations. Results are shown for the X|Y and Y|X methods and different datasets for the two relations.}
    \label{tab:H0_results_comp}
    \centering
    \begin{tabular}{cccccc}
        \hline
        \hline
        Scaling & Dataset & {Best-fit} & {Max. anisotropy} & {$H_0$} & {Maximum} \bigstrut[t]\\
        relation & used & {method} & {direction ($l,b$)} & {variation (\%)} & {significance ($\sigma$)} \bigstrut[b]\\
        \hline
        \multirow{4}{*}{$L_\mathrm{X}-\sigma_\mathrm{v}$}  & \multirow{2}{*}{eeHIFLUGCS} & {X|Y} & {$(225^\circ, -25^\circ)$} & {$28.3\pm 6.6\%$} & {$3.66\sigma$} \bigstrut[t]\\
        & & {Y|X} & {$(305^\circ, -20^\circ)$} & {$22.1\pm 5.2\%$} & {$3.79\sigma$} \bigstrut[b]\\
        & \multirow{2}{*}{MCXC} & {X|Y} & {$(280^\circ, -25^\circ)$} & {$27.0\pm 6.2\%$} & {$3.71\sigma$} \bigstrut[t]\\
        & & {Y|X} & {$(300^\circ, -30^\circ)$} & {$21.8\pm 4.6\%$} & {$4.13\sigma$} \bigstrut[b]\\
        \hline
        \multirow{6}{*}{$Y_\mathrm{SZ}-\sigma_\mathrm{v}$} &\multirow{2}{*}{$\mathrm{S/N} \geq 2$} & {X|Y} & {$(295^\circ, -35^\circ)$} & {$28.5\pm 6.4\%$} & {$4.13\sigma$} \bigstrut[t]\\
        & & {Y|X} & {$(310^\circ, -35^\circ)$} & {$21.8\pm 4.6\%$} & {$4.51\sigma$}  \bigstrut[b]\\
        &\multirow{2}{*}{$\mathrm{S/N} \geq 3$} & {X|Y} & {$(185^\circ, -40^\circ)$} & {$23.2\pm 5.1\%$} &  {$3.86\sigma$} \bigstrut[t]\\
        & & {Y|X} & {$(315^\circ, -30^\circ)$} & {$21.1\pm 4.9\%$} & {$4.19\sigma$}  \bigstrut[b]\\
        &\multirow{2}{*}{$\mathrm{S/N} \geq 4.5$} & {X|Y} & {$(185^\circ, -35^\circ)$} & {$22.5\pm 6.1\%$} & {$3.24\sigma$} \bigstrut[t]\\
        & & {Y|X} & {$(185^\circ, -40^\circ)$} & {$17.9\pm 4.41\%$} & {$3.66\sigma$}  \bigstrut[b]\\
        \hline
    \end{tabular}
\end{table*}

%---------------------------------------------------------------------------------------------
\subsection{Comparison with different data samples and best-fit methods}
\label{sec:comparison}
We compared our results with different data sets to check for apparent cosmic anisotropy. For $L_\mathrm{X}$, we used the MCXC catalogue data and compared it with the results obtained from the eeHIFLUGCS catalogue. For the $Y_\mathrm{SZ}$ data, we compared the results with different S/N cuts such as S/N $\geq 3$ and S/N $\geq 4.5$. The latter is chosen since it is the S/N cut chosen in the official PLANCKSZ2 cluster catalogue \citep{PSZ2_15}. To check if the results are consistent between the fitting methods, the results from both the X|Y and Y|X methods are presented. 
The general behaviour of these datasets for different best-fit analysis methods is shown in Table \ref{tab:extra_scaling_relations}.

Table \ref{tab:H0_results_comp} shows the direction of maximum anisotropy, its amplitude, and $\Delta H_0 \%$ for these datasets using both fitting methods. 
The results from different fitting methods and datasets return consistent results with each other. These results indicate that the general direction of the most deviating regions is similar for different datasets and analysis methods. The direction of anisotropy is found in the general direction of $l>180^\circ$ and $b<0^\circ$ for all datasets. The Y|X fitting method consistently yields less angular $H_0$ variation than the X|Y fitting method, possibly due to less scatter in the Y direction. The significances are calculated using the Eq. \ref{eq:sigma_map} and are mentioned only for a consistency check. As shown in Sect. \ref{sec:isotropic_MC_results}, these significances are overestimated due to a large scatter in the relation and isotropic MC simulations are required to quantify the true statistical significance of the observed anisotropy. 

%---------------------------------------------------------------------------------------------

\section{Discussion}
\label{sec:discussion}

\subsection{Comparison with M21}
\label{sec:comparison_M21}

\subsubsection{Probability of $T$ bias}
\label{sec:comp_sims}
In the work done by M21, the most deviating region found for the relations $L_\mathrm{X}-T$ and $Y_\mathrm{SZ}-T$ are ($l=270^\circ$, $b=-9^\circ$) and ($l=268^\circ$, $b=-16^\circ$) respectively. If the anisotropies in these relations are due to $T$ biases, then for a fixed $L_\mathrm{X}$ or $Y_\mathrm{SZ}$, $T$ in these regions should be biased high compared to the rest of the sky ($10.6\pm4.6\%$ and $13.2\pm4.3\%$ respectively). These are opposite to the results obtained in this work since we obtain $T$ bias of $\sim-9.3\%$ and $\sim-13.1\%$ in the same regions.
\begin{figure}[htbp]
    \centering
    \includegraphics[width=0.72\hsize]{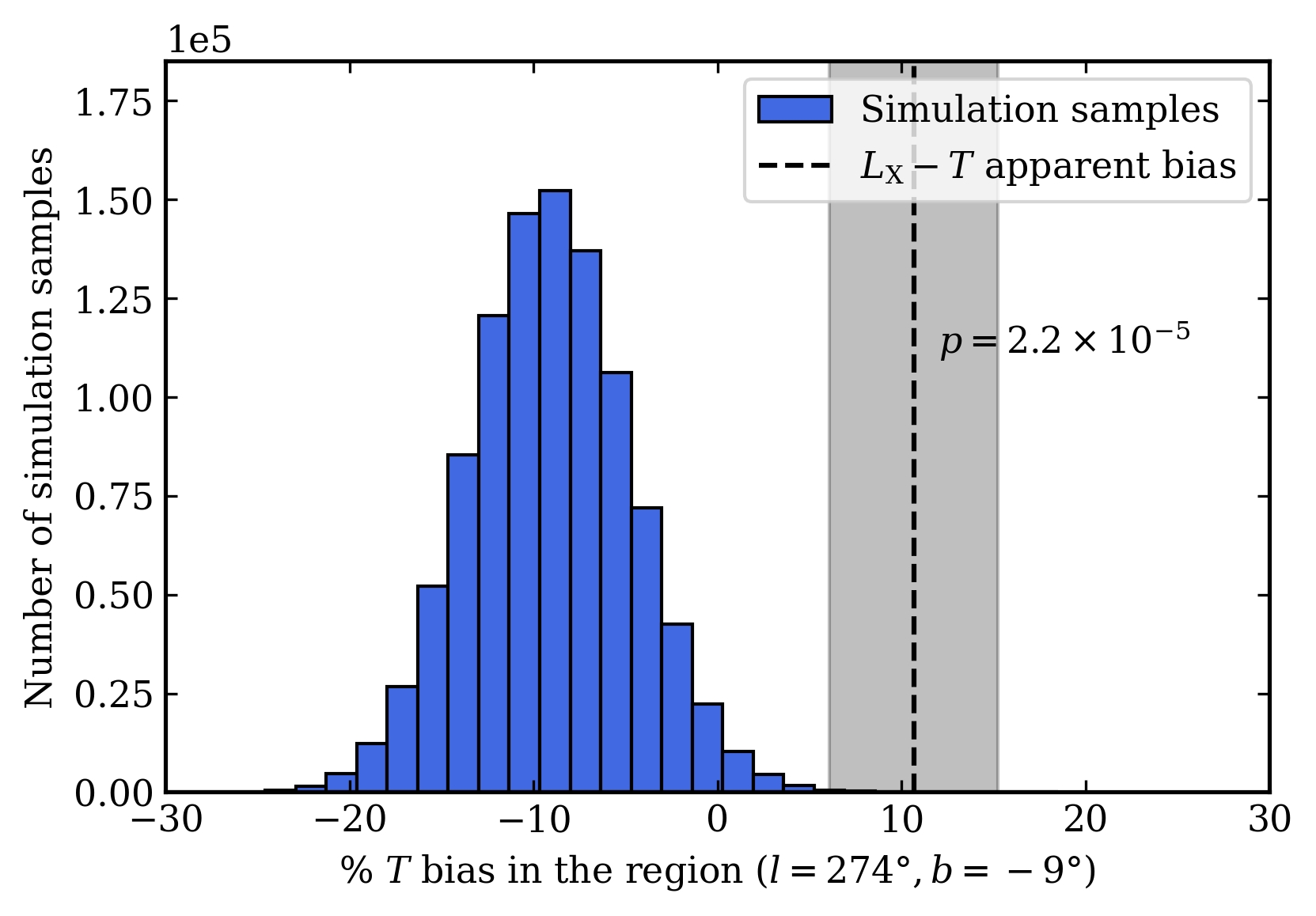}
    \includegraphics[width=0.72\hsize]{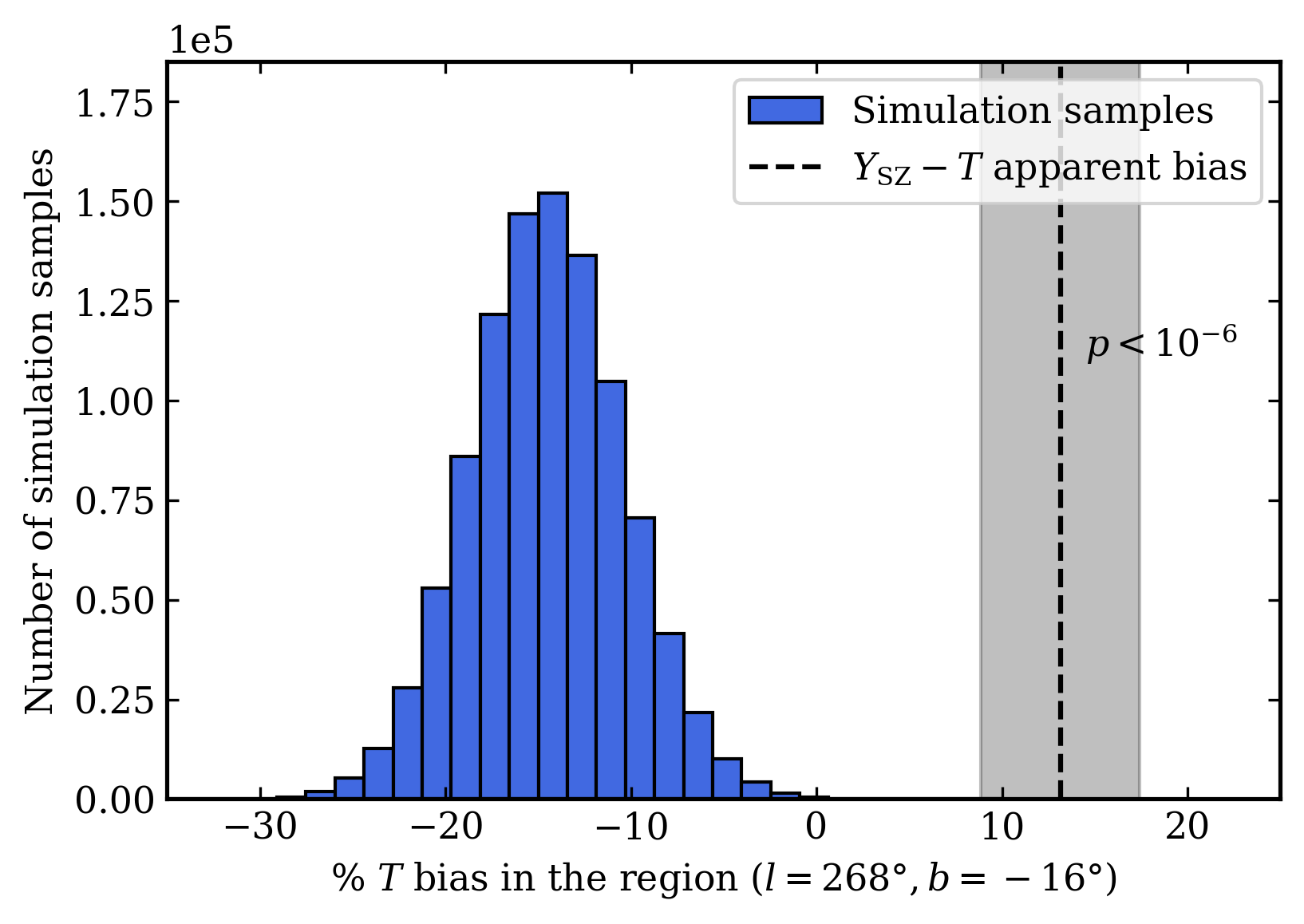}
    \caption{Distribution of $T$ bias in the $R_{LT}$ (\textit{top}) and $R_{YT}$ (\textit{bottom}) regions for 1 million simulated samples. The vertical black dashed line indicates the $T$ bias needed to account for the results of M21 ($\Delta T_{LT}$ and $\Delta T_{YT}$), and the shaded region represents its $1\sigma$ uncertainties. The $p$-value displayed next to it shows the probability of obtaining samples with a higher $T$ bias than $\Delta T_{LT}$ and $\Delta T_{YT}$, respectively.}
    \label{fig:Comp_sim_region}
\end{figure}
For ease of notation, the regions of maximum anisotropy for both relations are referred to as the $R_{LT}$ and $R_{YT}$ regions, respectively. Their $T$ biases are referred to as $\Delta T_{LT}$ and $\Delta T_{YT}$, respectively. We perform additional simulations to quantify how strongly the results from the analysis of $\sigma_\mathrm{v}-T$ confirm or reject the possibility of a systematic bias in the cluster $T$ measurements. 

We draw random values of $T$ from a log-normal distribution centred at the real $T$ measurement, and its standard deviation is given as the total scatter of the scaling relation. These simulations are similar to the isotropic simulations with the main difference being that the $T$ are drawn around their real values instead of those predicted from the scaling relation.
The $T$ bias for all simulated samples from the most deviating regions $R_{LT}$ and $R_{YT}$  are noted and are compared with $\Delta T_{LT}$ and $\Delta T_{YT}$. Fig. \ref{fig:Comp_sim_region} shows the distribution of $T$ bias in these regions obtained from one million simulated samples. Negative $T$ bias refers to an underestimation in $T$ compared to the rest of the sky.

The probability that the observed $L_\mathrm{X}-T$ and $Y_\mathrm{SZ}-T$ anisotropies in M21 are caused by an overestimation of $T$ is $2.2\times 10^{-5}$ ($4.24\sigma$) and $<10^{-6}$ ($>4.89\sigma$) respectively. The presence of a $T$ decrease in our work -- albeit non-significant -- strongly disfavours a $T$ bias as the cause of the M21 observed anisotropies.

\subsubsection{Joint result of cosmic anisotropy with M21}
\label{sec:Joint_M21}
The M21 study found a $(9.0\pm 1.7)\%$ angular variation in $H_0$ in the direction of $(l,b) = \left(280^{\circ +35^\circ}_{\ -35^\circ}, -15^{\circ +20^\circ}_{\ -20^\circ}\right)$. There is a strong agreement in the direction of maximum anisotropy between these findings and the results from $L_\mathrm{X}-\sigma_\mathrm{v}$ and $Y_\mathrm{SZ}-\sigma_\mathrm{v}$ analysis. One of the main distinctions between the two studies is the $(27.6\pm4.4)\%$ angular variation in $H_0$ from the joint analysis in this work, which is significantly higher than the variation found by M21 due to a large scatter.

\begin{figure}[htbp]
    \centering
    \includegraphics[width=0.495\hsize]{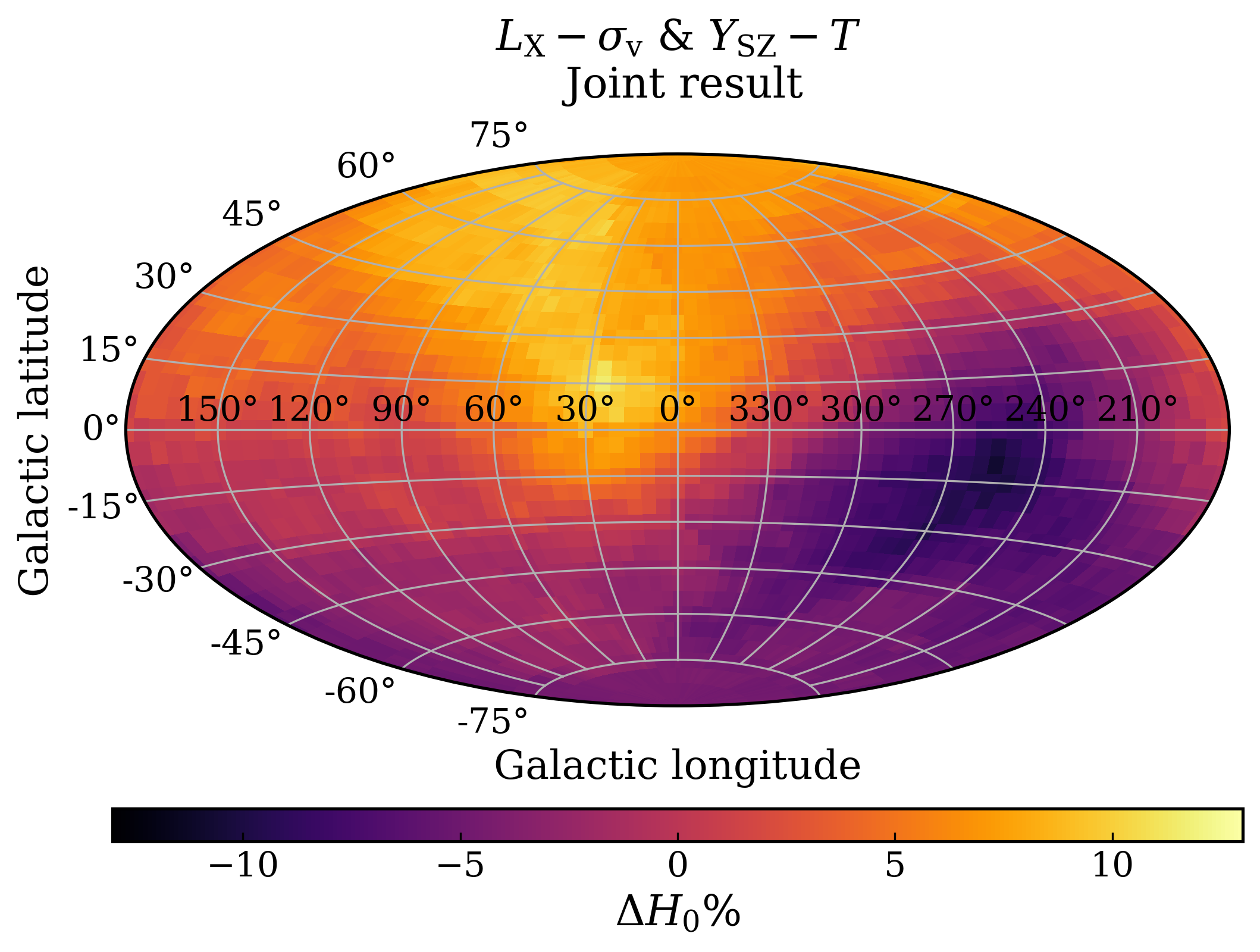}
    \includegraphics[width=0.495\hsize]{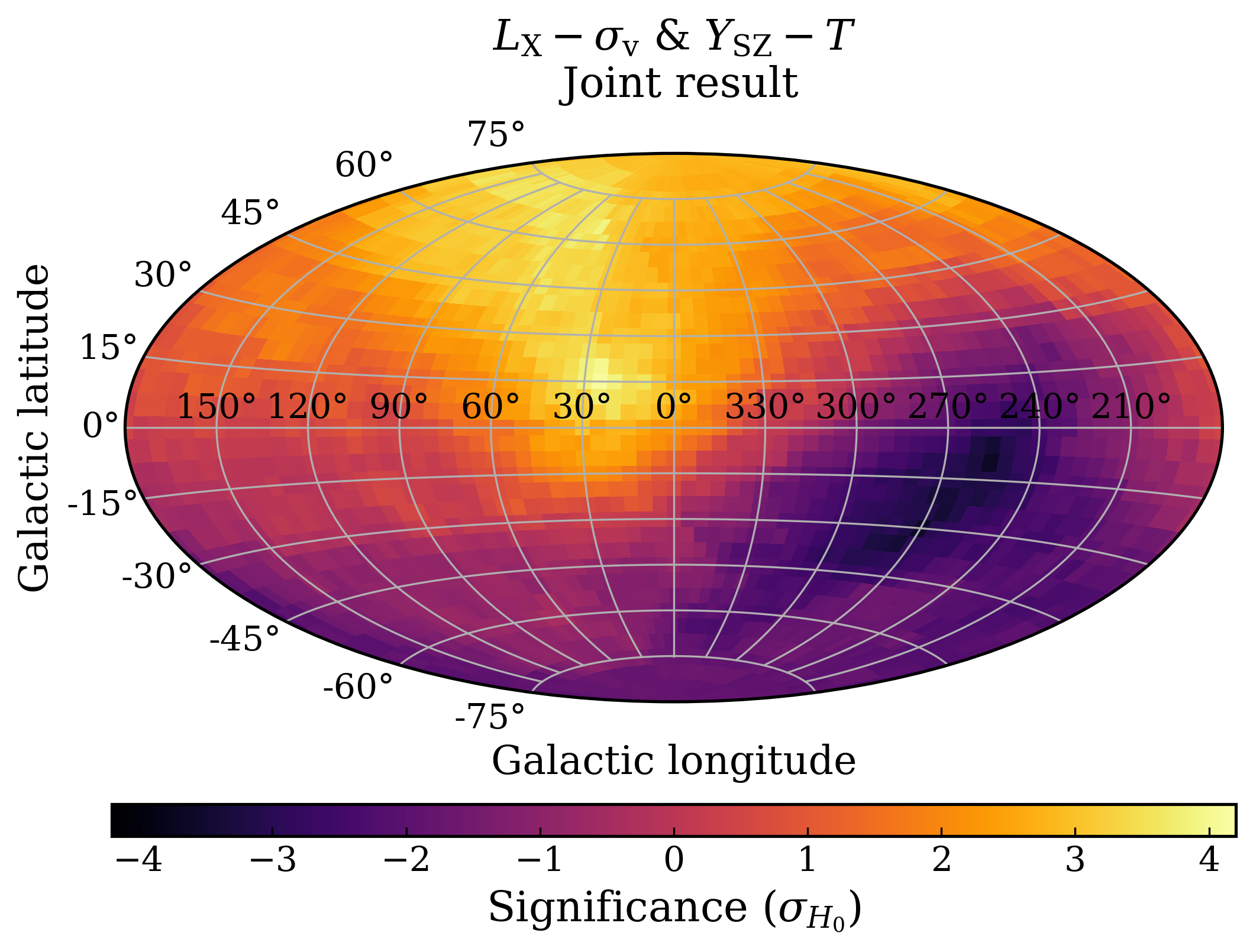}
    \includegraphics[width=0.495\hsize]{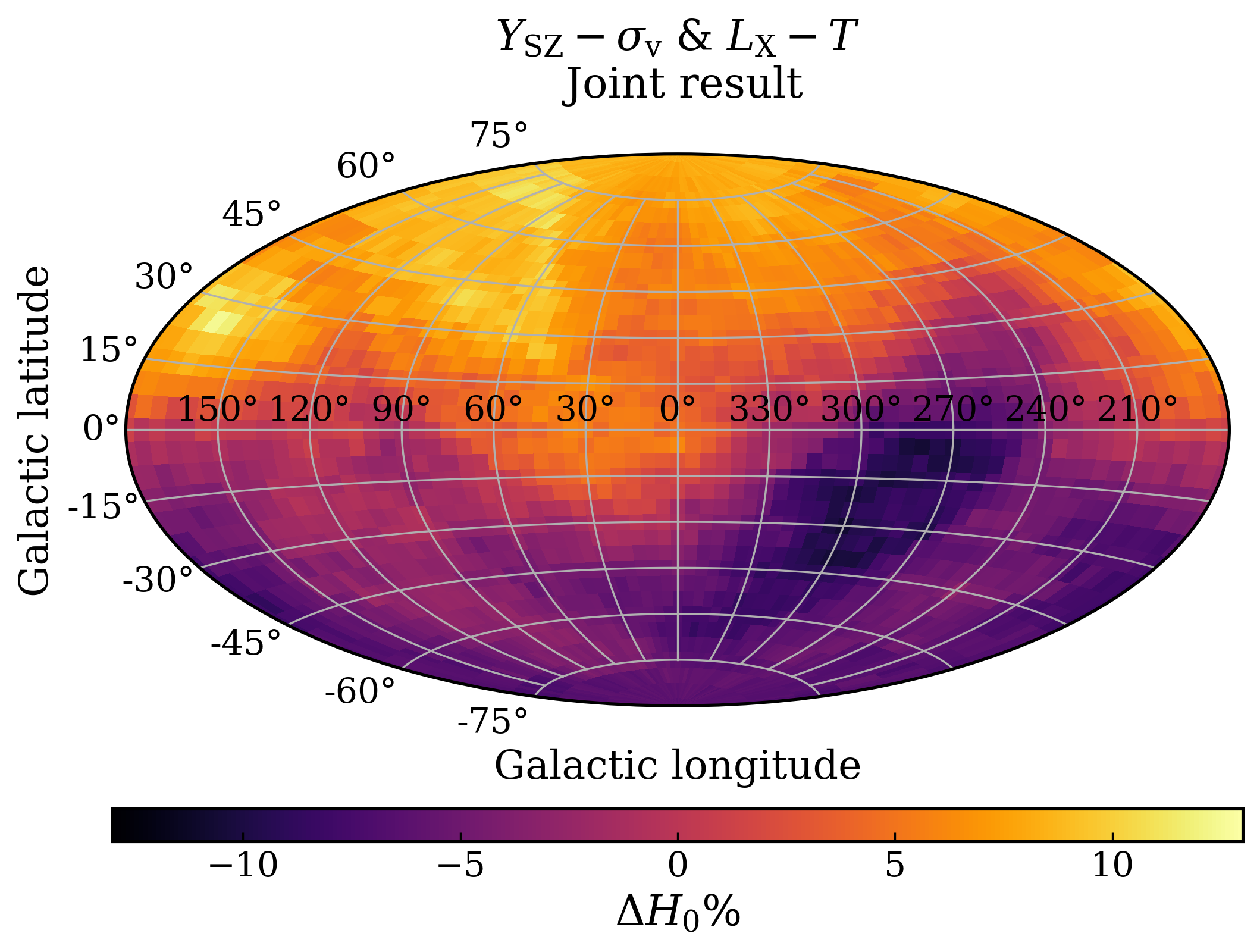}
    \includegraphics[width=0.495\hsize]{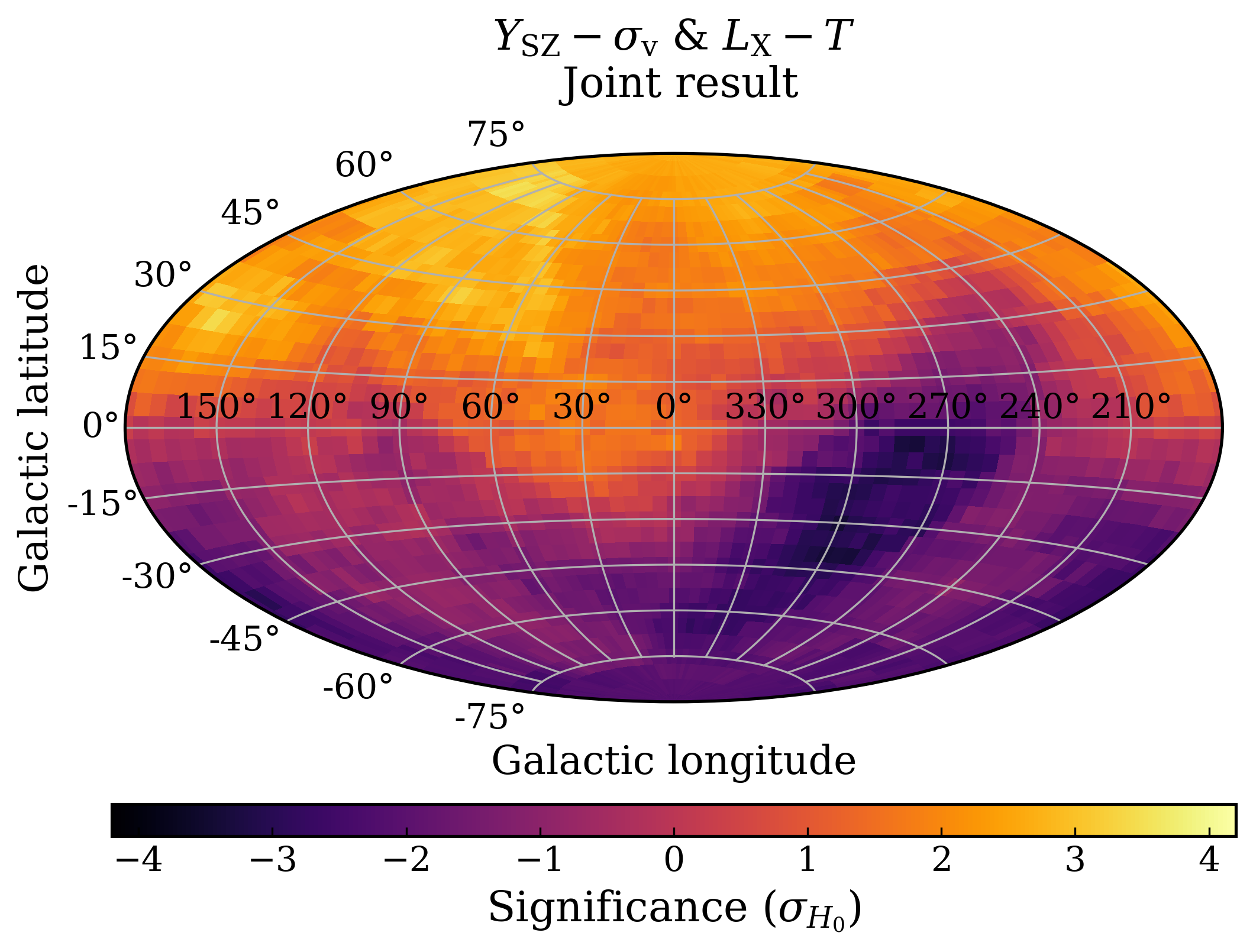}
    \caption{Map of $H_0$ angular variations (\textit{left}) and its significance (\textit{right}) for the joint analysis of relations $L_\mathrm{X}-\sigma_\mathrm{v}$ \& $Y_\mathrm{SZ}-T$ (\textit{top}) and $Y_\mathrm{SZ}-\sigma_\mathrm{v}$ \& $L_\mathrm{X}-T$ (\textit{bottom}).}
    \label{fig:Joint_M21}
\end{figure}

We perform a joint analysis with the results of M21 by pairing the datasets that are completely independent, i.e., pairing $L_\mathrm{X}-\sigma_\mathrm{v}$ and $Y_\mathrm{SZ}-\sigma_\mathrm{v}$ with M21's $Y_\mathrm{SZ}-T$ and $L_\mathrm{X}-T$ respectively. By multiplying the posteriors of $H_0$ for both relations, we create maps showing the $H_0$ angular variation and its significance (see Fig. \ref{fig:Joint_M21}).
The maximum $H_0$ angular variation decreased to $\sim 11\%$ in both joint results, which is similar to the results of M21. The maximum significance in the joint analysis of $L_\mathrm{X}-\sigma_\mathrm{v}$ and $Y_\mathrm{SZ}-T$ remains the same when compared to the results of $L_\mathrm{X}-\sigma_\mathrm{v}$ analysis. However, there is a decrease in significance in the joint analysis of other relations when compared to the results of $Y_\mathrm{SZ}-\sigma_\mathrm{v}$ analysis. A common trend in both maps is that the results of M21 dominate them due to having a low scatter in relations compared to our results.

We also check for any correlation between the residuals of the two relations that are being used in the joint analysis. We do not find any correlation between the $L_\mathrm{X}-\sigma_\mathrm{v}$ and $Y_\mathrm{SZ}-T$ relations. We can, therefore, treat the two relations as independent while analysing their joint results. However, we find a correlation of $\rho=-0.66$ between the residuals of the $Y_\mathrm{SZ}-\sigma_\mathrm{v}$ and $L_\mathrm{X}-T$ relations. Thus, the joint result of these relations could underestimate the significance.

\subsection{Isotropic samples based on random cluster positions}
\begin{figure}[htbp]
	\centering
	\includegraphics[width=0.65\hsize]{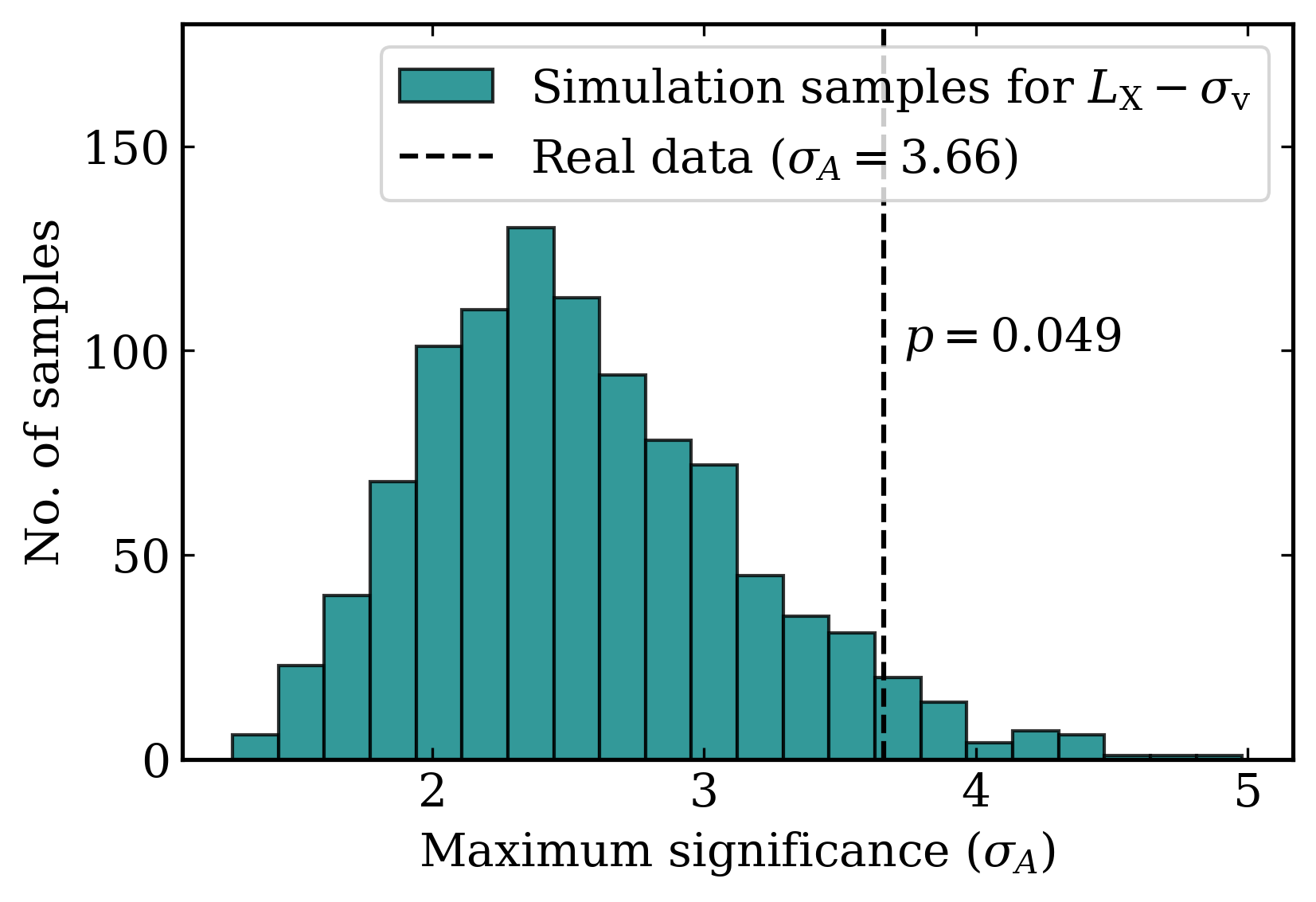}
    \includegraphics[width=0.65\hsize]{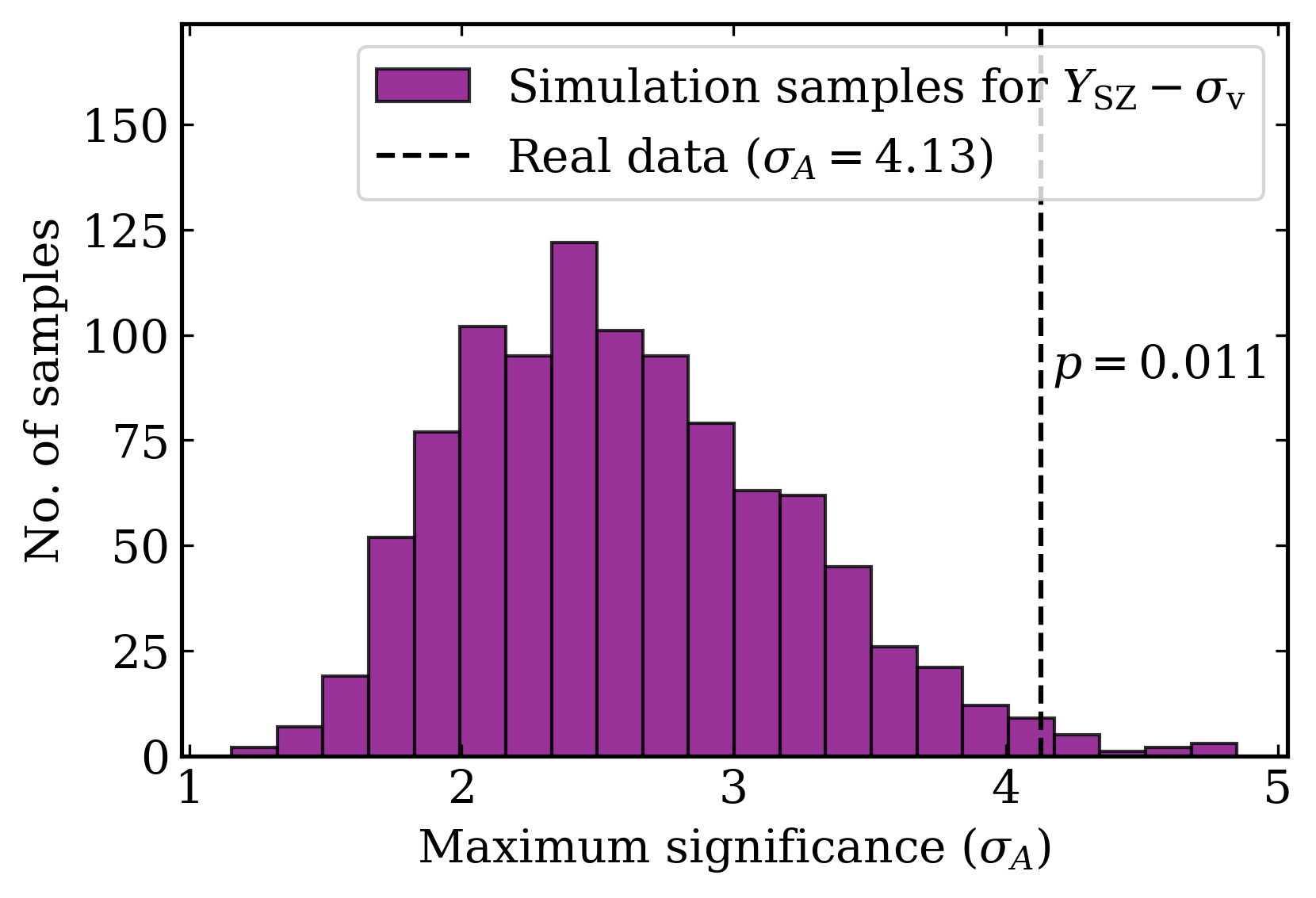}
	\caption{Distribution of maximum significance obtained from the 1000 isotropic samples created from the random shuffling of cluster positions for the $L_\mathrm{X}-\sigma_\mathrm{v}$ (\textit{top}), and $Y_\mathrm{SZ}-\sigma_\mathrm{v}$ (\textit{bottom}) relations. The dashed line represents the maximum significance obtained in the observed data, and the $p$-value next to it shows the probability of obtaining higher significance than the observed data.}
	\label{fig:isotropic_shuffle}
\end{figure}

Aside from using isotropic MC samples, we employ a commonly used method for generating isotropic samples, which involves shuffling the cluster positions randomly and calculating the probability of obtaining higher anisotropy in these samples compared to the observed data. We create 1000 such isotropic samples for both the relations, find their maximum significance values, and compare their distributions to the observed data (see Fig. \ref{fig:isotropic_shuffle}). Our results indicate that the probability of observing higher significance in an isotropic sample (based on random cluster positions) compared to the real data for the $L_\mathrm{X}-\sigma_\mathrm{v}$ and $Y_\mathrm{SZ}-\sigma_\mathrm{v}$ relations are $\sim 5\%$ and $\sim 1\%$ respectively.

When we compare these to the isotropic MC simulation results in Sect. \ref{sec:isotropic_MC_results}, we find that the probability remains unchanged for the $Y_\mathrm{SZ}-\sigma_\mathrm{v}$ relation. However, for the $L_\mathrm{X}-\sigma_\mathrm{v}$ relation, we note an increase in the $p$-value compared to isotropic MC simulation results. This decrease in significance can be attributed to our use of large cone sizes for the analysis. When we shuffle the cluster positions, it is likely that, at times, the clusters affected by the anisotropy might fall within the same cone or close to one another. As a result, the mock sample exhibits higher significance variations.

\subsection{Effects of best-fit slope variations}
\label{sec:slope_variations}
\begin{figure}[htbp]
	\centering
	\includegraphics[width=0.495\hsize]{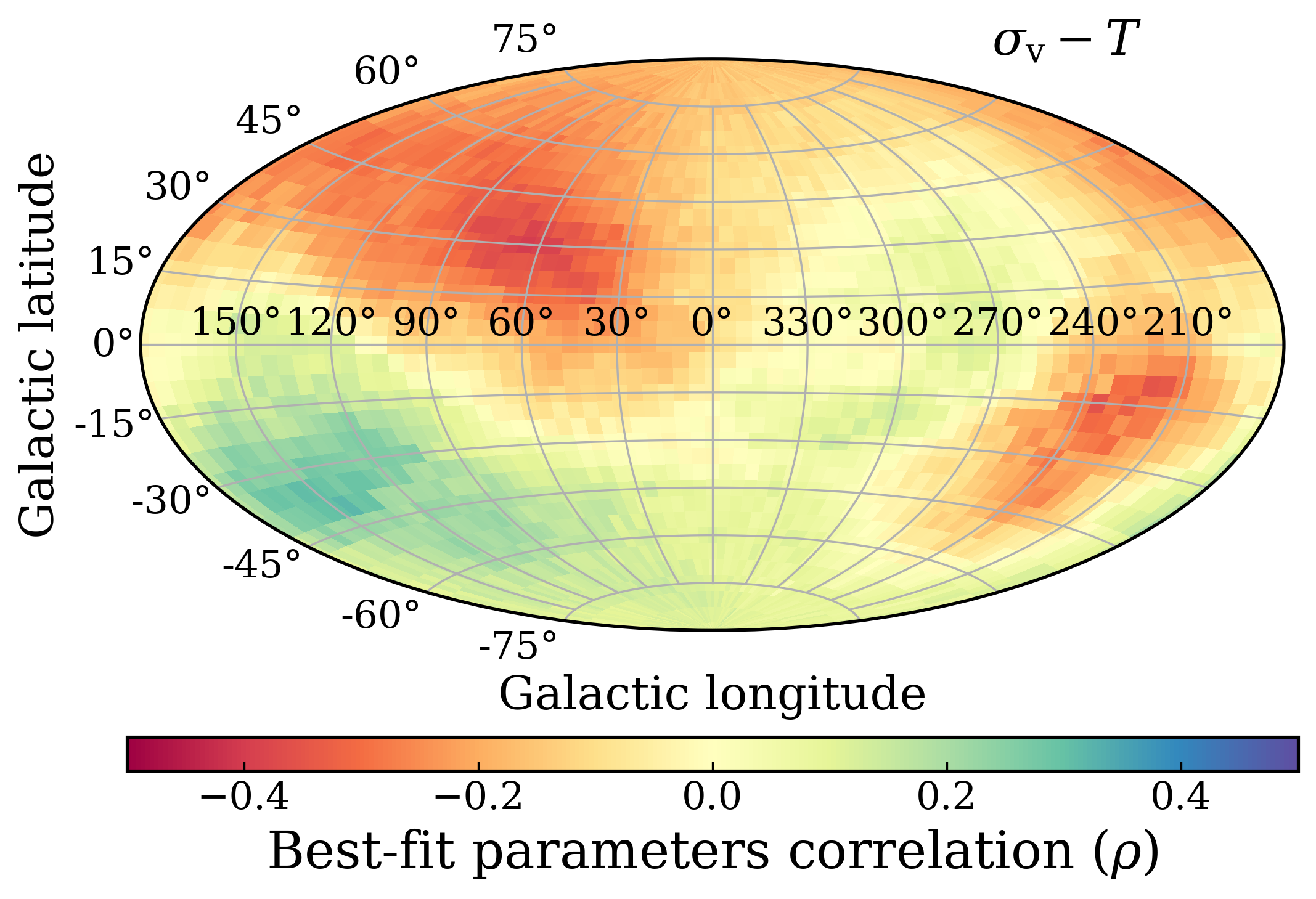}
    \includegraphics[width=0.495\hsize]{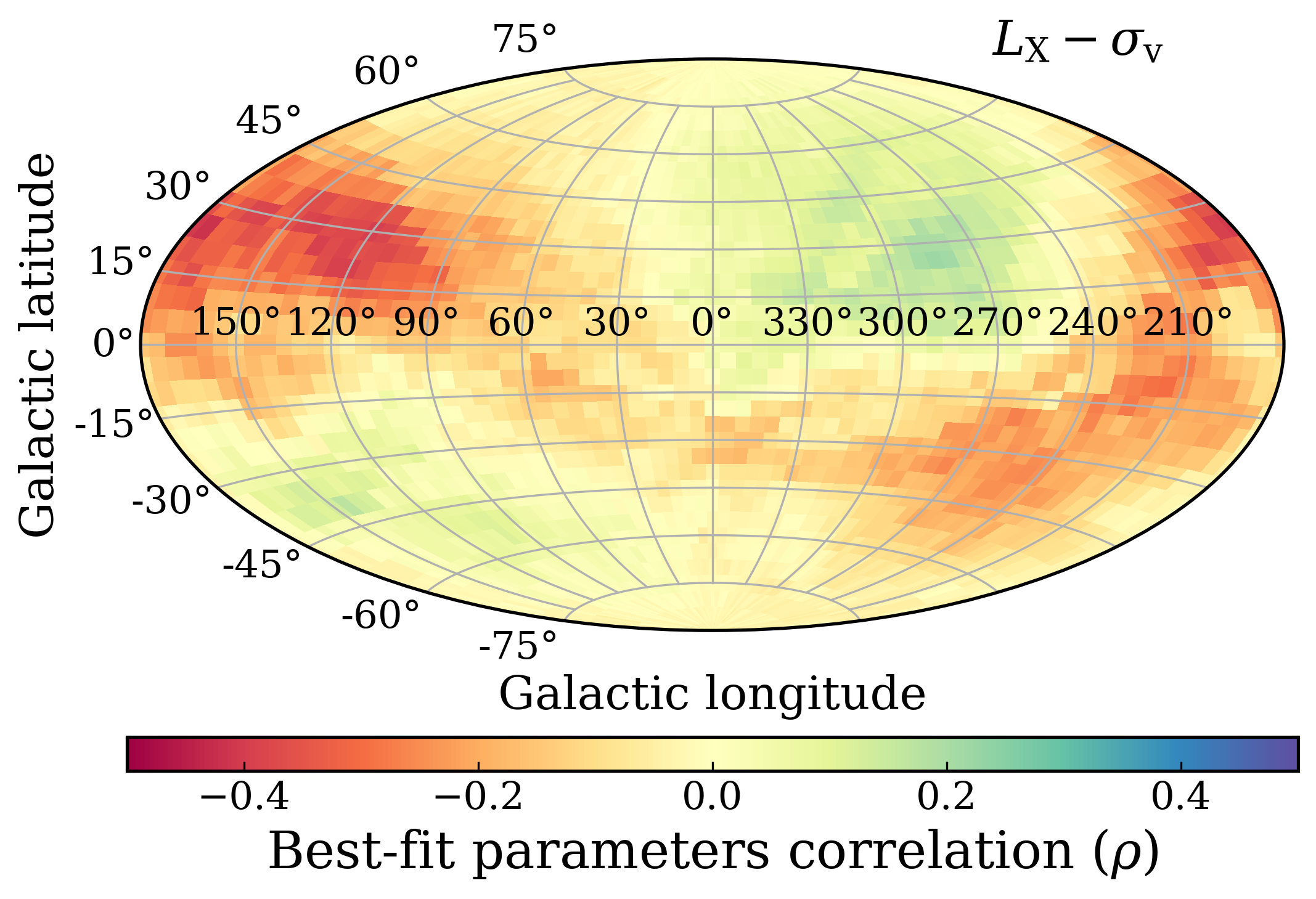}
    \includegraphics[width=0.495\hsize]{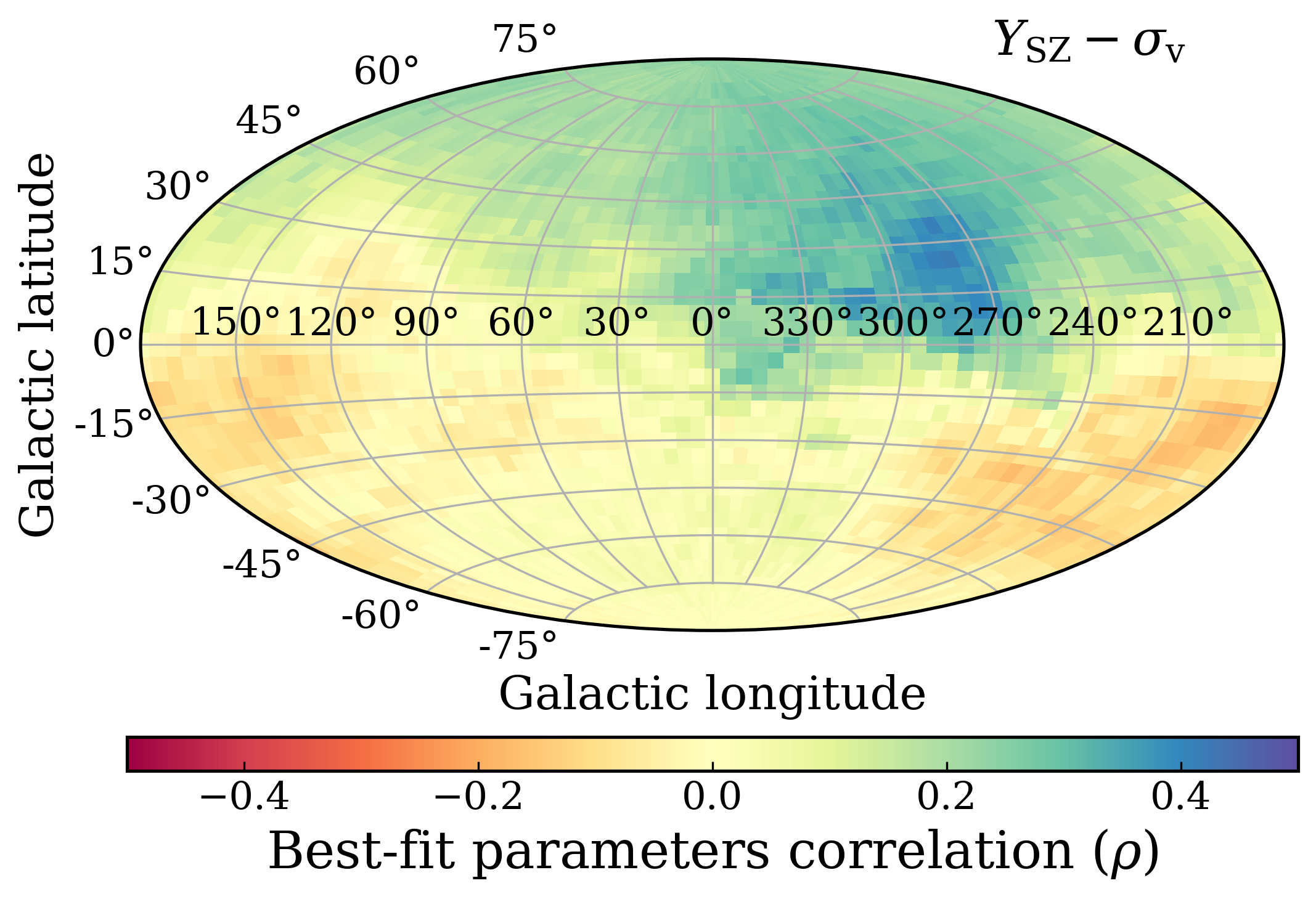}
	\caption{2-D Maps of the correlation coefficient between the best-fit parameters for the $\sigma_\mathrm{v}-T$ (\textit{top left}), $L_\mathrm{X}-\sigma_\mathrm{v}$ (\textit{top right}), and $Y_\mathrm{SZ}-\sigma_\mathrm{v}$ (\textit{bottom}) relations. All maps have the same colour scale ($-0.5$ to $+0.5$).}
	\label{fig:2D_corr}
\end{figure}

For all three relations, we treat the slope as a free parameter, and we find some variations in best-fit slopes across the sky (see Fig. \ref{fig:2d_anisotropies_Sigma-T} for $\sigma_\mathrm{v}-T$ and Appendix \ref{sec:free_vs_fixed} for $L_\mathrm{X}-\sigma_\mathrm{v}$ and $Y_\mathrm{SZ}-T$). We do not expect a significant correlation between the best-fit parameters because the pivot points are chosen close to the data median. However, some cones may have a strong correlation that could bias the observed results. To examine this, we calculate the correlation between $A$ and $B$ in each region and create a 2-D map based on this (Fig. \ref{fig:2D_corr}). For $\sigma_\mathrm{v}-T$, the correlation between the best-fit parameters in all the regions is less than $\pm0.4$. Similar results are also found for the $Y_\mathrm{SZ}-\sigma_\mathrm{v}$ relation. In the analysis of $L_\mathrm{X}-\sigma_\mathrm{v}$, we observe certain regions with a moderate correlation of $+0.5$, but these regions are distant from the anisotropy region, so we don't have any reason to suspect a bias due to parameter correlation.

We test our findings with the slope fixed to that of the full data to see the impact of the slight correlation between parameters (refer to Appendix \ref{sec:free_vs_fixed} for more details). We find no significant difference in the results when the slope is fixed as compared to the free slope analysis.

\subsection{Anisotropy at different $z$ scales and with different $z$ evolutions}
\label{sec:z_cuts_evolution}
We examine the distribution of redshift across different regions of the sky. If any region shows significantly higher or lower average redshift ($z$) values, it could indicate that we are comparing different scales, which is not ideal. We create the redshift distribution for clusters present in the region of maximum anisotropy and compare it with the rest of the sky for the $L_\mathrm{X}-\sigma_\mathrm{v}$ and $Y_\mathrm{SZ}-\sigma_\mathrm{v}$ relations (see Fig. \ref{fig:z_dist}). We find that, on average, the region with maximum anisotropy has lower $z$, whereas the rest of the sky has a large population of clusters at higher $z$. To ensure that our results are not biased by the population of clusters at different $z$ scales, we apply various $z$ cuts and compare these results with our initial findings.

We apply several lower cuts on $z$, such as $z>0.03$, $z>0.05$, and $z>0.07$ and an uppercut of $z<0.15$ to our data individually, and repeat our analysis. Redshift cuts higher than $z>0.07$ are not considered since we are limited by the number of clusters in different cones. The $L_\mathrm{X}-\sigma_\mathrm{v}$ relation has fewer clusters, and several cones for the cut of $z>0.07$ had fewer than 20 clusters, and thus, we do not include this cut for the relation.

\begin{figure}[htbp]
    \centering
    \includegraphics[width=1\hsize]{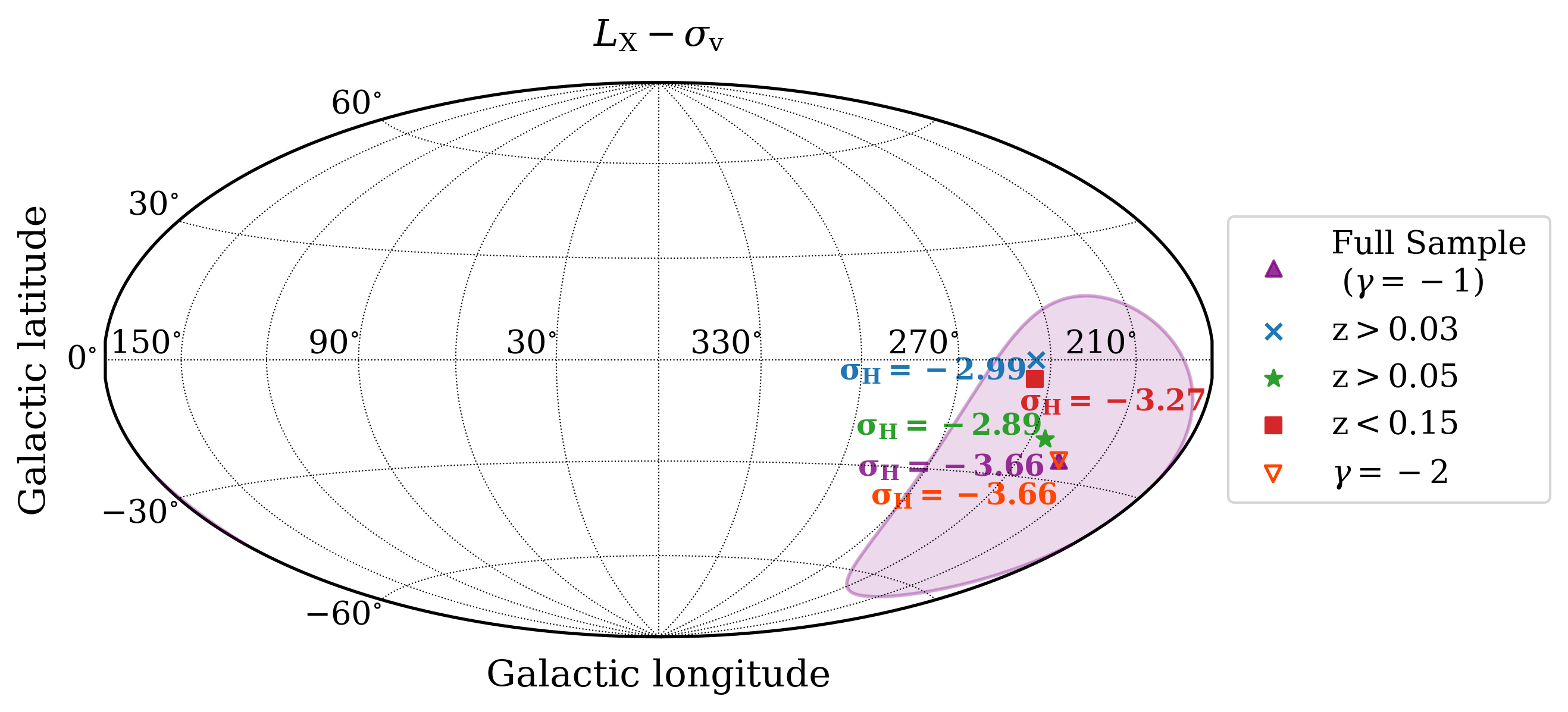}
    \includegraphics[width=1\hsize]{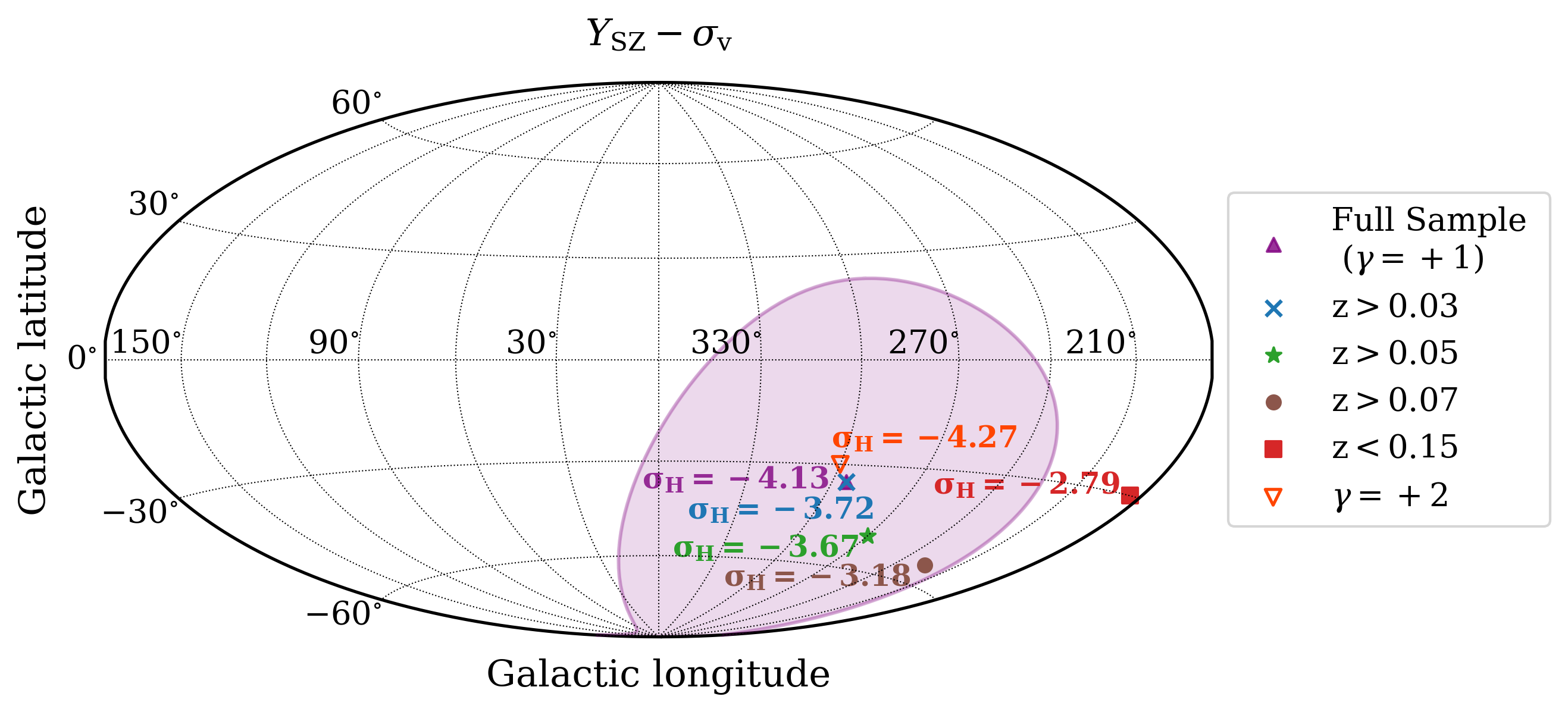}
    \caption{Positions and amplitude of maximum anisotropy detected for several $z$ cuts and $z$ evolution power for $L_\mathrm{X}-\sigma_\mathrm{v}$ (\textit{top}), and $Y_\mathrm{SZ}-\sigma_\mathrm{v}$ (\textit{bottom}) relations. The shaded regions represent the directional uncertainties of the full sample.}
    \label{fig:z_cuts}
\end{figure}

We plot the positions and amplitude of regions with maximum anisotropy obtained from these cuts for both relations, including the results obtained with the full dataset in Fig. \ref{fig:z_cuts}. These cuts return consistent results with the full sample, and we see that our results are not strongly biased due to cluster populations at different $z$ scales. Along with these results, we also plot the results with different $z$ evolution power ($\gamma$). 

We observe distinct trends in the 2-D map of median $z$ used within each cone, with the northern galactic sky showing clusters at higher $z$ compared to the southern half (see Fig. \ref{fig:2D_z}). A similar galactic north-south divide is also present in the 2-D maps of both relations (see Fig. \ref{fig:2d_norm_Lx-Sigma} and Fig. \ref{fig:2d_norm_Ysz-Sigma}), due to the northern galactic sky having higher $A$ compared to the southern half. Regions with higher median $z$ are strongly affected by incorrect redshift evolution assumptions. Therefore, it is crucial to test if the assumed $\gamma$ for the two relations results in a bias and the observed galactic north-south divide in our significance maps. As mentioned in the Table \ref{tab:scaling_relations}, $\gamma$ for the $L_\mathrm{X}-\sigma_\mathrm{v}$ and $Y_\mathrm{SZ}-\sigma_\mathrm{v}$ relations are $-1$ and $+1$ respectively. We change $\gamma$ to $-2$ and $+2$, respectively, to see the effect of choosing a strong redshift evolution on the anisotropy detected.  

The results for $L_\mathrm{X}-\sigma_\mathrm{v}$ indicate almost no difference in significance and in the direction of anisotropy. In the $Y_\mathrm{SZ}-\sigma_\mathrm{v}$ relations, we observe minor differences from $\gamma=+1$ due to clusters with higher $z$ in the data sample. Even when we consider the most extreme value of $\gamma$ for the two relations ($-10$ and $+10$, respectively), we are unable to eliminate the north-south divide in the significance maps. Therefore, we can conclude that our results are not biased due to incorrect $z$ evolution.

%--------------------------------------------------------------------
\section{Summary}
\label{sec:summary}

In this work, we investigate the isotropy of the local universe using galaxy cluster scaling relations between cosmology-dependent cluster properties and a cosmology-independent property. We utilise a cosmology-independent variable that has not been used before: the velocity dispersion of a galaxy cluster. Previous studies by M21 used cluster $T$ as their cosmology-independent quantity and found an apparent spatial variation of approximately $9\%$ in the Hubble constant, $H_0$, across the sky.

We examined the $\sigma_\mathrm{v}-T$ relation across the sky to check if a position-dependent systematic bias of $T$ measurements causes an overestimation of apparent $H_0$ variations in the work of M21. We obtain no significant anisotropies across the sky in the $\sigma_\mathrm{v}-T$ relation. The probability of obtaining the observed variations for $\sigma_\mathrm{v}-T$ relation in an isotropic Universe is $0.72$. The region with the most variations in $\sigma_\mathrm{v}-T$ relation is found in a similar direction as M21 but with a $T$ underestimation of $(16.3 \pm 7.1)\%$, hinting at the possibility that the significance of the M21 results might even be underestimated. A positive $T$ bias needed to explain the results of M21 can be rejected at a high probability $(p\sim10^{-5})$.

We utilised the $L_\mathrm{X}-\sigma_\mathrm{v}$ and the $Y_\mathrm{SZ}-\sigma_\mathrm{v}$ relations to probe the (an)isotropy of the universe. From the joint analysis, we find the most anisotropic region in the direction of $(295^\circ\pm71^\circ, -30^\circ\pm71^\circ)$ at $3.64\sigma$. The statistical significance is obtained using the isotropic Monte Carlo simulations. The direction of maximum anisotropy is similar to the results of M21 $(l,b) = \left(280^{\circ +35^\circ}_{\ -35^\circ}, -15^{\circ +20^\circ}_{\ -20^\circ}\right)$. These results are in disagreement with the standard cosmological model. The results obtained from both these analyses further strengthen the results of M21.

%--------------------------------------------------------------------

\begin{acknowledgements}
L.L. acknowledges the financial contribution
from the INAF grant 1.05.12.04.01.
\end{acknowledgements}

%-------------------------------------------------------------------

\bibliographystyle{aa}
\bibliography{refs}

%-------------------------------------------------------------------

\begin{appendix}

\section{Choice for the cone and interval sizes}
\label{sec:cone_interval_sizes}
\begin{figure}[htbp]
    \centering
    \includegraphics[width=0.49\hsize]{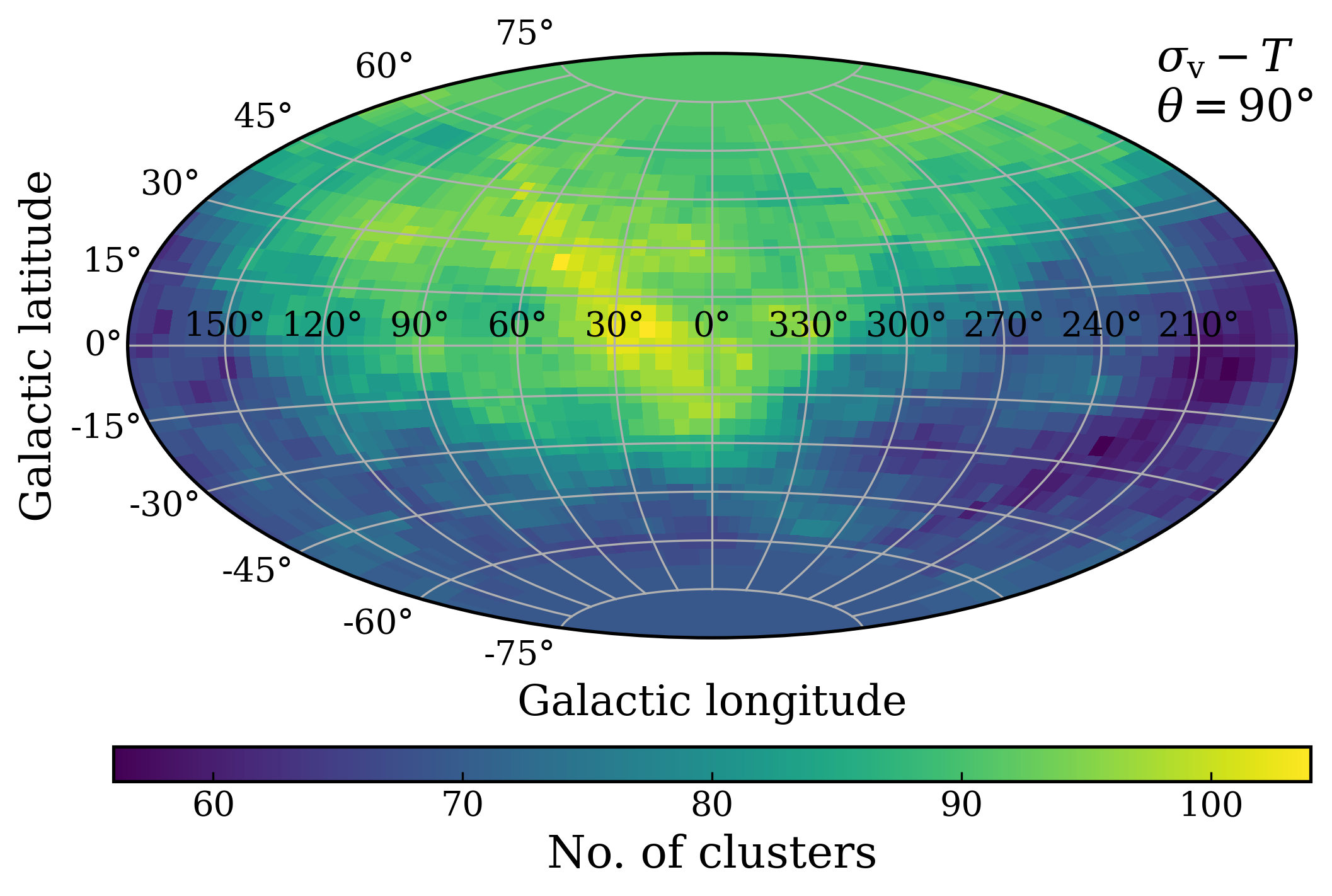}
    \includegraphics[width=0.49\hsize]{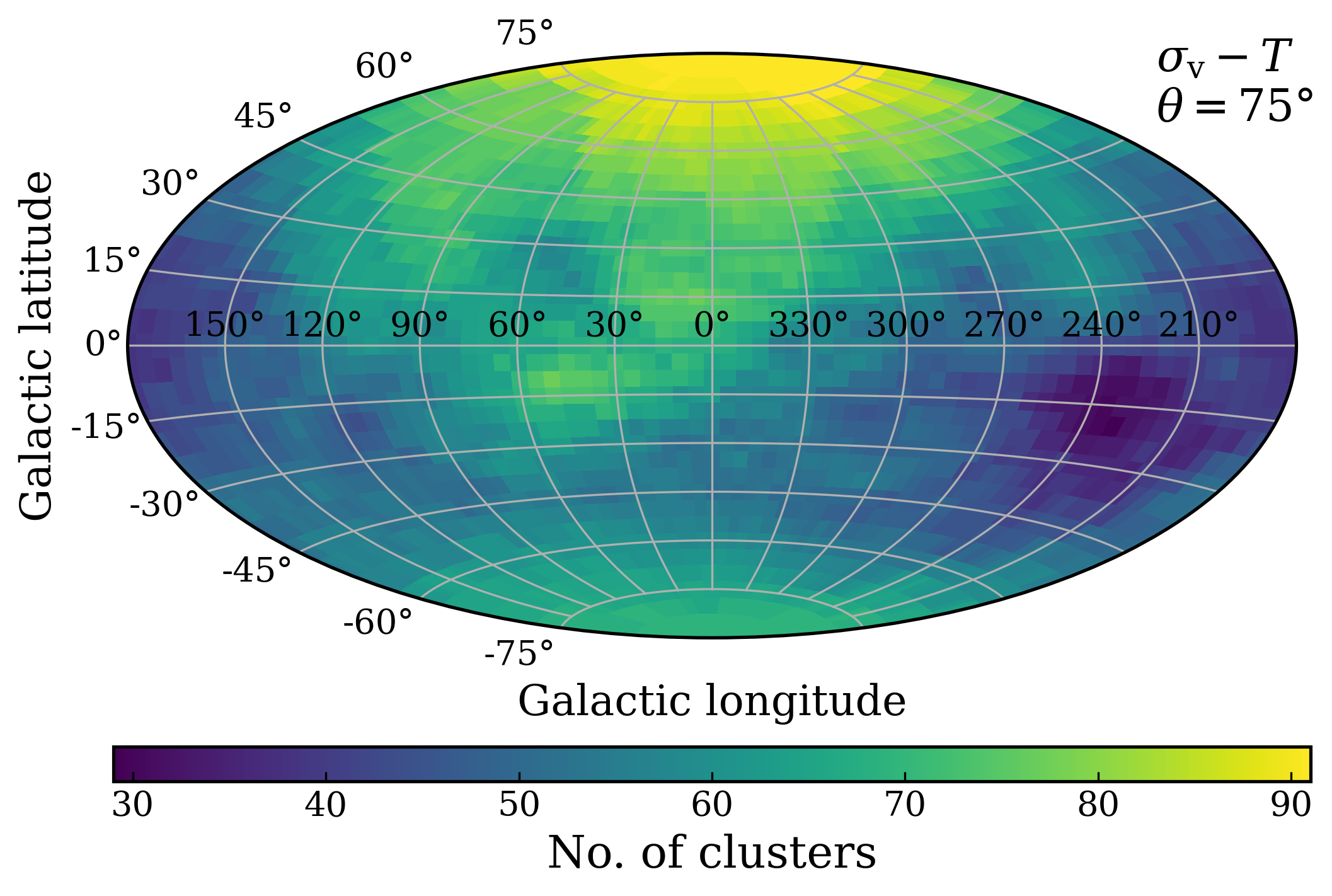}
    \includegraphics[width=0.49\hsize]{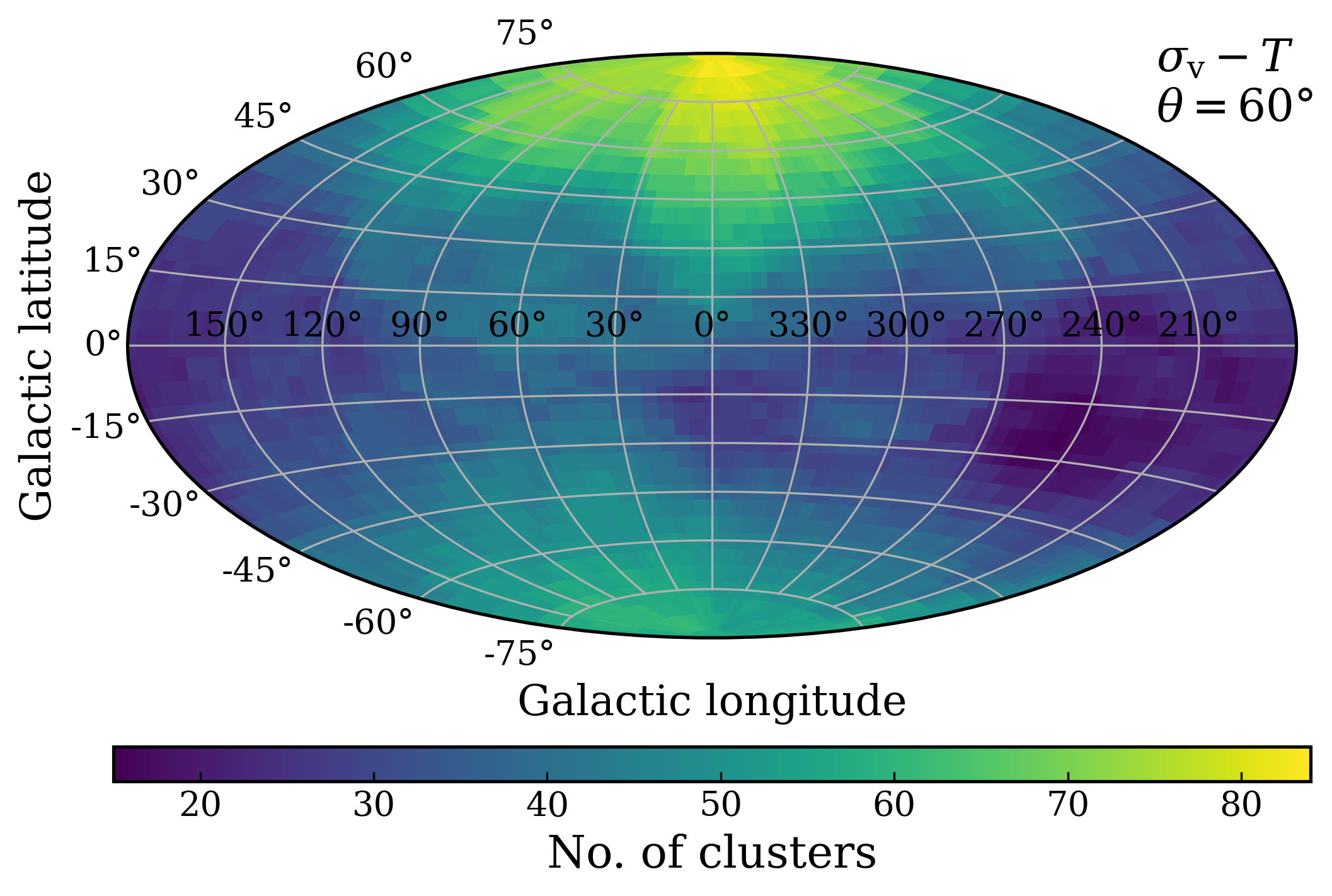}
    \includegraphics[width=0.49\hsize]{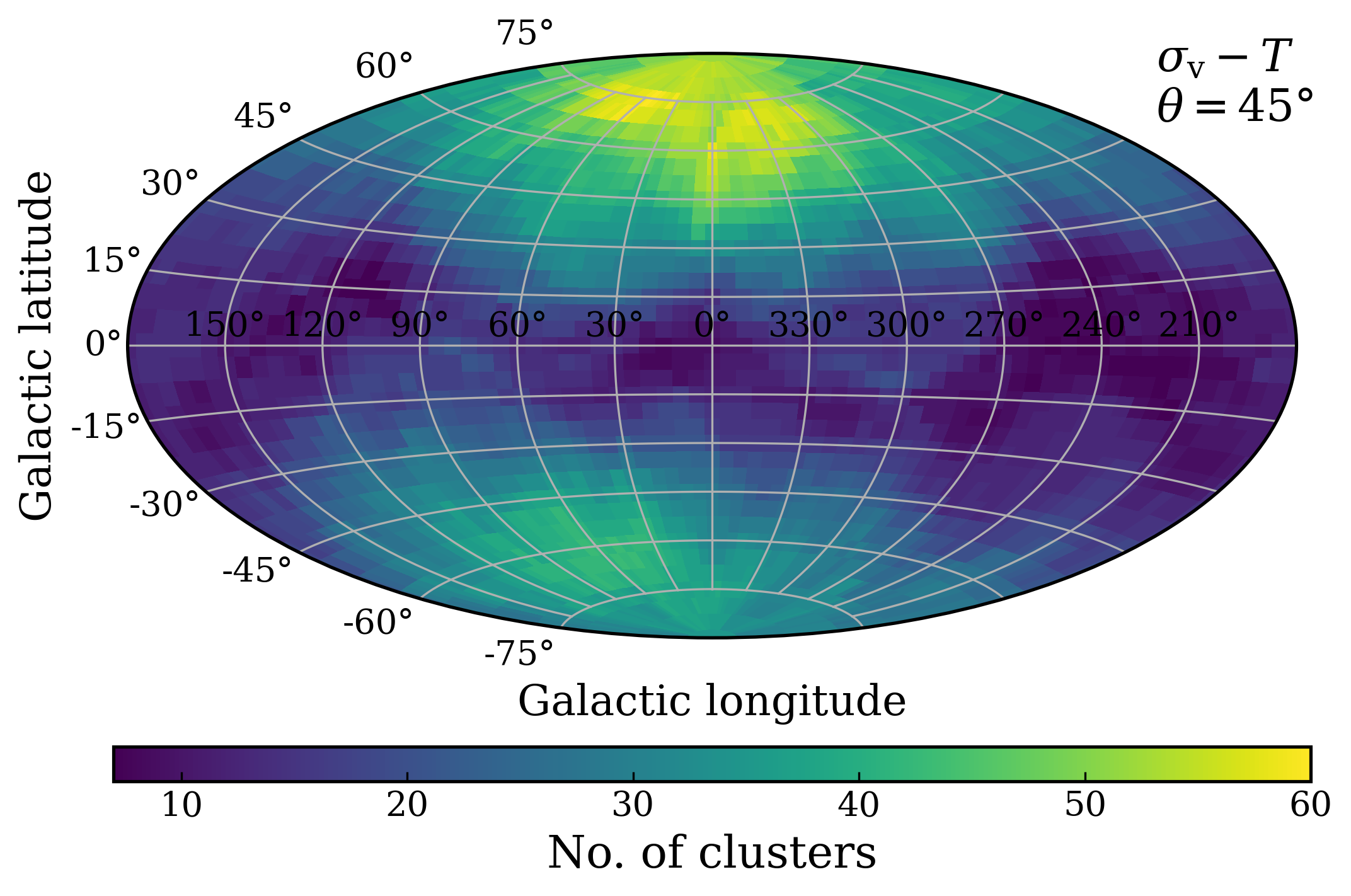}
    \caption{Maps of the number of clusters for different cone sizes in the relation $\sigma_\mathrm{v}-T$. The cone sizes are $90^\circ$ (\textit{top left}), $75^\circ$ (\textit{top right}), $60^\circ$ (\textit{bottom left}), and $45^\circ$ (\textit{bottom right}). Note that the colour scale is different for the four plots.}
    \label{fig:cone_size}
\end{figure}

The cone size is chosen such that the number of clusters inside the cone is as large as possible without covering too large an area.
Four cone sizes are considered to decide the best size for the analysis. 
The radii of these cones are $90^\circ$, $75^\circ$, $60^\circ$, and $45^\circ$. Fig. \ref{fig:cone_size} shows the number of clusters inside a cone for the relation $\sigma_\mathrm{v}-T$ for different cone sizes.

The cone size of $45^\circ$ covers a small area, and the number of clusters inside the cone is also small. There are less than 10 clusters in the region of the Galactic belt.
The cone size of $90^\circ$ covers a large area, and the number of clusters inside the cone is also large. This creates a significant overlap between the cones.
The cone size of $60^\circ$ has a high number of clusters for most regions, but certain regions have less than 20 clusters.
The cone size of $75^\circ$ strikes a good balance between the area covered and the number of clusters inside the cone. Thus, the cone size of $75^\circ$ is chosen for the analysis.

\begin{figure}[htbp]
    \centering
    \includegraphics[width=0.49\hsize]{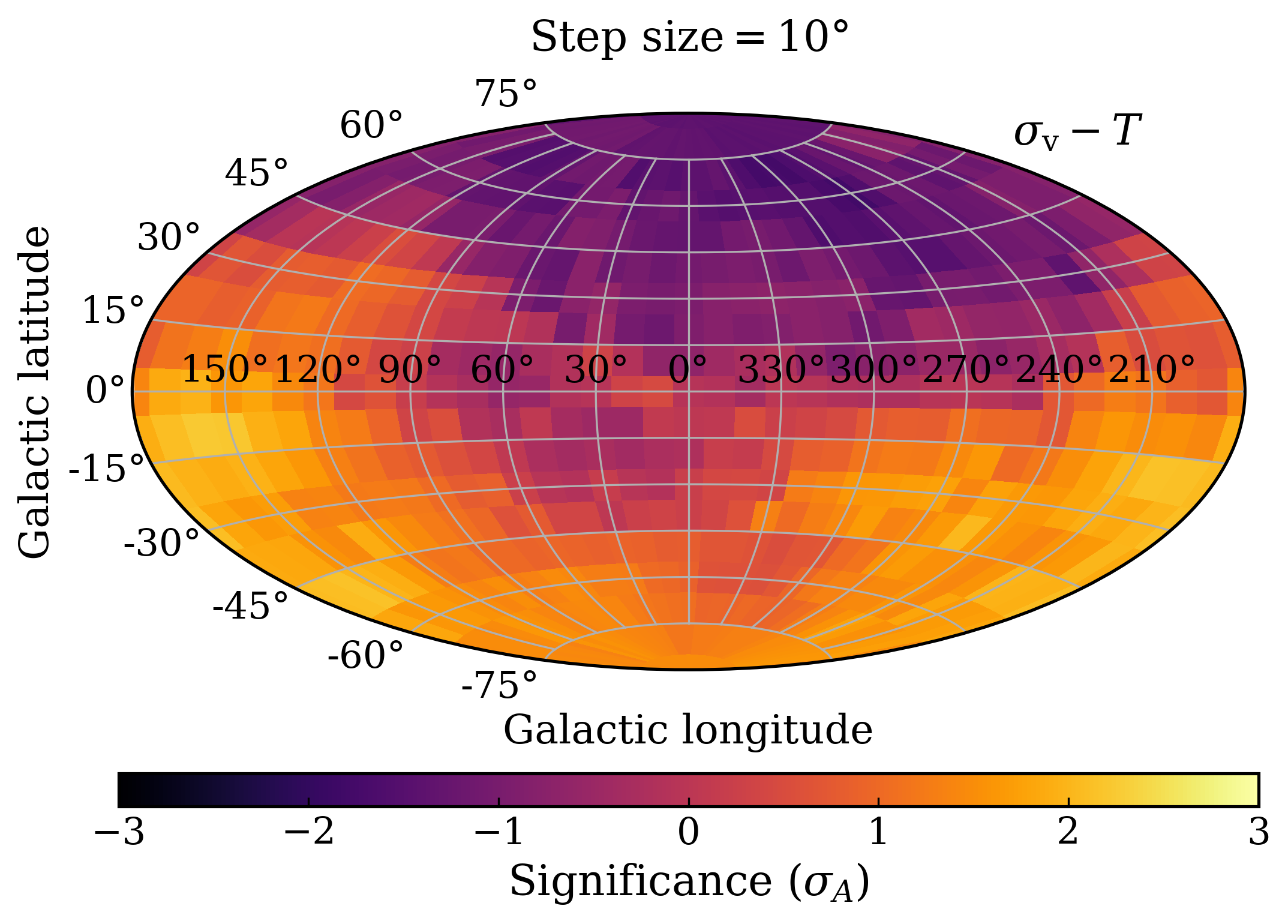}
    \includegraphics[width=0.49\hsize]{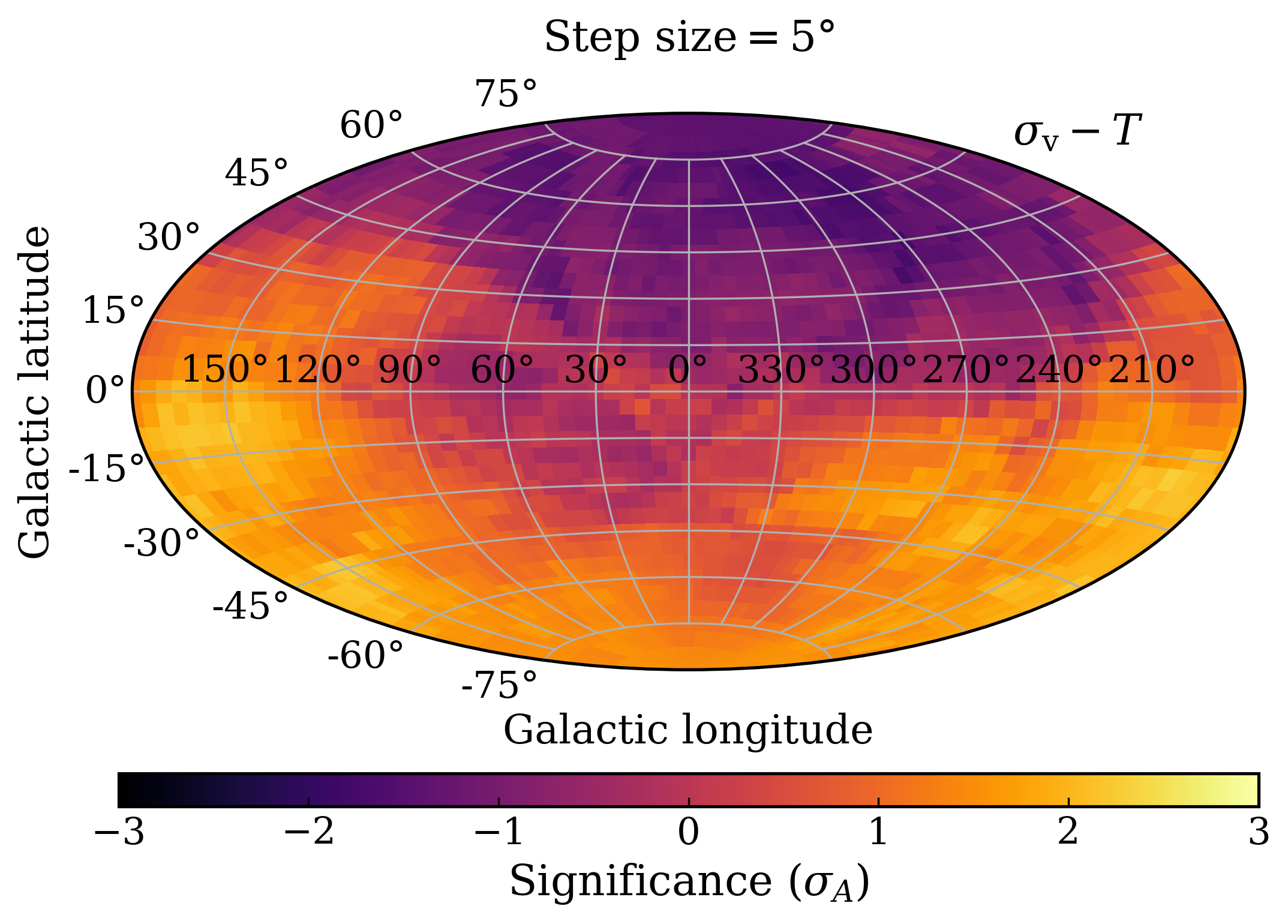}
    \includegraphics[width=0.49\hsize]{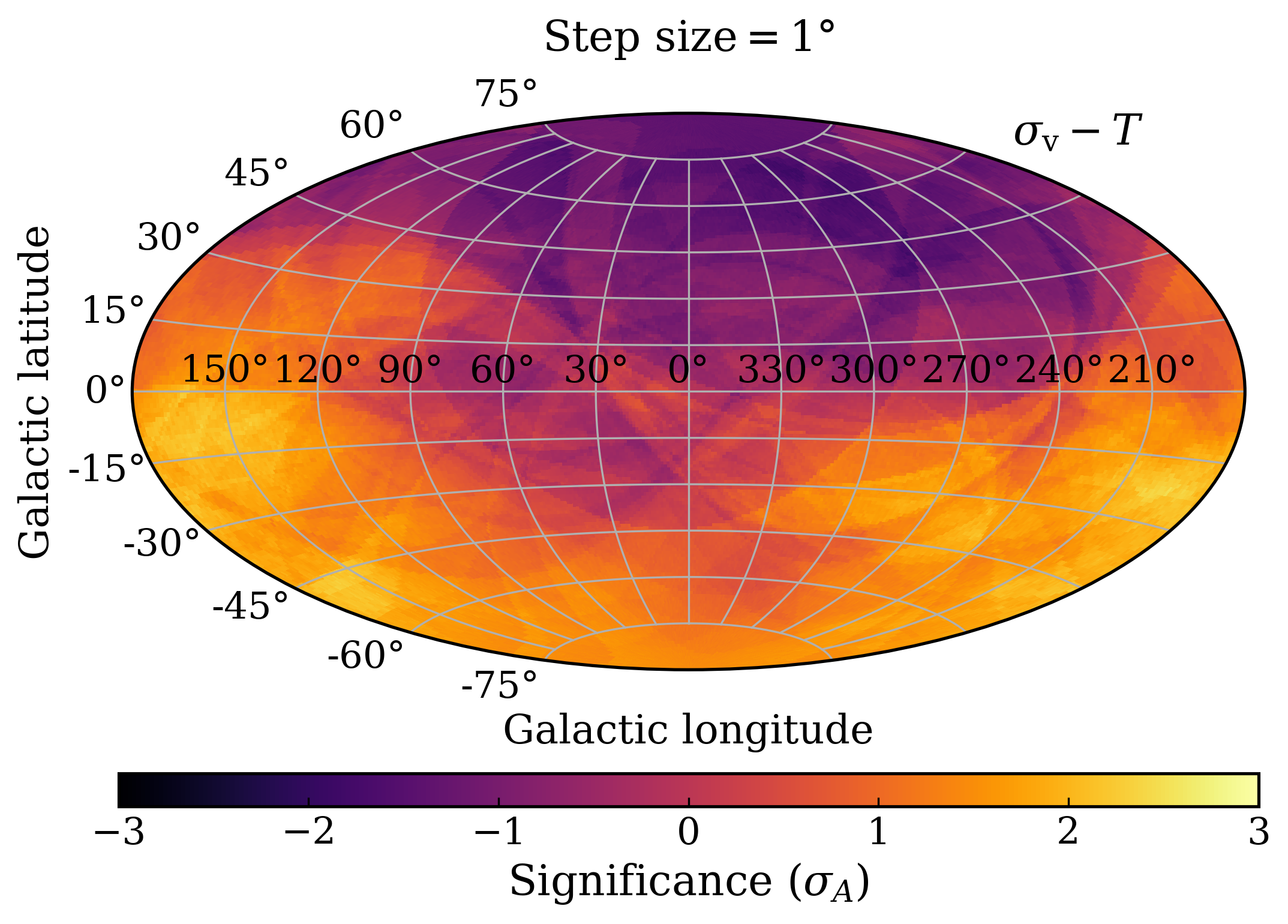}
    \caption{Sigma map of the $A$ from the $\sigma_\mathrm{v}-T$ relation for different step sizes.
    The step sizes are $10^\circ$ (\textit{top left}), $5^\circ$ (\textit{top right}), and $1^\circ$ (\textit{bottom}).}
    \label{fig:step_size}
\end{figure}

The centre for these cones is taken at a step size of $5^\circ$.
This is mainly done to reduce the computation time. Fig. \ref{fig:step_size} shows the significance of the best-fit $A$ for the relation $\sigma_\mathrm{v}-T$ for different step sizes. The results show that the choice of step size does not affect the results significantly.

In the significance map with a step size of $1^\circ$, some artefacts are seen as curved lines. These artefacts can be seen in the significance map with a step size of $5^\circ$, but they are not as prominent.
These artefacts are due to strong outliers present in a particular region. These clusters are not considered outliers when considering the entire sample. All regions with this strong outlier will produce similar best-fit results, creating the artefacts.

\section{Redshift trends in the residuals}
\label{sec:z_trends}
\begin{figure}[htbp]
    \centering
    \includegraphics[width=0.495\hsize]{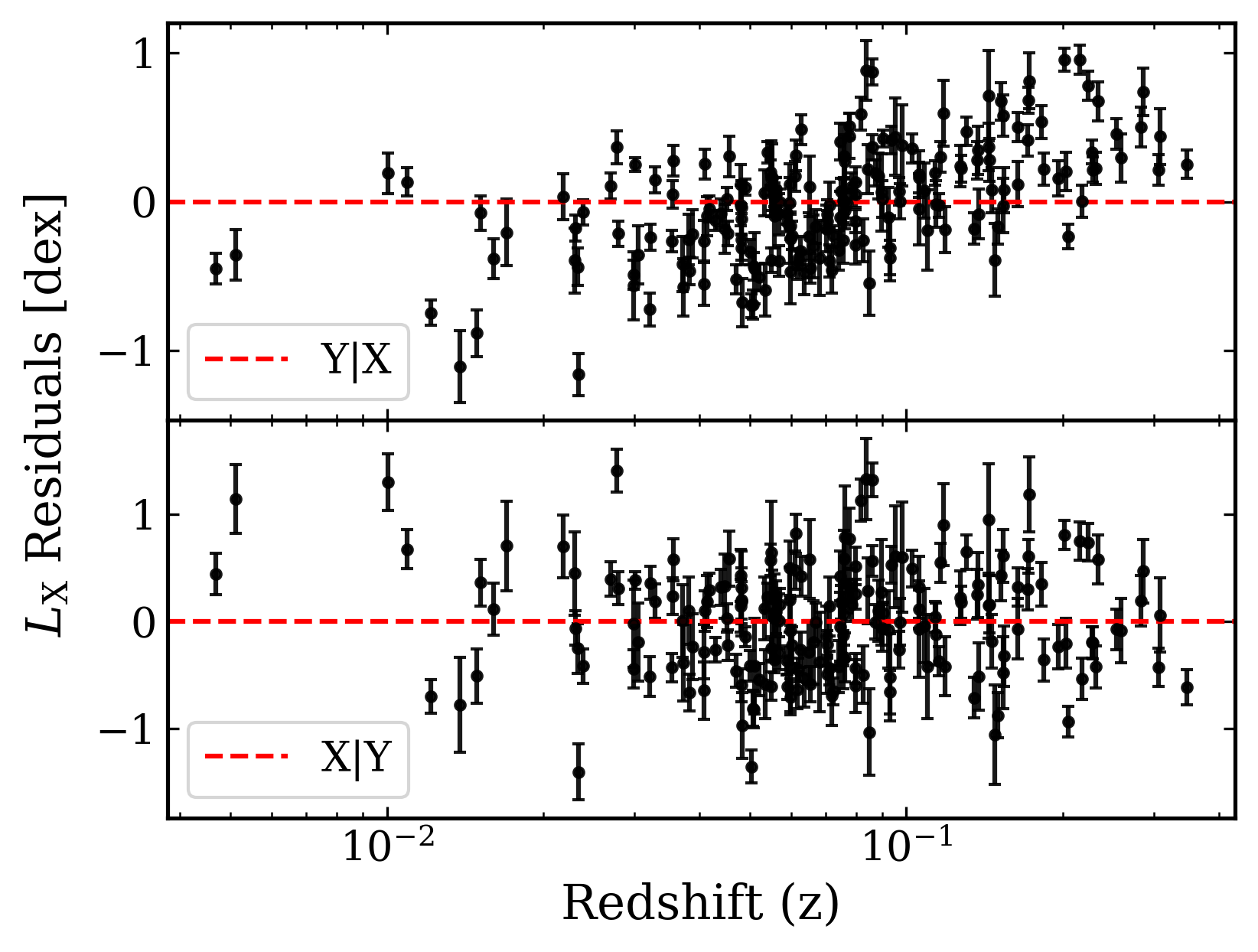}
    \includegraphics[width=0.495\hsize]{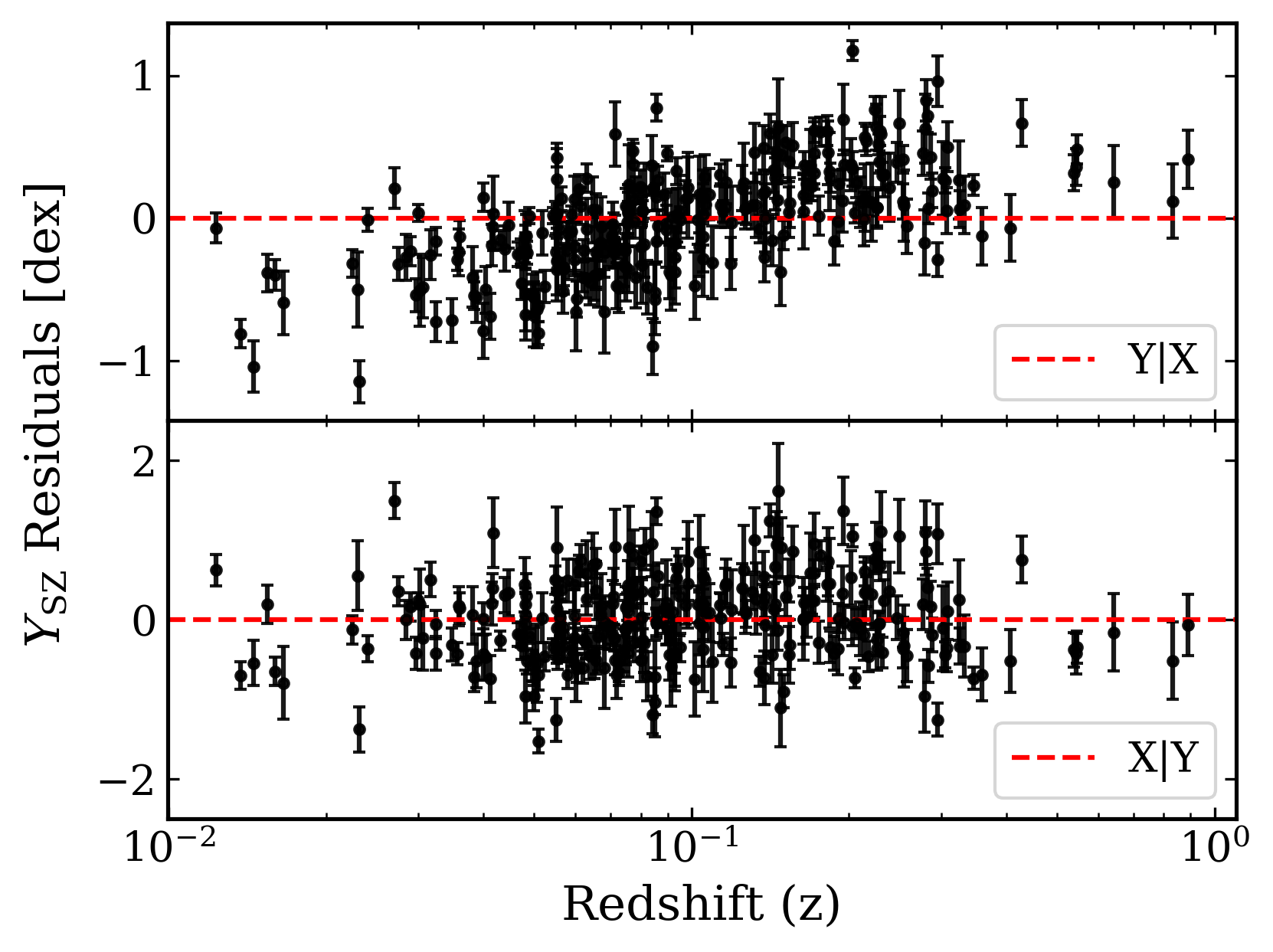}
    \caption{Y-axis residual trends with respect to cluster redshift for the $L_\mathrm{X}-\sigma_\mathrm{v}$ (\textit{left}) and $Y_\mathrm{SZ}-\sigma_\mathrm{v}$ (\textit{right}) relations. The top panel displays the residuals using the Y|X minimisation, and the bottom panel displays the residuals using the X|Y minimisation method.}
    \label{fig:residual_trends}
\end{figure}

Using the Y|X best-fit minimisation method, strong $L_\mathrm{X}$ (and $Y_\mathrm{SZ}$) residual trends are observed as a function of cluster redshift.
Fig. \ref{fig:residual_trends} shows the $y$-axis residuals for the relations $L_\mathrm{X}-\sigma_\mathrm{v}$ and $Y_\mathrm{SZ}-\sigma_\mathrm{v}$ as a function of $z$ using the Y|X and X|Y minimisation methods.

It is generally known that when there are significant uncertainties on the $x$-axis while minimising on the $y$-axis, the best-fit results can show strong biases. Therefore, we adopt the X|Y minimisation for this scaling relation as these trends are not observed in this method.

\section{Comparison between free slope and fixed slope}
\label{sec:free_vs_fixed}

\begin{table*}
    \caption{Comparison of best-fit parameters for the $L_\mathrm{X}-\sigma_\mathrm{v}$ and $Y_\mathrm{SZ}-\sigma_\mathrm{v}$ scaling relations with different datasets.
    Results are shown for both X|Y and Y|X methods. The intrinsic scatter mentioned here is measured in the direction in which the scatter is minimised.}
    \label{tab:extra_scaling_relations}
    \centering
    \begin{tabular}{cccccccc}
        \hline
        \hline
        Dataset & {$C_Y$} & {$C_X$} & {Method} & {$A$} & {$B$} & {$\sigma_\mathrm{int}$ (dex)} & {$N$}\bigstrut\\
        \hline
        \multicolumn{8}{c}{$L_\mathrm{X}-\sigma_\mathrm{v}$ ($\gamma=-1$)}  \bigstrut\\
        \multirow{2}{*}{eeHIFLUGCS} & \multirow{2}{*}{$10^{44}\ \unit{erg\ s^{-1}}$} & \multirow{2}{*}{$750\ \unit{km\ s^{-1}}$} &  X|Y & {$1.515^{+0.134}_{-0.122}$} & {$5.988^{+0.452}_{-0.394}$} & {$0.080^{+0.005}_{-0.005}$} & {195} \bigstrut\\
        &  &  & Y|X & {$1.538^{+0.103}_{-0.097}$} & {$2.722^{+0.214}_{-0.213}$} & {$0.381^{+0.022}_{-0.021}$} & {200} \bigstrut\\
        & & & & & & &\\
        \multirow{2}{*}{MCXC} & \multirow{2}{*}{$10^{44}\ \unit{erg\ s^{-1}}$} & \multirow{2}{*}{$750\ \unit{km\ s^{-1}}$} &  X|Y & {$1.383^{+0.112}_{-0.105}$} & {$6.536^{+0.410}_{-0.374}$} & {$0.082^{+0.004}_{-0.004}$} & {316} \bigstrut\\
        &  &  & Y|X & {$1.434^{+0.081}_{-0.077}$} & {$2.910^{+0.186}_{-0.189}$} & {$0.384^{+0.018}_{-0.017}$} & {319} \bigstrut\\
        \hline
        \multicolumn{8}{c}{$Y_\mathrm{SZ}-\sigma_\mathrm{v}$ ($\gamma=+1$)} \bigstrut\\
        \multirow{2}{*}{$\mathrm{S/N} \geq 2$} &  \multirow{2}{*}{$30\ \unit{kpc^2}$} & \multirow{2}{*}{$800\ \unit{km\ s^{-1}}$} & X|Y & {$1.490^{+0.112}_{-0.106}$} & {$6.419^{+0.423}_{-0.384}$} & {$0.073^{+0.004}_{-0.004}$} & {284} \bigstrut\\
        &  &  & Y|X & {$1.341^{+0.068}_{-0.069}$} & {$2.915^{+0.194}_{-0.195}$} & {$0.342^{+0.018}_{-0.017}$} & {285} \bigstrut\\
        & & & & & & &\\
        \multirow{2}{*}{$\mathrm{S/N} \geq 3$} & \multirow{2}{*}{$30\ \unit{kpc^2}$} & \multirow{2}{*}{$800\ \unit{km\ s^{-1}}$} & X|Y & {$1.413^{+0.118}_{-0.106}$} & {$6.643^{+0.499}_{-0.427}$} & {$0.071^{+0.004}_{-0.004}$} & {256} \bigstrut\\
        &  &  & Y|X & {$1.391^{+0.075}_{-0.070}$} & {$2.892^{+0.208}_{-0.211}$} & {$0.334^{+0.019}_{-0.017}$} & {256} \bigstrut\\
        & & & & & & &\\ 
        \multirow{2}{*}{$\mathrm{S/N} \geq 4.5$} & \multirow{2}{*}{$30\ \unit{kpc^2}$} & \multirow{2}{*}{$800\ \unit{km\ s^{-1}}$} & X|Y & {$1.343^{+0.126}_{-0.119}$} & {$6.727^{+0.593}_{-0.519}$} & {$0.071^{+0.005}_{-0.004}$} & {201} \bigstrut\\
        &  &  & Y|X & {$1.520^{+0.089}_{-0.083}$} & {$2.702^{+0.233}_{-0.229}$} & {$0.325^{+0.020}_{-0.019}$} & {201}\bigstrut\\
        \hline
    \end{tabular}
\end{table*}

\begin{figure}[htbp]
    \centering
    \includegraphics[width=0.495\hsize]{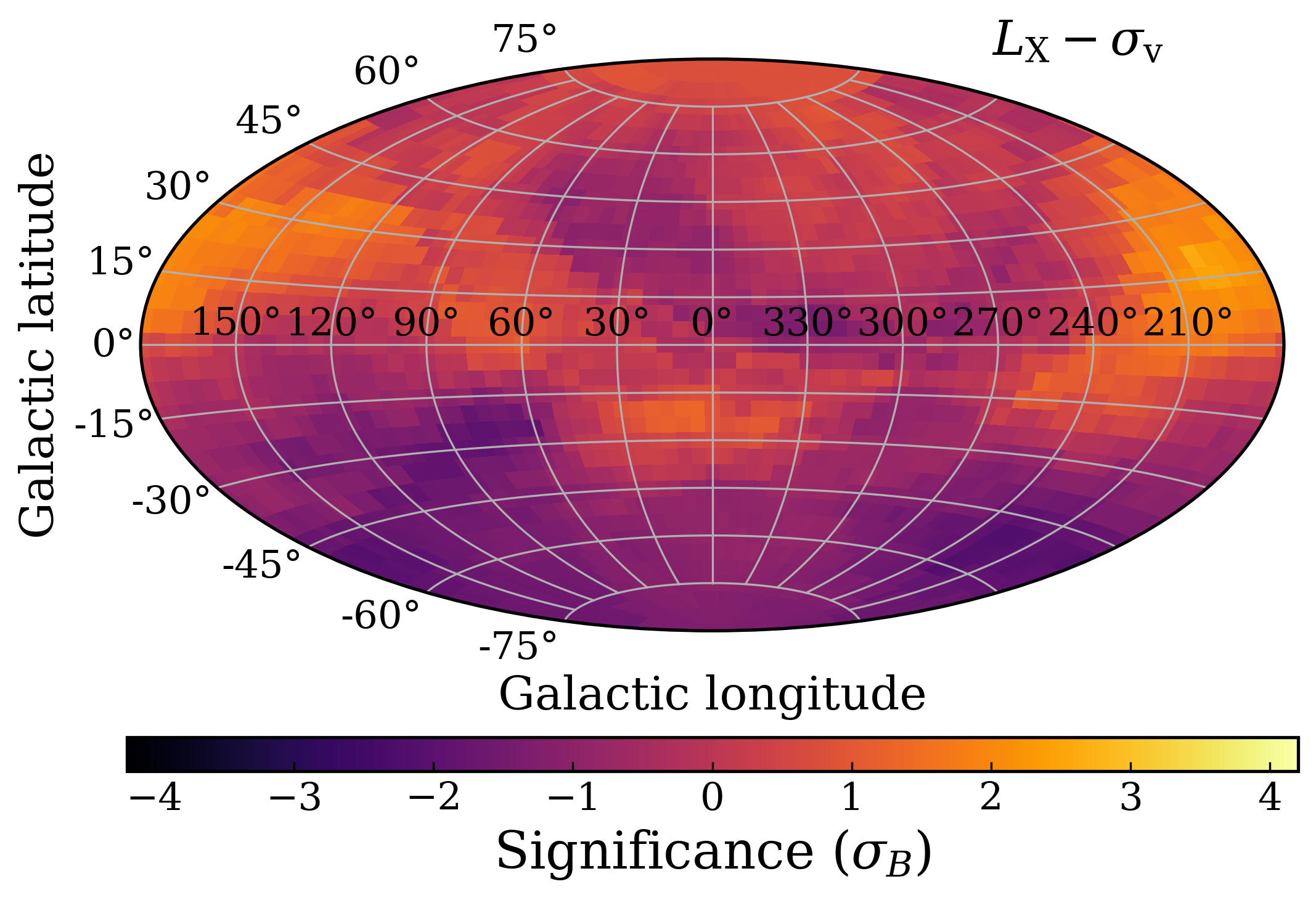}
    \includegraphics[width=0.495\hsize]{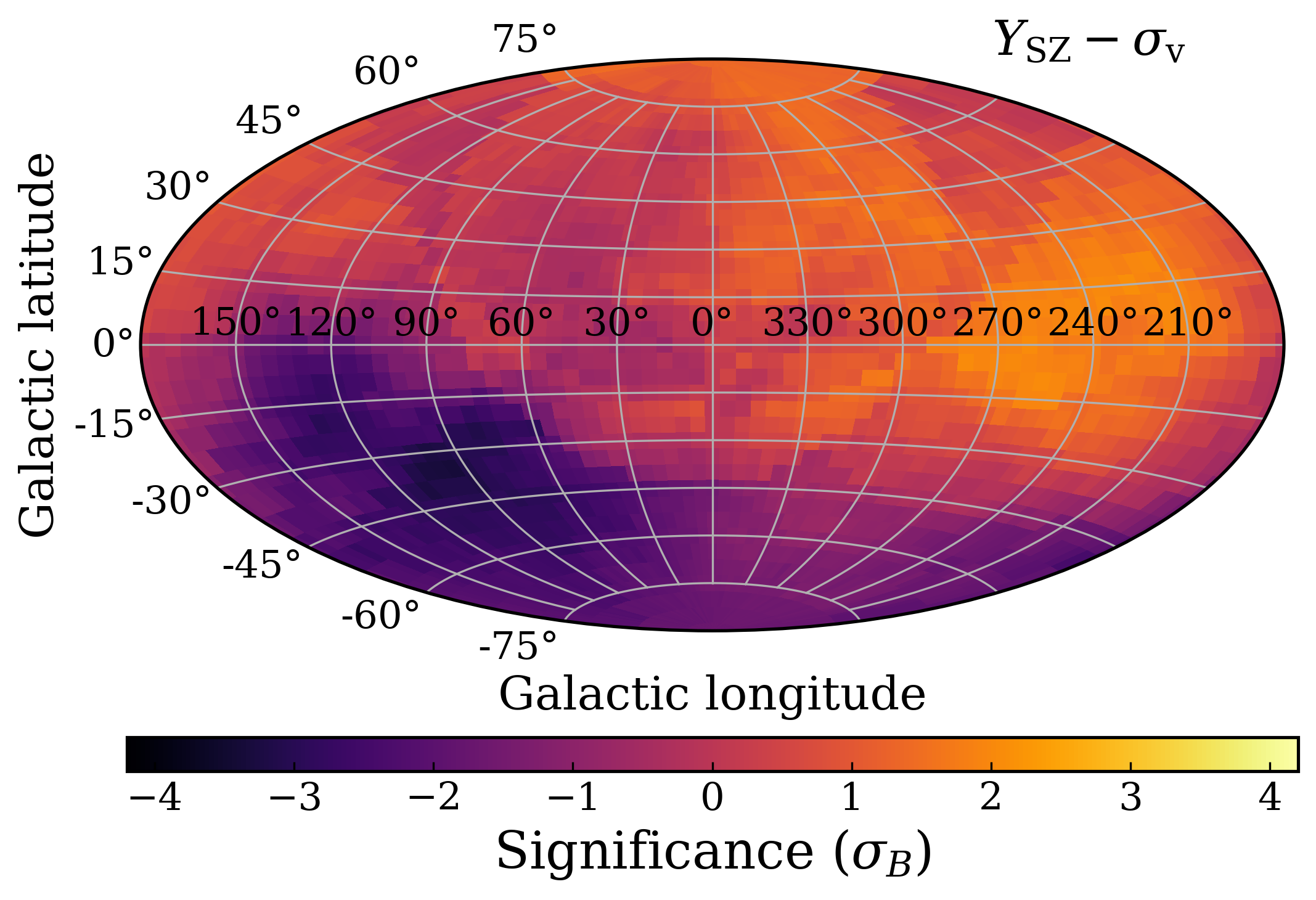}
    \caption{Significance maps of best-fit $B$ for $L_\mathrm{X}-\sigma_\mathrm{v}$ (\textit{left}), and $Y_\mathrm{SZ}-\sigma_\mathrm{v}$ (\textit{right}) relations.}
    \label{fig:slope_sigma}
\end{figure}

We observe variations in the best-fit slope in certain regions for all three relations (see Fig. \ref{fig:2d_anisotropies_Sigma-T} and Fig. \ref{fig:slope_sigma}). Although there are no strong correlations between the best-fit parameters, additional tests are conducted by keeping the slope fixed at the value obtained from the full sample.

\begin{figure}[htbp]
    \centering
    \includegraphics[width=0.495\hsize]{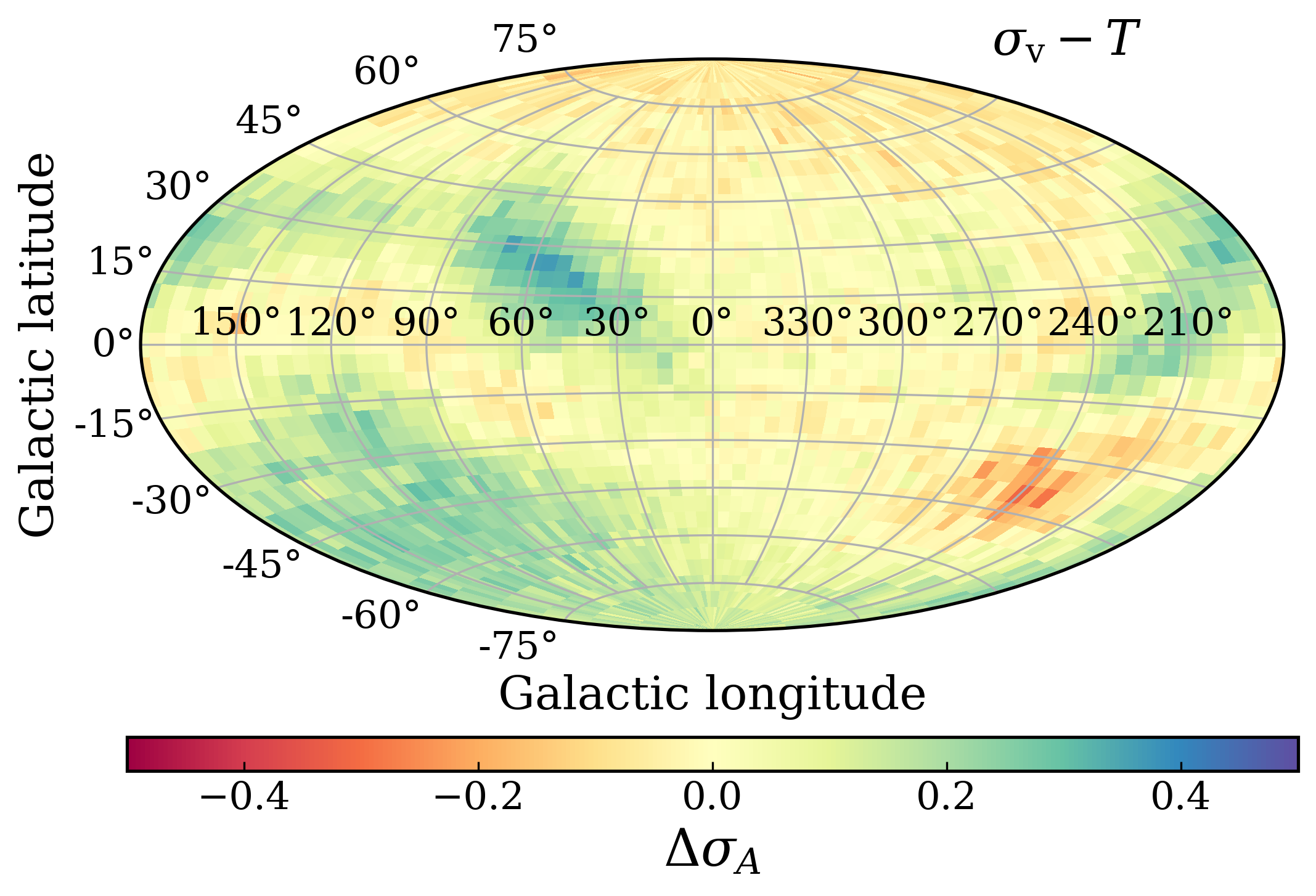}
    \includegraphics[width=0.495\hsize]{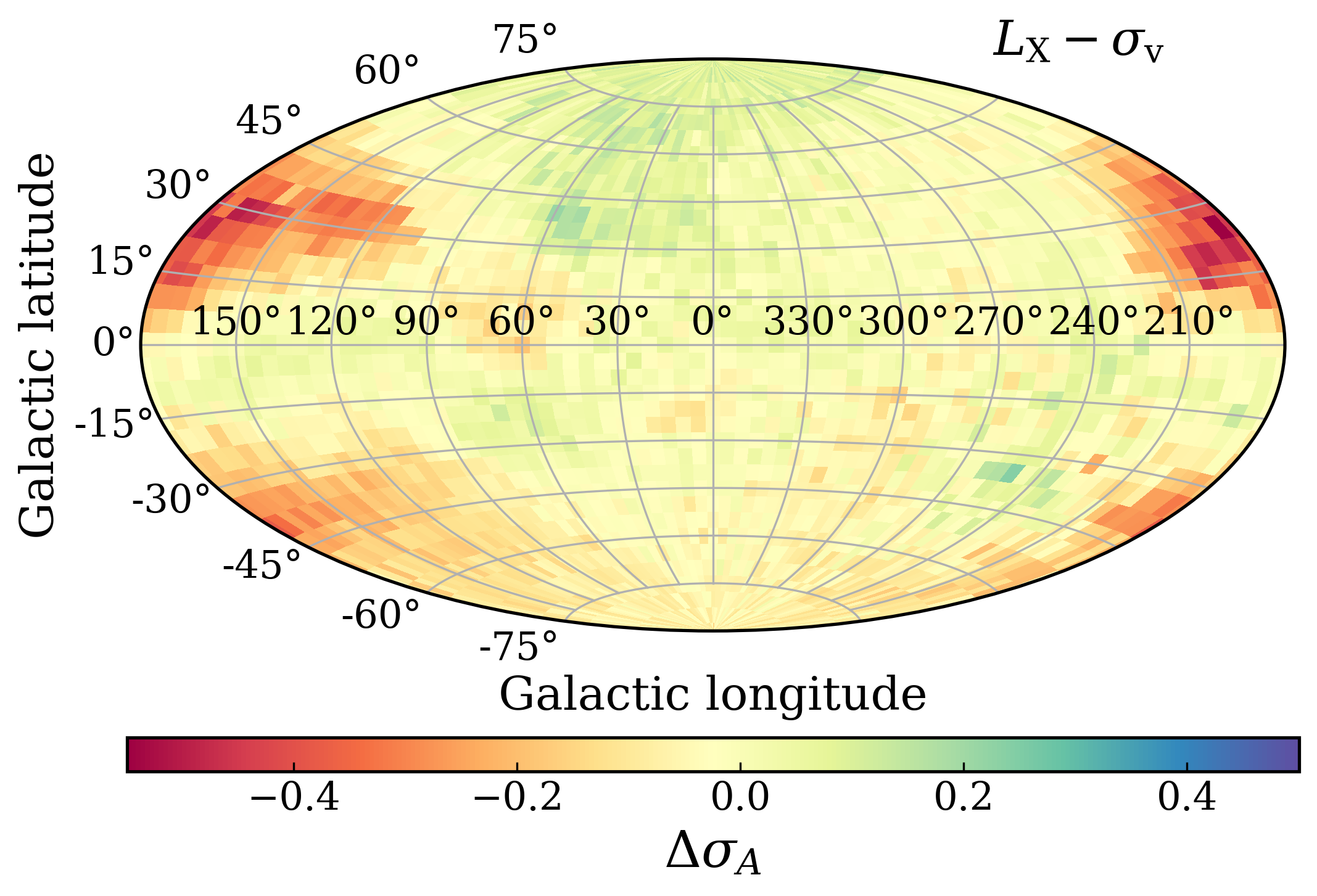}
    \includegraphics[width=0.495\hsize]{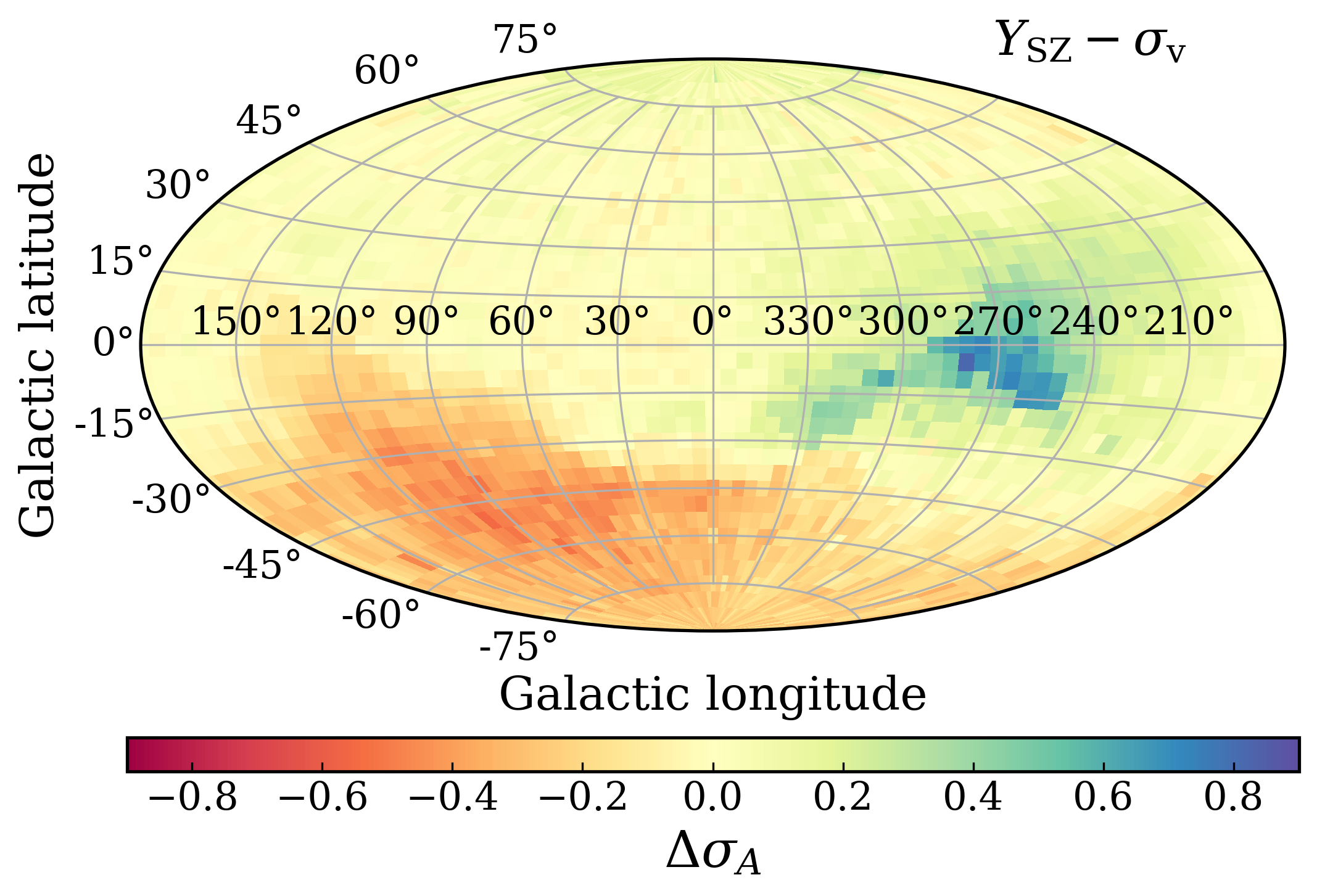}
    \caption{Difference between the significance of $A$ obtained from fixed slope and free slope analysis for the $\sigma_\mathrm{v}-T$ (\textit{top left}), $L_\mathrm{X}-\sigma_\mathrm{v}$ (\textit{top right}), and $Y_\mathrm{SZ}-\sigma_\mathrm{v}$ (\textit{bottom}) relations. Top panel maps have the same colour scale ($-0.5$ to $+0.5$), whereas the bottom panel map has a colour scale of $-0.9$ to $+0.9$ to visualise higher variations.}
    \label{fig:fixed_slope}
\end{figure}

Figure \ref{fig:fixed_slope} compares the results from free and fixed slope analysis for all three relations. We calculate $\Delta \sigma_A$ as the difference of $\sigma_A$ obtained from the fixed and free slope analysis. For the $\sigma_\mathrm{v}-T$ and $L_\mathrm{X}-\sigma_\mathrm{v}$ relations, these differences are less than $\pm 0.5 \sigma$. However, for the $Y_\mathrm{SZ}-\sigma_\mathrm{v}$, there are somewhat higher differences ($\sim 0.8 \sigma$) in certain regions. Nonetheless, the direction and magnitude of the maximum significance obtained for this relation remains unchanged. Therefore, we can infer that the variations in the best-fit slope are not a major concern as they do not impact the conclusions drawn from the analysis as expected, given the rather weak correlation observed between slope and normalisation.

\section{General behaviour of scaling relations for different datasets and fitting methods}
\label{sec:general_behaviour_extra}
The Table \ref{tab:extra_scaling_relations} presents the best-fit parameters for the relations $L_\mathrm{X}-\sigma_\mathrm{v}$ and $Y_\mathrm{SZ}-\sigma_\mathrm{v}$ obtained using different datasets and fitting methods. The best-fit parameters are consistent across the different datasets. Notably, the X|Y fitting method consistently yields a higher slope than the Y|X minimisation method for the various datasets.

\section{$z$ distributions}
\begin{figure}[htbp]
    \centering
    \includegraphics[width=0.495\hsize]{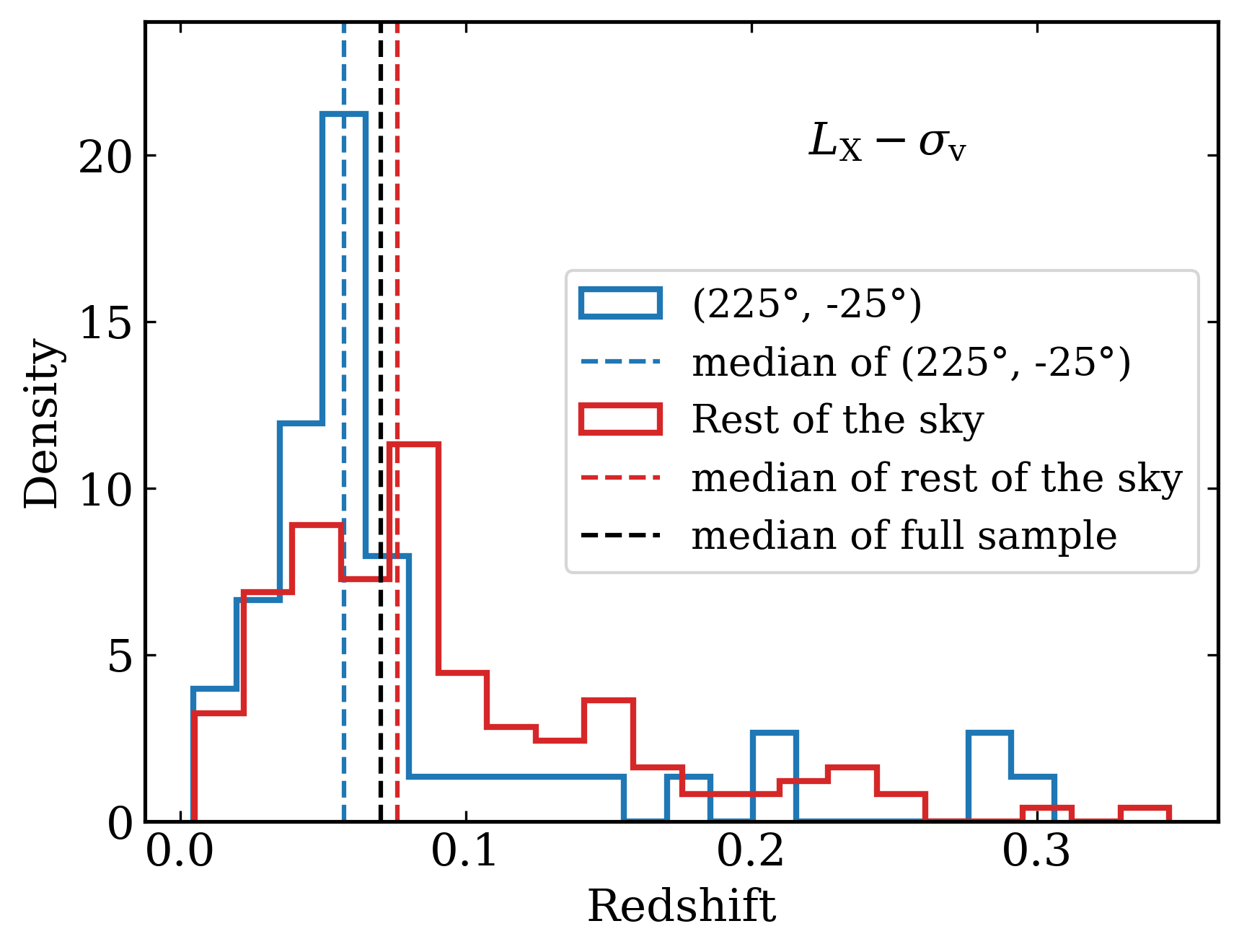}
    \includegraphics[width=0.495\hsize]{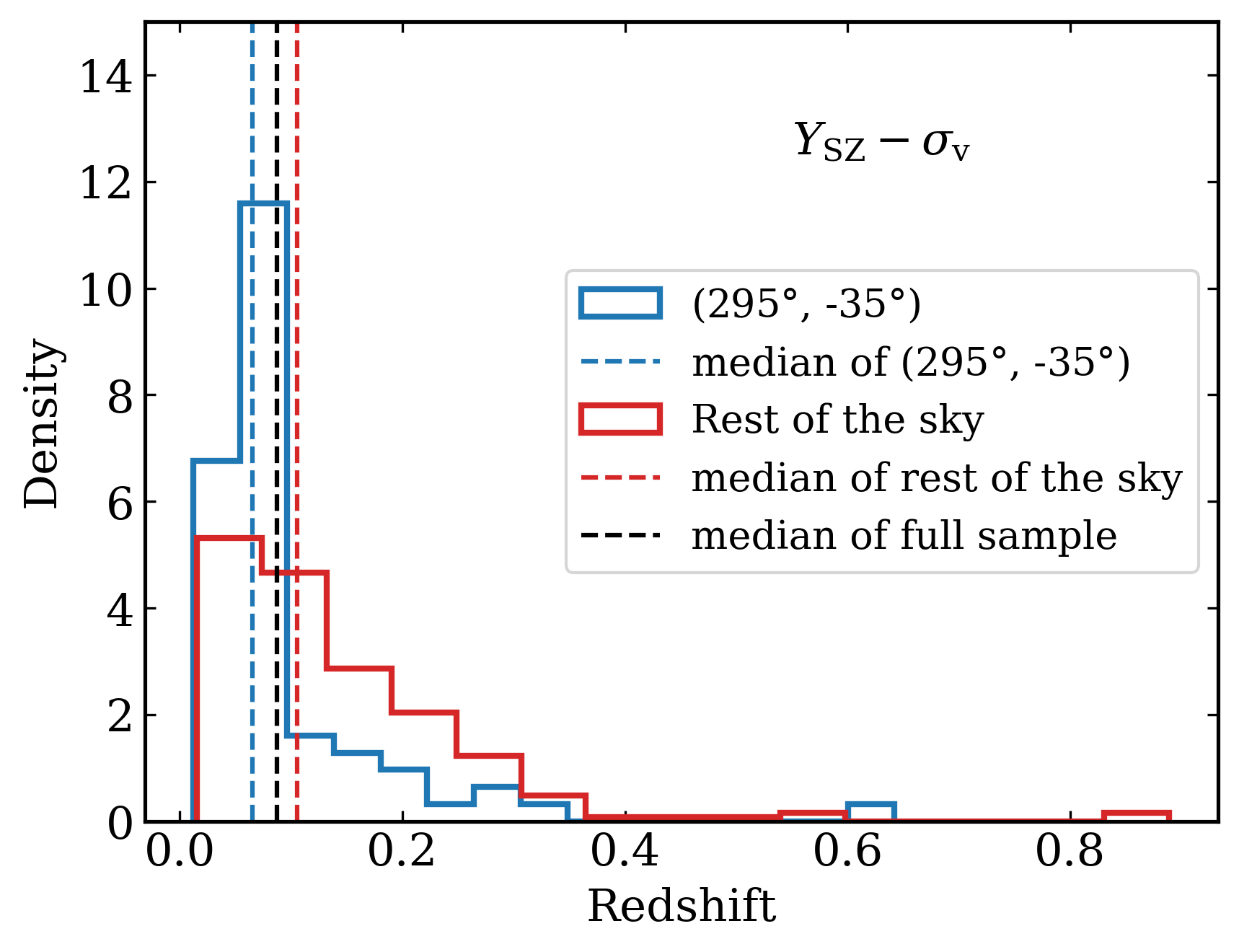}
    \caption{Distribution of redshifts for the clusters inside the cone centred at maximum anisotropy and rest of the sky for $L_\mathrm{X}-\sigma_\mathrm{v}$ (\textit{left}), and $Y_\mathrm{SZ}-\sigma_\mathrm{v}$ (\textit{right}) relations}
    \label{fig:z_dist}
\end{figure}

The Fig. \ref{fig:z_dist} shows the distribution of cluster redshift in the region of maximum anisotropy and the rest of the sky for the two relations $L_\mathrm{X}-\sigma_\mathrm{v}$ and $Y_\mathrm{SZ}-\sigma_\mathrm{v}$. We observe some differences between the distributions of the two regions. On average, clusters in the most anisotropic regions have lower $z$.

\begin{figure}[htbp]
    \centering
    \includegraphics[width=0.495\hsize]{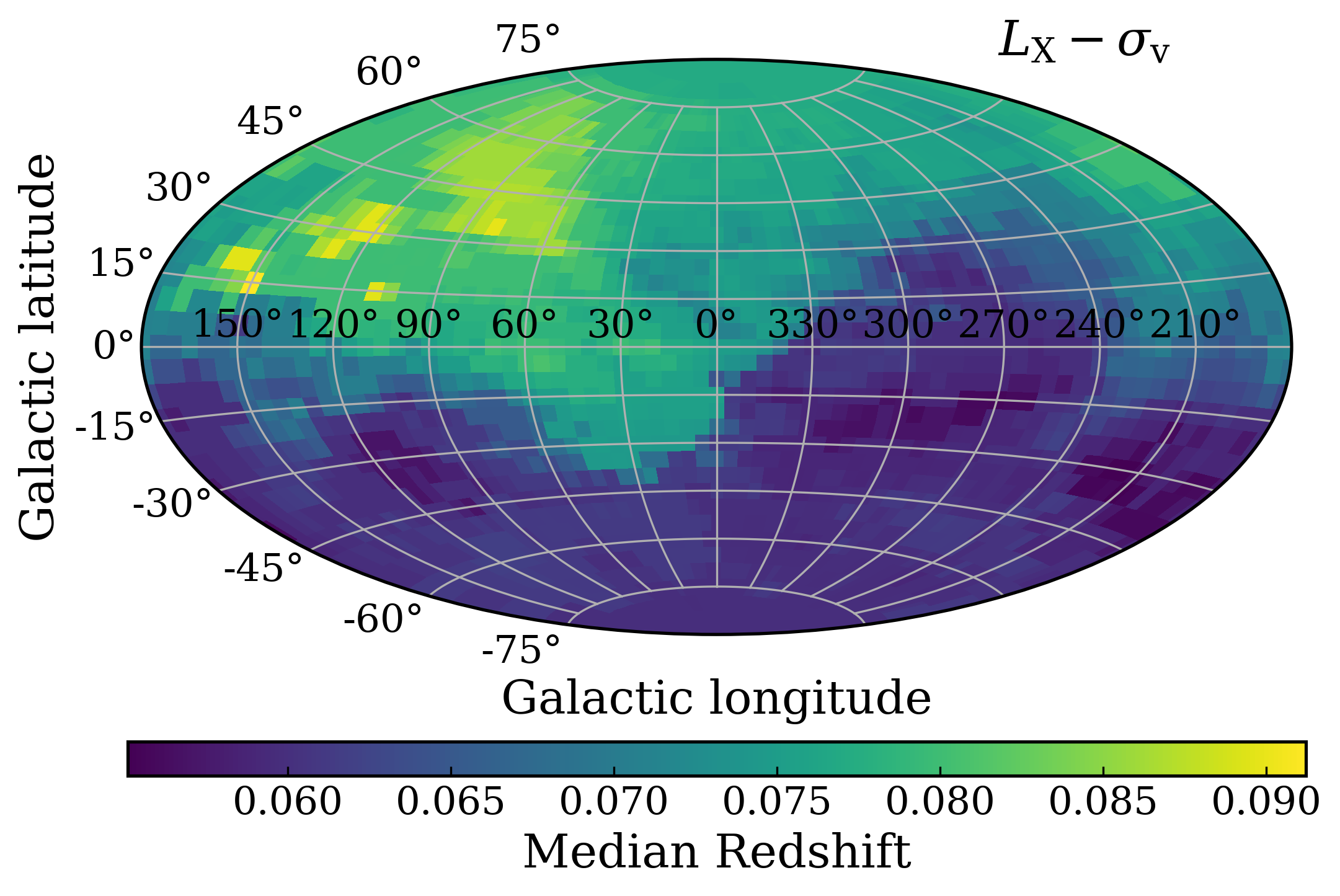}
    \includegraphics[width=0.495\hsize]{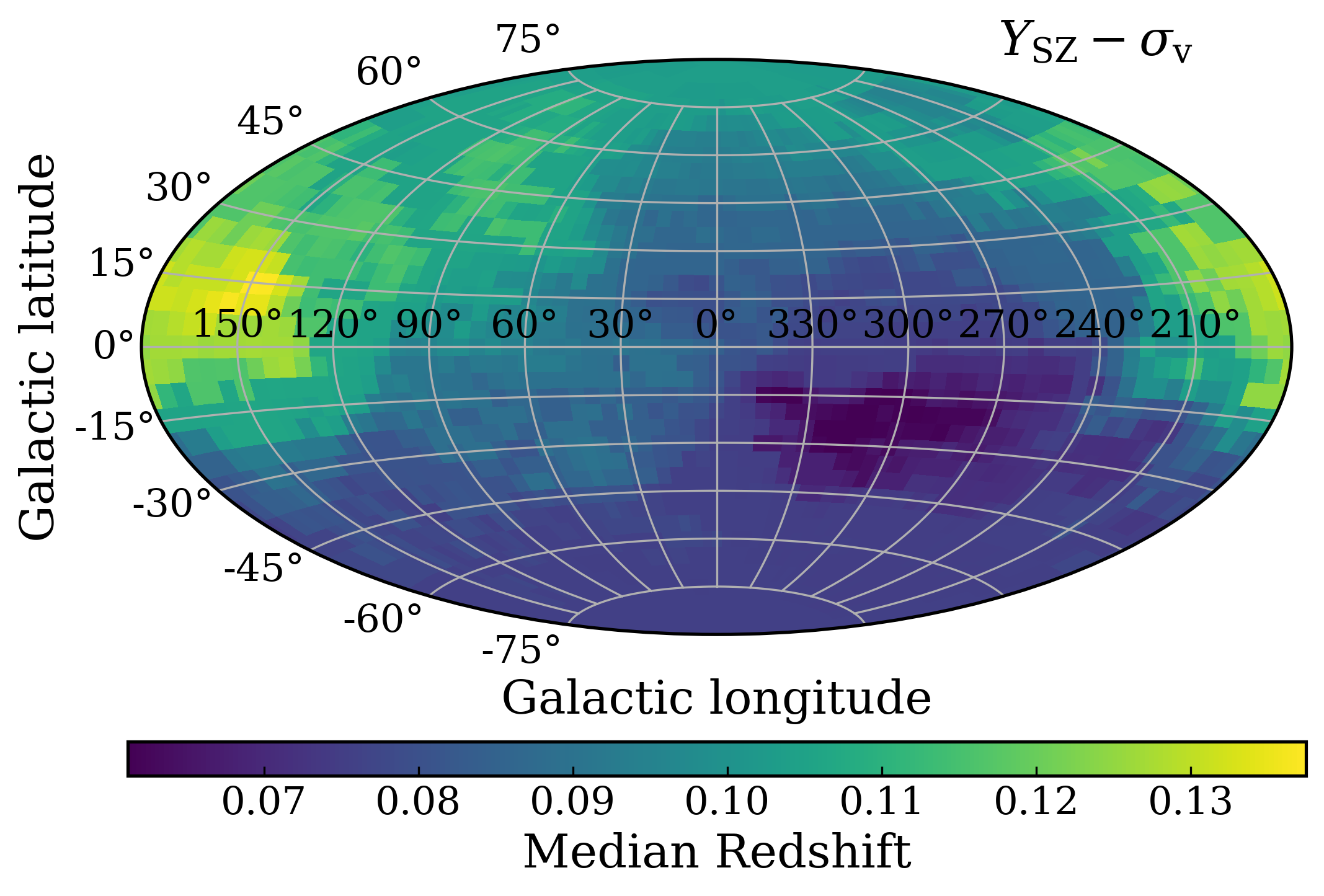}
    \caption{2-D map of median redshifts of the clusters inside a cone for $L_\mathrm{X}-\sigma_\mathrm{v}$ (\textit{left}), and $Y_\mathrm{SZ}-\sigma_\mathrm{v}$ (\textit{right}) relations}
    \label{fig:2D_z}
\end{figure}

We also create a 2-D map of the median $z$ within each cone to identify other $z$ trends in our data. The most significant trend observed is the galactic north-south divide for the $L_\mathrm{X}-\sigma_\mathrm{v}$ relation. On average, clusters are found at higher $z$ in the northern galactic sky compared to the southern half.

\section{Relation between $A$ and $H_0$}
\label{sec:A_to_H0}

The best-fit normalisation is related to the Hubble parameter, and variations in the best-fit normalisation can be converted to variations in the Hubble parameter. For the relation $L_\mathrm{X}-\sigma_\mathrm{v}$ ($L_\mathrm{X} = A\times \sigma_\mathrm{v}^B $),

\begin{equation}
    f_\mathrm{X}K_\mathrm{corr} \times 4\pi (D_\mathrm{L}(z))^2 = A\times \sigma_\mathrm{v}^B\,.
\end{equation}
Here, $f_\mathrm{X}$ is the observed X-ray flux, $K_\mathrm{corr}$ is the K correction used to convert $L_\mathrm{X}$ from the rest frame of the observer to the cluster's rest frame, and $D_\mathrm{L}(z)$ is the luminosity distance, which is given by 
\begin{equation}
    D_\mathrm{L}(z) = \frac{c}{H_0}(1+z)\int_{0}^{z}\frac{dz'}{E(z')}\,.
\end{equation}
Thus, the equation becomes
\begin{equation}
    f_\mathrm{X} K_\mathrm{corr} \times 4\pi \left(c(1+z)\int_{0}^{z}\frac{dz'}{E(z')}\right)^2 = H_0^2A\times \sigma_\mathrm{v}^B\,.
\end{equation}

This is also true for the $Y_\mathrm{SZ}-\sigma_\mathrm{v}$ relation since the quantity $Y_\mathrm{SZ}$ depends on the integrated Compton parameter $Y$ and square of the angular diameter distance $(D_\mathrm{A}(z))$. Thus, the equation $Y_\mathrm{SZ} = A\times \sigma_\mathrm{v}^B $ transforms to 
\begin{equation}
    Y \left(\frac{c}{(1+z)}\int_{0}^{z}\frac{dz'}{E(z')}\right)^2 = H_0^2A\times \sigma_\mathrm{v}^B
\end{equation}

This shows that absolute constraints can only be put on the quantity $H_0^2A$. To obtain the value of $H_0$ in different regions of the sky from $A$, equation \ref{eq:H0_conversion} is used.

\end{appendix}

\end{document}